\documentclass{aa}  

\newcommand{\rc}{\bf}
\usepackage{graphicx}
\usepackage{txfonts}
\usepackage{color,soul}
\usepackage{hyperref}
\usepackage{amstext}

\begin{document}

   \title{The {\em EXTraS} Project: Exploring the X-ray transient and variable sky}


\author{A. De Luca\inst{1,2} 
          \and
R. Salvaterra\inst{1}
\and
A. Belfiore\inst{1}
\and
S. Carpano\inst{3}
\and
D. D'Agostino\inst{4}
\and
F. Haberl\inst{3}
\and
G.~L. Israel\inst{5}
\and
D. Law-Green\inst{6}
\and
G. Lisini\inst{7}
\and
M. Marelli\inst{1}
\and
G. Novara\inst{7}
\and
A.~M. Read\inst{6}
\and
G. Rodriguez-Castillo\inst{5}
\and
S.~R. Rosen\inst{8,6}
\and
D. Salvetti\inst{1}
\and
A. Tiengo\inst{7,1,2}
\and
G. Vianello\inst{9}
\and 
M.~G. Watson\inst{6}
\and
C. Delvaux\inst{3}
\and
T. Dickens\inst{6}
\and
P. Esposito\inst{7,1}
\and
J. Greiner\inst{3}
\and
H. H\"ammerle\inst{3}
\and
A. Kreikenbohm\inst{10}
\and
S. Kreykenbohm\inst{10}
\and
M. Oertel\inst{10}
\and
D. Pizzocaro\inst{1}
\and
J.~P. Pye\inst{6}
\and
S. Sandrelli\inst{11}
\and
B. Stelzer\inst{12,13}
\and
J. Wilms\inst{10}
\and
F. Zagaria\inst{7,14}
          }

   \institute{INAF, Istituto di Astrofisica Spaziale e Fisica Cosmica Milano, via E.\ Bassini 15, I-20133 Milano, Italy
\and
INFN, Sezione di Pavia, via A. Bassi 6, I-27100 Pavia, Italy
\and
Max-Planck-Institut f\"ur extraterrestrische Physik, Gie{\ss}enbachstra{\ss}e 1, D-85748 Garching, Germany
\and
CNR, Istituto di Elettronica e di Ingegneria dell'Informazione e delle Telecomunicazioni, via de Marini 6, I-16149 Genova, Italy
\and
INAF, Osservatorio Astronomico di Roma, Via Frascati 33, I-00078 Monteporzio Catone, Italy
\and
School of Physics \& Astronomy, University of Leicester, Leicester, LE1 7RH, UK
\and
Scuola Universitaria Superiore IUSS Pavia, Piazza della Vittoria 15, 27100, Pavia, Italy
\and
European Space Astronomy Center (ESA/ESAC), Operations Department, Vilanueva de la Cañada, E-28692 Madrid, Spain
\and
Department of Physics, Stanford University, 382 via Pueblo Mall, Stanford, CA 94305-4013 USA
\and
Dr. Karl Remeis-Observatory and Erlangen Centre for Astroparticle Physics, Universit\"at Erlangen-N\"urnberg, Sternwartstr. 7, D-96049 Bamberg, Germany
\and
INAF, Osservatorio Astronomico di Brera, via Brera 28, I-20121 Milano, Italy
\and
Institut f\"ur Astronomie \& Astrophysik, Eberhard-Karls Universit\"at T\"ubingen, Sand 1, 72076, T\"ubingen, Germany
\and
INAF - Osservatorio Astronomico di Palermo, Piazza del Parlamento 1, 90134, Palermo, Italy
\and
Institute of Astronomy, University of Cambridge, Madingley Road, Cambridge CB3 0HA, UK
              }

\offprints{andrea.deluca@inaf.it}

   \date{Received ; accepted}

 
\abstract{Temporal variability in flux and spectral shape is ubiquitous in the X-ray sky and carries crucial information about the nature and emission physics of the sources. The EPIC instrument on board the XMM-Newton observatory is the most powerful tool for studying variability even in faint sources. Each day, it collects a large amount of information about hundreds of new serendipitous sources, but the resulting huge (and growing) dataset is largely unexplored in the time domain.
The project called Exploring the X-ray transient and variable sky (EXTraS) systematically extracted all temporal domain information in the XMM-Newton archive. This included a search and characterisation of variability, both periodic and aperiodic, in hundreds of thousands of sources spanning more than eight orders of magnitude in timescale and six orders of magnitude in flux, and
a search for fast transients that were missed by standard image analysis. All results, products, and software tools have been released to the community in a public archive. A science gateway has also been implemented to allow users to run the EXTraS analysis remotely on recent XMM datasets.
We give details on the new algorithms that were designed and implemented to perform all steps of EPIC data analysis, including data preparation, source and background modelling, generation of time series and power spectra, and search for and characterisation of different types of variabilities. We describe our results and products and give information about their basic statistical properties and advice on their usage. We also describe available online resources.
The EXTraS database of results and its ancillary products is a rich resource for any kind of investigation in almost all fields of astrophysics. Algorithms and lessons learnt from our project are also a very useful reference for any current and future experiment in the time domain.}



   \maketitle
%
%

\section{Introduction}
Variability pervades the cosmos. Almost all astrophysical sources, from stars in the vicinity of the solar system to supermassive black holes in the cores of very distant galaxies, have a distinctive variability. Their flux and spectral shape change within a range of timescales. This is especially the case in the high-energy range of the electromagnetic spectrum.  In the X-rays and gamma-rays, the sky is extremely dynamic and observation of a peculiar variability led to the discovery of new classes of objects, some of them completely unexpected, in the past decades.

We may mention different examples of transient or highly variable high-energy sources.
(i) Gamma-ray bursts (GRBs) are the most powerful cosmic explosions for electromagnetic output. They are likely produced by the collapse of massive stars into black holes or by the coalescence of two neutron stars.
(ii) Soft gamma-ray repeaters (SGRs) are X-ray sources that are believed to be powered by magnetars, that is, by neutron stars with the strongest magnetic fields in the Universe.
(iii) (Transient) X-ray binaries are black holes, neutron stars, or white dwarfs that accrete matter from their stellar companion.
(iv) Stellar flares are X-ray flares from magnetically active, late-type stars that are either isolated or in binary systems.
(v) Blazar flares are gamma-ray flares that are produced by the jets of supermassive black holes at the centres of galaxies.
(vi) Tidal disruption events are the gravitational capture and disruption of a star by a supermassive black hole.
(vii) Supernova X-ray flashes are produced by the supernova shock that emerges from the exploding star.

Crucial information is often carried by periodic variability that arises from the rotation of a (compact) star or from the orbital motion in a binary system. Examples of high-energy pulsators are
(i) spinning up and down, accreting, magnetic neutron stars in binary systems;
(ii) spinning down young neutron stars, whose emission  is powered by the dissipation of rotational, thermal, or even magnetic energy, as in the cases of classical radio pulsars, the so-called Magnificent Seven neutron stars \citep{haberl2007}, and magnetars;
(iii) accreting magnetic white dwarf systems, such as polars and intermediate polars;
(iv) orbital modulations (including periodic dips and eclipses) of the X-ray flux in various classes of X-ray binaries with accreting neutron stars, black holes, or white dwarfs (especially if seen from a high inclination).

Variability is key to understanding the nature and physics of the sources. It is plainly impossible to summarise the range of science topics in a few lines that can be accessed and addressed by time-domain investigations in the X-ray range. X-ray variability yields unique insights into accretion physics (e.g. radiation efficiency of accretion flows, mechanisms for generating winds and jets) and strong gravity physics (e.g. conditions in the inner disk) through observations of 
active galactic nuclei, tidal disruption events, and gamma-ray bursts (marking the birth of a black hole). We can learn about the mechanisms of massive star explosions, and about the progenitors of supernovae, by observing supernova shock breakout events (which would also enable more sensitive searches for the long sought-after associated gravitational waves and neutrinos). X-ray variability allows us to focus on the physics of magnetic field generation and dynamics in compact objects (e.g. through observations of violent and less violent events related to the extreme magnetic fields of magnetars) and in normal stars (observation of stellar flares and coronal emission). The latter point holds great promise for our understanding of planetary system formation and evolution (the effects of flares on protoplanetary disks and on the habitability of planetary systems), and for understanding our own Sun.

Most of the variable phenomena described above have been discovered with instruments with a large field of view (FoV)
such as the All-Sky Monitor (ASM) on board the Rossi X-ray Timing Explorer, the Imager on Board the INTEGRAL Satellite (INTEGRAL/IBIS), the Burst Alert Telescope (BAT) on board the Neil Gehrels Swift observatory, and the Monitor of All-sky X-ray Image (MAXI) on the International Space Station,  which, constantly observing large fractions of the sky, can also detect relatively rare events. In the soft X-ray energy range (0.2-12 keV), focusing telescopes are much more sensitive than wide-field instruments. The current generation of space observatories each day collect a very large amount of data about serendipitous sources located within their FoV, including a huge amount of information regarding their variability. Data archives from these telescopes have great potential for studying variability of (serendipitous) X-ray sources, which in principle is only limited by photon statistics and by the intrinsic time resolution of the instruments. However, this information remains mostly unused.

In particular, the European Photon Imaging Camera (EPIC) instrument  on board the European Space Agency mission XMM-Newton \citep{Jansen2001}, consisting of two MOS cameras \citep{Turner2001} and of a pn detector \citep{Strueder2001}, is the most powerful tool for studying the variability of faint X-ray sources because the combination of large effective area, good angular, spectral, and temporal resolution, and large FoV is unprecedented. More than 20 years after its launch, EPIC is still fully operational, and its very rich archive of data continues to grow. The serendipitous content in XMM data is being explored within large projects. The catalogue of serendipitous sources extracted from EPIC observations
is indeed the largest and most sensitive compilation of X-ray sources ever produced before the realisation of the eROSITA all-sky survey\footnote{See \url{https://www.mpe.mpg.de/eROSITA}. See also \citet{predehl21}.}.
Its most recent release (2019 December) at the time of drafting this paper, dubbed 4XMM-DR9\footnote{\url{http://xmmssc.irap.omp.eu/Catalogue/4XMM-DR9/4XMM_DR9.html}} \citep{webb20},
lists more than 810,000 detections of more than 550,000 unique sources over more than 1,150 square degrees of the sky. The median flux of these sources is $\sim 5.3\times10^{-15}$ erg cm$^{-2}$ s$^{-1}$ and $\sim 1.2 \times 10^{-14}$ erg cm$^{-2}$ s$^{-1}$ in the 0.5-2 keV and 2-12 keV energy ranges, respectively.

About $\sim20,000$ sources have been detected in the so-called XMM Slew Survey \citep[XSS,][]{Saxton2008}, using data that were collected during manoeuvres to reorient the telescope between targets.
The data have a shallower sensitivity, but cover more than 70\% of the sky. The XSS provides significantly better sensitivity (limiting flux $\sim 3\times10^{-12}$ erg cm$^{-2}$ s$^{-1}$) than any all-sky survey currently available to the community. In the soft 0.2-2 keV band, the XSS is almost as sensitive (limiting flux $\sim6\times10^{-13}$ erg cm$^{-2}$ s$^{-1}$) as the ROSAT All-Sky Survey (RASS). 

The time-domain information on such a rich sample of sources remains largely unexplored. 
The 4XMM catalogue incorporates light curves of the top $\sim 36\%$ brightest sources. These light curves are generated with a time bin of 20 times the frame time for the pn camera (resulting in time binning at 1.46 s in most cases), or with a time bin yielding at least (on average) 20 counts per bin, with a minimum bin time (for bright sources) of 10 s for the MOS cameras. A simple test for time variability (a $\chi^2$ test) is automatically performed on these light curves (pn light curves are rebinned at this stage to have at least 20 counts per bin), and a variability flag is assigned. A catalogue from stacked data (4XMM-DR9s) is also generated for overlapping observations, providing information on the long-term variability of sources between different detections.
Systematic investigations of variability are not carried out by the catalogue team. 
The XSS (and new slew data, which are routinely collected) provides the best opportunity at present,
compared to the RASS,  for discovering extremely rare high-variability objects. A number of such objects (novae, tidal disruption events, etc.) have indeed been selected \citep[e.g.][]{Saxton2012}. However, no systematic dedicated study and cataloguing of the variability has yet been performed.

We describe in this paper the main features of the  project called Exploring the X-ray variable and transient sky (EXTraS),  which was carried out in 2014-2016. It produced the most thorough investigation of temporal properties of XMM-Newton and EPIC sources ever performed. 
All results and products of EXTraS have been available since the end of the project through a public data archive, which describes the variability of more than 400,000 sources spanning more than eight orders of magnitude in timescale and six orders of magnitude in flux. Applications range from the search for rare events to population studies, with an impact on the study of virtually all astrophysical source classes. 

The paper is organized as follows: In Sect.~\ref{sect:extras} we give a concise overview of the EXTraS project, and in Sect.~\ref{sect:stv}--\ref{sect:ltv} we describe details of the EPIC data analysis that was carried out in different research lines. We describe new algorithms that were designed and implemented within the project to 
deal with
the peculiar highly variable background noise of the EPIC instrument, and to search for and characterise different types of variability. We also report details of our main products and results, including basic statistical properties and advice for their usage. In Sect.~\ref{sect:resources} we describe the web resources that were made available to the community. 
In Sect.\ref{sect:summary} we briefly summarise. Appendices include further details of the data analysis and products. 

%

%

\section{The EXTraS project}\label{sect:extras}

The EXTraS\footnote{EXTraS (Exploring the X-ray Transient and variable Sky) is a collaborative effort of six European partners: Istituto Nazionale di Astrofisica (INAF, Italy, coordinator); Scuola Universitaria Superiore IUSS Pavia (Italy), Consiglio Nazionale delle Ricerche (CNR, Italy); University of Leicester (UK); Max Planck Gesellschaft zur Foerderung der Wissenschaften - Max Planck Institut f\"ur extraterrestrische Physik (MPG-MPE, Germany); Friedrich-Alexander Universitat Erlangen-Nuremberg - Erlangen Center for Astroparticle Physics (ECAP, Germany). EXTraS was funded (2014-2016) by the European Union within the Seventh Framework Programme (FP7-Space). See the project web site \url{www.extras-fp7.eu} for further details on the team and contact information.}
 project was aimed at fully investigating and disclosing the serendipitous content of the EPIC database in the time domain and to make it available and easy to use to the whole community. 
 EXTraS includes four different lines of EPIC data analysis:

\begin{enumerate}
\item {\bf Short-term, aperiodic variability (STV)}, aimed at detecting and  characterising aperiodic variability in the largest possible number of sources from the XMM serendipitous source catalogue on all timescales ranging from the instrument time resolution to the duration of an observation (see Sect.~\ref{sect:stv}).
\item {\bf Search for coherent pulsations}, aimed at detecting and characterising the largest possible number of X-ray pulsators in a period range from $\sim0.2$ s up to the highest value allowed by the duration of the observation (see Sect.~\ref{sect:pulsators}).
\item {\bf Search for transients}, aimed at detecting  the largest possible sample of new, faint X-ray transients. These sources are only above detection threshold for a very short time interval and thus are missed by standard image analysis and are not listed in the XMM serendipitous source catalogue (see Sect.~\ref{sec:transients}).
\item {\bf Long-term variability (LTV)}, aimed at detecting and characterising long-term variability, taking advantage of the large number of overlapping observations performed at different epochs, using both pointed and slew data, combining detections and upper limits in long-term light curves spanning up to 15 years (see Sect.~\ref{sect:ltv}).
\end{enumerate}

All EXTraS products and results together with new software tools have been released to the community in 2017 March through a public archive (see Sect.~\ref{sect:resources}). This includes (i) a database of all results, describing temporal properties of $\sim400,000$ EPIC sources on timescales ranging from $\sim0.1$ s to $\sim10$ years and in flux ranges spanning from $\sim10^{-9}$ to $\sim10^{-15}$ erg cm$^{-2}$ s$^{-1}$ in the 0.2-10 keV energy range, and (ii) about $\sim20$ millions of ancillary files (light curves, hardness ratios, power spectra, etc.). A science gateway was also implemented (see Sect.~\ref{sect:resources}) to allow users to run EXTraS pipelines on any dataset from the XMM Science Archive. 

As a part of the project, multiwavelength characterisation of sources based on available catalogues and phenomenological classification of sources using machine-learning algorithms were also implemented. These activities are not described in this paper, which focus on EPIC data analysis. We refer to \citet{gatuzz18} for details.

%
%

\section{Short-term, aperiodic variability (STV)}\label{sect:stv}
\subsection{Aims and scope}
The goal is to provide users with a thorough characterisation of any type of short-term variability, ideally, on all timescales ranging from the instrument time resolution to the duration of an observation for the largest possible number of sources included in the XMM-Newton serendipitous source catalogue. This extends the basic temporal analysis of  bright sources included in the production of the XMM catalogue in several ways: (i) we study a  larger fraction of sources, down to much fainter fluxes, (ii) we use all EPIC data, including time intervals affected by soft proton flares, (iii) we study variability at the shortest timescales even in faint sources, overcoming limitations of uniformly binned time curves with large bins, (iv) we perform an energy-resolved analysis, and we also study spectral variability, and finally, (v) we compute a full set of quantitative parameters to describe variability patterns and properties. 

Our analysis builds on the 3XMM-DR4 source catalogue, which is the most recent release of the XMM serendipitous source catalogue available at the start of the EXTraS project. It includes 7437 observations performed between 2000 February and 2012 December.  We excluded 420 observations collected in mosaic mode because processing pipeline subsystem (PPS) products (see next section) are not available. Our analysis is performed for each camera and for each exposure separately. Multiple exposures collected within a specific observation by a specific camera 
are studied independently. Following the 3XMM selection, we considered only exposures taken in imaging mode and discarded those taken in small window by the pn camera.  The small field of view precludes our approach for the characterisation of the background.

\subsection{Data preparation and filtering}\label{sect:dataprep}
For the MOS cameras, we used event files from the  PPS products. For the pn camera, we were faced with a known bug in the pipeline used to generate the PPS products, in which improper management of counting mode occurrences can result in incoherent time tagging of events within an exposure, preventing a consistent temporal analysis. 
This problem affected the data sets of observations in the PPS archive that were
processed with the XMM-Newton Science Analysis Software (SAS) versions earlier
than 13.5 (see Appendix \ref{appendix:SAS_issue}).
We reprocessed all pn data starting from observation data files (ODF) using SAS v14.0, where the issue had been fixed (PPS files in the current archive should be free from this problem as a result of the recent bulk reprocessing of data performed in 2019 December).

We selected good events by applying the same quality filters as were used for the production of the 3XMM catalogue (e.g. we excluded time periods with an attitude change $>$3'). As an important difference, we also considered time intervals affected by high particle background, which are generally discarded in 3XMM processing. This resulted in our recovering a major fraction of XMM-Newton exposure time, more than 20\%, for scientific exploitation.

We selected photons in the 0.2--12 keV energy range. An energy-resolved analysis in the 0.2--1 keV (super-soft), 1--2 keV (soft) and 2--12 keV (hard) energy ranges was also performed, as described in Sect \ref{sect:energyresolved}. We considered all the flags as in 3XMM-DR4. 
Barycentric corrections were applied to all events and GTIs using the SAS  task \texttt{barycen}\footnote{We rely on the JPL DE405 planetary ephemeris, see \url{http://iau-comm4.jpl.nasa.gov/README}
}.


\subsection{Selection of 3XMM sources}\label{sect:selection}

We only considered point-like sources, excluding all those marked as possibly extended by the 3XMM analysis (3XMM parameter EP\_EXTENT\_ML$>$4 and extension larger than $12''$,  rejecting 52168 out of 531261 3XMM detections, corresponding to 9.8\%). This choice is aimed at preserving uniformity of the analysis. Extended sources require a different background treatment. We also excluded from the analysis all sources below a minimum number of ten expected source events per  camera  (3XMM parameter  $PN\_8\_CTS>10$,  $M1\_8\_CTS>10$,  $M2\_8\_CTS>10$). This left 418,387 source detections (81.6\% of detections in 3XMM-DR4). A further selection was made at a later stage based on the number of actually observed events in the optimised source region for each specific exposure, camera, and energy band under analysis (see Sect\ref{sect:sourceregion}).
To identify selected 3XMM sources at the single exposure and camera level, we cross-correlated PPS source lists with the catalogue. 


\subsection{Source regions}\label{sect:sourceregion}

For each source, we optimised a circular extraction region. As a figure of merit, we used the signal-to-noise ratio (S/N) according to the following definition: 

\begin{equation*}
S/N(r) = \frac{E[src|r]}{\sqrt{E[src+bkg|r]}} \sim 
\frac{E[src|r]}{\sqrt{max(C_0, E[src|r], O[src+bkg|r]}},
\end{equation*}

where $E[X]$ represents the expectation value of the quantity $X$ and $O[X]$ is the value that is observed. In this case, X is the number of counts from the source ($src$) or from background ($bkg$), including the leakage from other sources, in the circle defined by radius $r$.  $C_0$ is a small constant (10$^{-6}$, chosen to be much smaller than the other terms in any case) that is introduced to numerically manage the cases where no events are expected or observed.  
This approach allows us to compute the S/N without having to model the background in advance. 
Selecting the maximum value in the set of three quantities in the square root in the right-hand term of the equation takes care of cases where very few counts are expected and none are observed. 
Expected counts from the source as a function of the extraction radius were computed based on the information provided by 3XMM. Assuming the 3XMM count rate in the overall energy range (band 8, from 0.2 to 12 keV),  we produced a map at the source position, for which we multiplied the instrument point-spread function (PSF), which is described by its King function parametrisation encoded in the CCF, by the CCD-dependent exposure map (computed using the SAS task eexpmap). A resolution of $0\farcs05$ was used to properly account for the effect of bad pixels and columns and of CCD borders. The observed counts as a function of the radius were directly evaluated from the cleaned event file. 

As a first step, the source extraction radius was optimised according to our figure of merit. Then we screened all nearby sources (within $5'$) that might  contaminate our source region. For each of these sources, we optimised an exclusion radius according to our figure of merit. These steps were iterated. First we refined $r$ and then the excluded region for each contaminating source, until the maximum of our figure of merit was reached. Last, we counted how many counts from the source were left in the resulting region. All sources with fewer than ten photons were not considered any further in our analysis.

\subsection{Background modelling}
\label{sec:bkgmod}
The background noise of the EPIC cameras is the sum of different components with different spatial and temporal properties. A proper treatment of this background is of paramount importance for characterising the variability of faint sources, especially during high-background periods. We implemented a new approach that is substantially different from common practice in EPIC data analysis.

In our analysis, we considered as background anything that was not listed as a point source (with 3XMM extension parameter $SC\_EXTENT$ smaller than $12''$) in 3XMM: extended sources, unresolved sources, and cosmic X-ray background and instrumental background (particle-induced and electronic noise). It has been common practice in X-ray imaging studies to extract the background from a background region that was independent of the source region, but had supposedly similar background properties. However, the photon background, which in our analysis includes extended sources, is far from flat; moreover, the particle-induced background, including soft protons, has a different vignetting with respect to the photon component \citep[e.g.][]{Kuntz2008}. Therefore we decided to model the background over the entire FoV to deduce its properties in the source region. 
We adopted a heuristic approach, considering the overall background as the sum of two components: one variable as a function of the time, and the other constant. Each component was assumed to have its own spatial distribution that is not known {\em \textup{a priori}} and was assumed not to vary in shape within a single exposure. 

To produce a model for the steady background component, we proceeded as follows: (i) We adopted the definition of good time intervals (GTIs) for the non-flaring background that is used by 3XMM.  (ii) We generated a raw counts map by applying 3XMM GTIs to the event file. (iii) We removed point-like sources by excluding circular regions centred on their positions.\ To do this, we adopted a cut-out surface brightness level of 0.05 cts/square arcsec.
(iv) We extended the map to the whole FoV. This operation does not rely on standard spline-fitting algorithms because they often incur  large systematics at the edges of the map (CCD edges, borders of the FoV) and for low statistics. Instead we smoothed the map by preserving the overall normalisation and filled the holes at the positions of removed sources by 2D linear interpolation. (v) Finally, the resulting map was divided by the exposure map (all exposure information was taken into account on a CCD-by-CCD basis.). Points (i) to (v) were repeated using the simulated image obtained through source modelling (see Sect.\ref{sect:sourceregion}).
Then, this was subtracted from the map of the steady background component to subtract the tails of the PSF. 

To produce the model for the variable background, we extracted a raw counts map by applying bad time intervals (i.e. complementary to GTIs in the exposure) to the event file and then repeated steps (iii), (iv), and (v) as above, and PSF tail subtraction. The resulting map includes the variable background component and the steady component (which is by definition always present).  To produce a map of the variable component alone, we then subtracted an exposure-rescaled version of the steady background map. 

Using source models together with the two background maps, we can recover the map of counts we expect for the entire exposure. 
We verified that the residuals obtained by subtracting the actually observed counts and normalising by the square root of the expected counts are distributed like a Gaussian.

\subsection{Background region}
\label{sec:bkgreg}
We define as a background region the entire detector. From this, we cut out optimised circles around sources.
As a figure of merit, we used the error bar we would obtain on an estimate of the background,

\begin{equation*}
FoM = - \frac{\epsilon\,C(x) + \sqrt{B(x)}}{B(x)},
\end{equation*}

where $C(x)$ is the overall expected number of source photons leaking into the background region, $x$ is the maximum number of leaked photons per source, and $B(x)$ is the number of expected photons in the background region. According to this definition, the error bar has two components: a statistical one due to the Poisson fluctuations, and a systematic one due to the leakage from sources into the background itself. The two components are combined linearly  
through a factor $\epsilon$ that weights their contributions. 
Setting $\epsilon$ to 0 would ignore source leakage and consider the entire detector as background. Setting $\epsilon$ to 1 would instead ignore statistical uncertainties and exclude all sources out to 5 arcmin (for technical reasons, we assumed that all the photons from a source fall within this distance, although this is not the case for XMM). We calibrated the value of $\epsilon$ in order to balance the need of minimising leakage from sources into the background and the risk of increasing Poisson uncertainties on the background in crowded fields by running tests on a set of 200 exposures (including~10,000 detections with a large variety in FoV content and background level). The most robust behaviour is obtained when $\epsilon=0.5.$ With this choice, the background has enough statistics (B$>$9000) in all cases, it is a good representation of the detector background (B/B$_{tot}>$0.15, where B$_{tot}$ is B evaluated over the entire detector), and the source contamination is minimal (C/B$<$0.06).

Optimisation of the figure of merit was obtained as follows.
The number of leaked photons as a function of the exclusion radius was computed for each source based on source models. The exclusion radius for each source ranged from 0 to 5 arcmin. By construction, we required 
an equal number of leaked photons for each source ($x$), which yielded a set of radii that  correspond to an overall source photon leakage.  Background counts were estimated based on background maps. Minimisation of the figure of merit as a function of the collective leakage of photons from sources yields the optimised background region.


\subsection{Light curves with uniform time binning}\label{sect:uniformbin}
Events were selected from the optimised source region, and a raw light curve was generated with uniform time binning. We generated a background light curve from the optimised background region with the same bins. Then we exploited our knowledge of the spatial (background maps) and temporal (background light curve) background distributions to predict the constant and variable background contributions inside the source region. The counts expected from each component were corrected for GTIs on a CCD-per-CCD basis. Source counts were then corrected for the PSF tails outside the extraction region, and for spatial vignetting. For each exposure and camera, for all detections passing the filter described above, a background-subtracted light curve was produced with a bin time of 500 s, 5000 s, and optimal uniform binning, which is a source-specific binning with (on average) at least 25 counts per bin (enough for the counts to approximate a Gaussian distribution). If a source is expected to produce fewer than 50 net counts, an optimal bin light curve was generated with two bins. To limit the number of bins for the brightest sources, the optimal bin size was always larger than 5s. We also produced light curves with 10 s bin size. These are not released, but were used 
as an input for the analysis in the frequency domain
(see Sect.\ref{sect:FFT}). 

The error bars were obtained by propagating Poisson uncertainties in the expected background and source components in each time bin. 
In particular, because  the background accounts for all or almost all the observed counts, we cannot assume that the observed
excess counts $x$ coincide with the expected excess counts $\mu$, otherwise the Poisson uncertainty would be null. 
Instead, we assumed that $\mu = 0.375 + max(x,0)$, and the associated uncertainty $dx = \sqrt{\mu} = \sqrt{0.375 + max(x,0)}$.
This solution to the Poisson bias is intermediate between the standard assumption $dx=\sqrt{x}$ and that introduced by \citet{mighell99}
$dx=max(\sqrt{x}, 1)$. See also \citet{anscombe48}.

The cumulative distribution of the rates, that is, the fraction of time spent by the source below a fixed rate, as a function of the rate itself was also computed for each light curve as a histogram with error bars, with a step along the y axis (Fractional time) for each bin in the original light curve.

\begin{figure*}[ht]
\centering
\includegraphics[height=5cm]{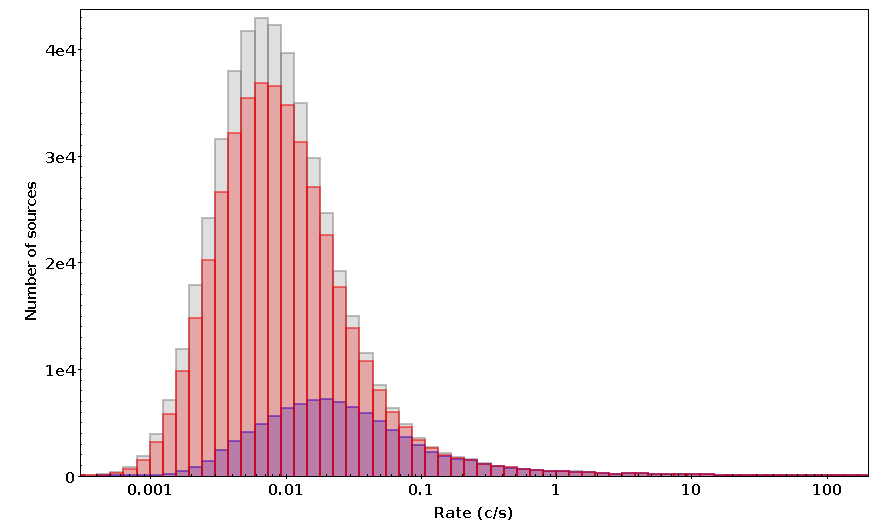} 
\includegraphics[height=5cm]{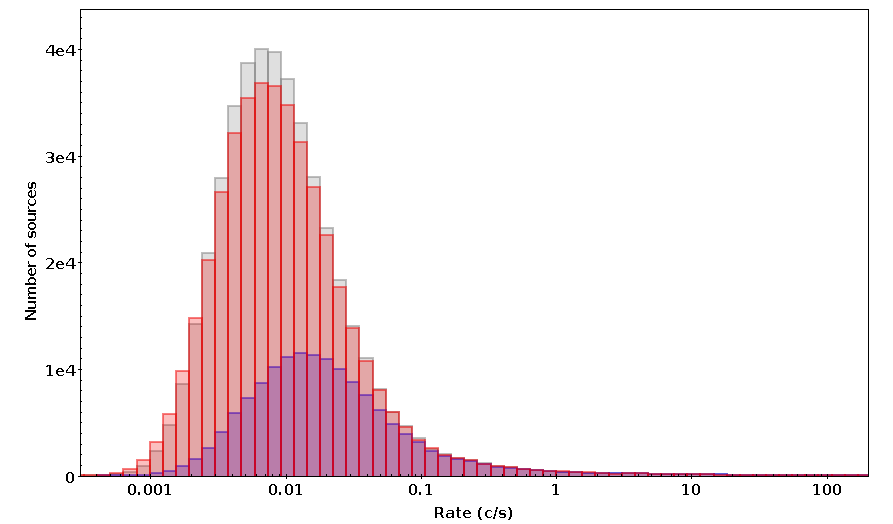} 
\includegraphics[height=5cm]{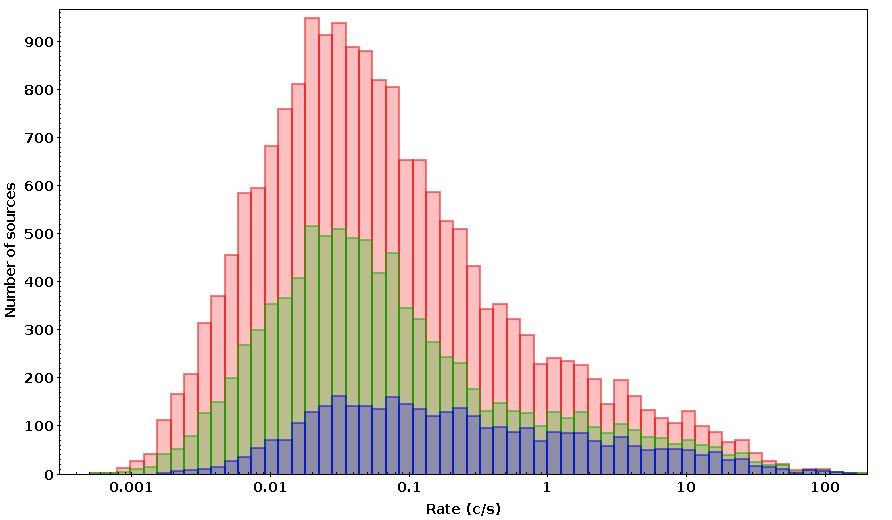} 
\includegraphics[height=5cm]{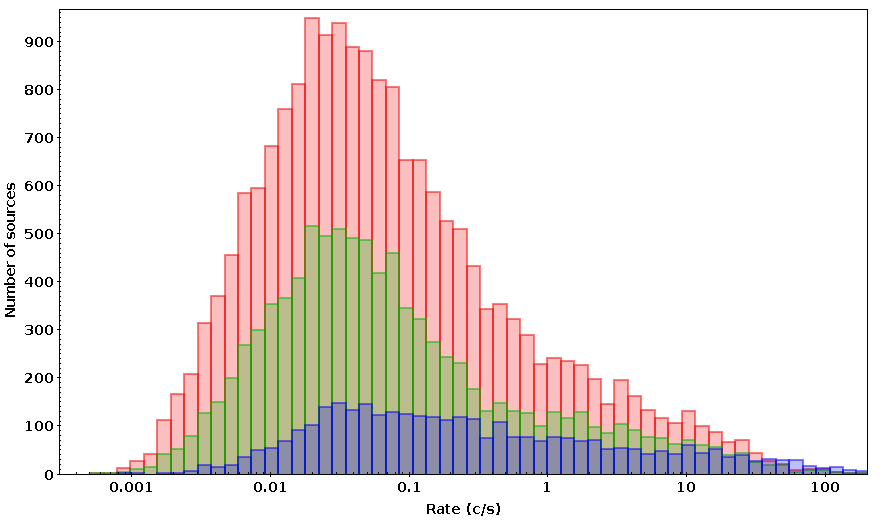} 
\caption{{\it Top left}: Distribution in count rate of point-like sources characterised in the time domain. Red: EXTraS/STV analysis. Blue: 3XMM; the total source sample is shadowed. {\it Top right}: Same as top left, EXTraS/STV vs. 4XMM.
{\it Bottom left}: Distribution in count rate of point-like sources displaying variability. Red: EXTraS/STV analysis (p-value $p<10^{-5}$, according to results of the fit of a constant model on 500s or optimal time bin, or with more than one Bayesian block in the sensitive segmentation). Green: EXTraS/STV analysis, based on uniform bin light curves only. Blue: 3XMM timing analysis (p-value $p<10^{-5}$, according to results of the fit of a constant model). {\it Bottom right}: Same as bottom left, EXTraS/STV vs. 4XMM. In all panels, all bar items start from the baseline.
}

\label{fig:stvvs3xmm}
\end{figure*}

\subsection{Bayesian block light curves}\label{sect:bb}
The Bayesian block algorithm \citep{Scargle2013} is designed to provide an optimal representation of a time series as a sequence of segments over which the underlying signal is constant to within the observational errors. Its application to EPIC data is challenging because of the high variability of the background as a function of time. Possible solutions for incorporating background subtraction in the Bayesian block algorithm were investigated by \citet{worpel15} with the specific aim of detecting transient or eclipsing sources in EPIC data. We adopted a different approach in which the variable background count rate was marginalised over.

We implemented a discrete application of the algorithm by \citet{Scargle2013}. Discreteness is introduced through the definition of an initial set of cells that represent the finest segmentation that could be achieved by the algorithm. We set an articulated trigger to define cells: We need at least 50 counts in the source region, or $50\times k$ photons in the background region, where $k$ is the ratio between the number of counts in the background and in the source region, and at the same time, the cell duration must be longer than the frame time. These criteria balance the need of enough photons for background subtraction in the Gaussian regime and the time resolution that allows detecting narrow features. The finest resolution that can be achieved uses the detector frames as initial time cells, but this requires a careful rethinking of background subtraction in the low-counts regime.  

The initial set of cells is processed by joining in blocks the cells that have compatible source rates. 
The positions of edges between neighbouring blocks is also optimised. 
Optimisation is performed according to a figure of merit (fitness function) additive over the blocks, assuming a prior distribution for the number of blocks.
As a fitness function, we use the logarithm of the likelihood of the source count rate, summed over blocks. The log-likelihood of the source count rate is marginalised over the distribution of the variable background rate, given the measures of the number of counts in the source region and in the background region, and knowing (from background modelling) the spatial distribution of steady and variable background and the count rate of the steady background. A Gaussian approximation for the likelihood profile is used both for the variable background rate and for the source rate, so that the marginalised likelihood profile is another Gaussian whose width can be obtained by simple error propagation.
We adopted a geometric prior on the number of blocks, $P(N_b) = P_{0} \times \gamma^{N_b}$ where $0<\gamma \leq1$, assigning a lower probability to a larger number of blocks.  
The standard prior for Bayesian blocks \citep{Scargle2013} is global, being related to the number of blocks in the optimal representation. In our implementation, the value of $\gamma$ was fixed in order to locally reflect a sigma cut in the separation of blocks: two blocks were separated if their rates were not consistent within $n$ sigma.
Depending on the threshold for this separation, we generated two sets of Bayesian block representations, one more sensitive to variability (at the cost of a higher number of spurious blocks), and the other more robust. The low prior (sensitive) and the high prior (robust) correspond to a nominal difference at $3\sigma$ and $4\sigma$, respectively, in source rate between neighbouring blocks. We calibrated them through Monte Carlo simulations of constant sources to evaluate the number of expected false blocks. As expected, false blocks are only due to statistical fluctuations (and therefore only depend on the number of initial cells and on the prior).
As in the case of uniformly binned light curves, the cumulative distribution of the rates was  computed for each Bayesian block light curve

The Bayesian block representation of the light curve does not allow distinguishing 
whether the change in count rate of the source between neighbouring blocks was sharp or smooth. Therefore, we introduced a parameter that we call slope (S). It is the minimum rate of change in the source count rate between two adjacent blocks. To compute S, we shrank
each of the blocks until their associated rates, $R_1$ and $R_2$, were exactly different at the $3\sigma$\text{ or }$4\sigma$ level in the sensitive or robust representation, respectively. We assumed that the uncertainty in the rates, $\delta R_1$ and $\delta R_2$, depends on time as $T^{-1/2}$, as expected for Poisson events. Assuming that the source count rate had changed linearly for the duration of the two blocks, $T_1 + T_2$, compatibly with the two rates, we obtained
\begin{equation*}
S=\frac{2}{9}\frac{\left ( R_2 - R_1 \right )^3}{\left ( \delta R_1 \times \sqrt{T_1} + \delta R_2 \times \sqrt{T_2} \right )^2}.
\end{equation*}
For similar blocks separated at the  $n\sigma$ level (as is the case for a source undergoing a linear trend in flux, with no background flares), the previous relation reduces to
\begin{equation*}
S \simeq 2 \left ( \frac{n}{3} \right )^2 \frac{R_2 - R_1}{T_1 + T_2}.
\end{equation*}

We also generated a Bayesian block light curve for the background. In this case, the Bayesian blocks algorithm reduces to the standard \citet{Scargle2013} implementation, with Poisson likelihood and a single scalar time series. The segmentation into blocks is very different from the one that we obtain for each source, and it is unique for the entire exposure.

\subsection{Energy-resolved analysis}\label{sect:energyresolved}

Starting from the event files and source models used for the full-band analysis, we extracted new event files and generated source models in three sub-bands: super-low (0.1-1.0 keV, SL), low (1.0-2.0 keV, LO), and high (2.0-12.0 keV, HI). As for the full-band analysis, all sources expected to have fewer than ten counts in a specific exposure and energy band were disregarded. We were left with 356,984, 338,869, and 322,281 detections in the SL, LO, and HI band, respectively. All steps of the energy-resolved analysis are fully similar to those described above for the full band. In each energy band, we generated four kinds of light curves for each source: (i) uniformly binned, with 500 s bin size; (ii) uniformly binned, optimal bin size; (iii) Bayesian blocks, sensitive separation level; and (iv) Bayesian blocks, robust separation level. We analysed 305,403 sources in more than one sub-band and 147,316 in all energy bands. For all sources that were kept in more than one energy band, we produced hardness ratio light curves starting from uniformly binned light curves with 500 s binning. Hardness ratios were defined as an estimator of the ratio of the difference between the net rates in two bands and their sum,
\begin{equation*}
HR = \frac{R_1 - R_2}{R_1 + R_2},
\end{equation*}
where 1 corresponds to the harder and 2 to the lower energy bands. We used a Monte Carlo simulation to estimate each single hardness ratio and
its uncertainty, taking the error bars in the two rates into account. We defined the hardness ratio estimator and its uncertainty as the midpoint
and half-width of the smallest interval with a coverage of 68\%.

\begin{figure*}[htb]
\centering 
\includegraphics[height=4.5cm]{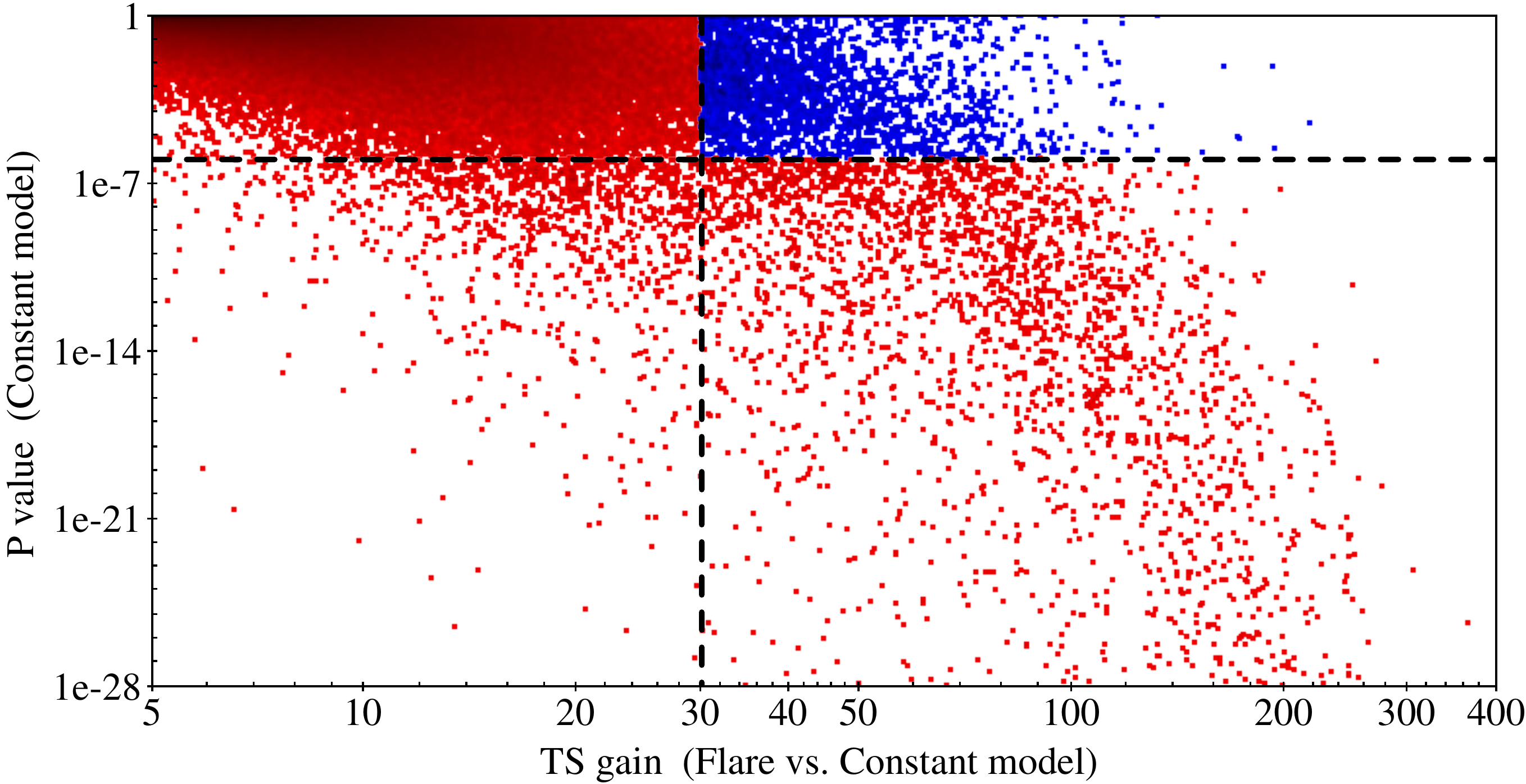} 
\includegraphics[height=4.5cm]{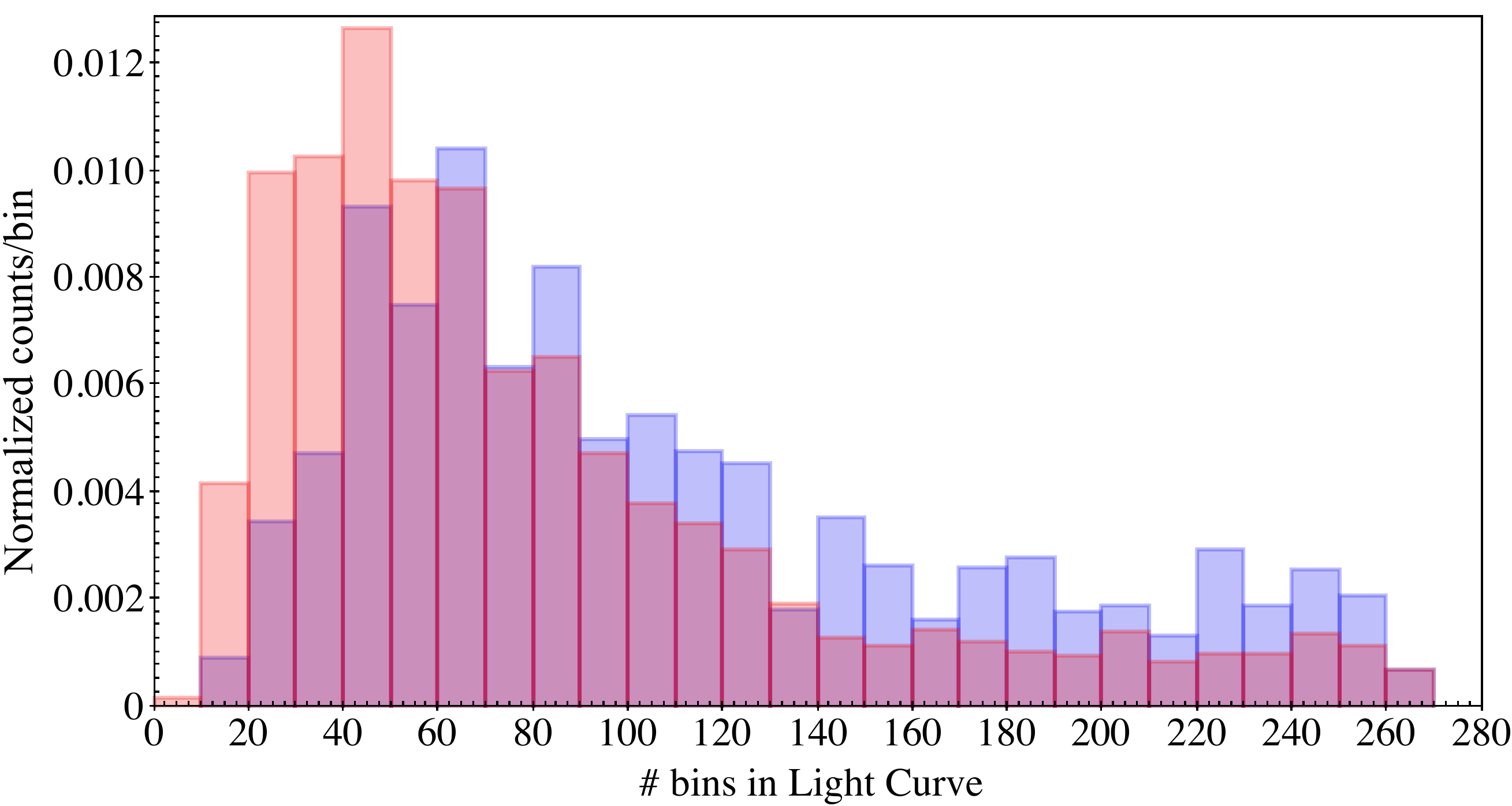} 
\caption{Left panel: Gain in TS by adding a flare to a constant vs. the p-value associated with a constant model for 500s bin light curves. The two lines indicate a threshold of p-value $=10^{-6}$ in both axes: the blue points are light curves that are overall compatible with a constant model, but for which a flare improves the fit significantly. Right panel: Histogram of the number of bins for the light curves, colour-coded as in the left panel (area of histograms normalised to unity; all bar items start from the baseline). 
} 
\label{fig:ub500flarevsconst}
\end{figure*}

\begin{figure*}[htb]
\centering 
\includegraphics[height=4.5cm]{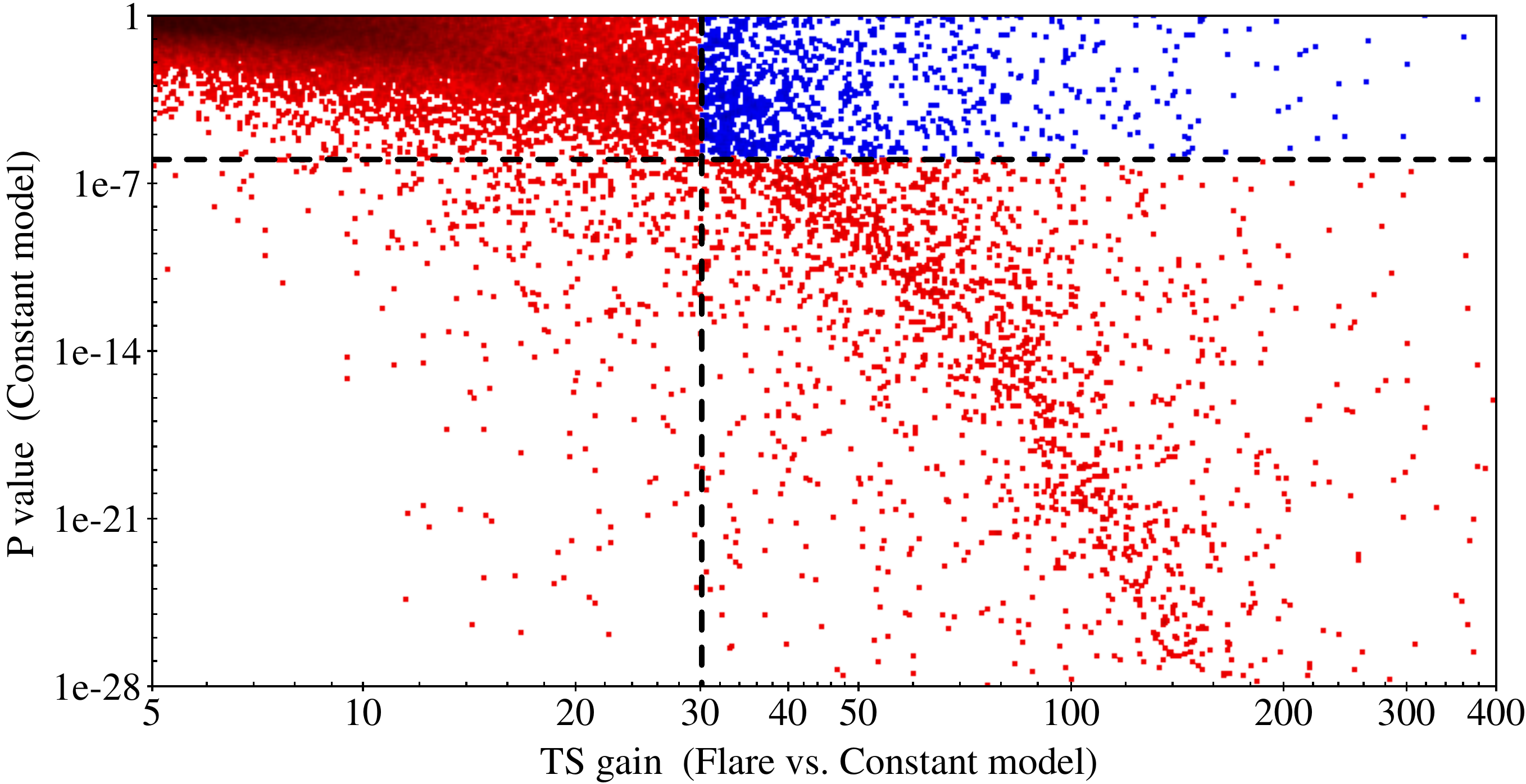}
\includegraphics[height=4.5cm]{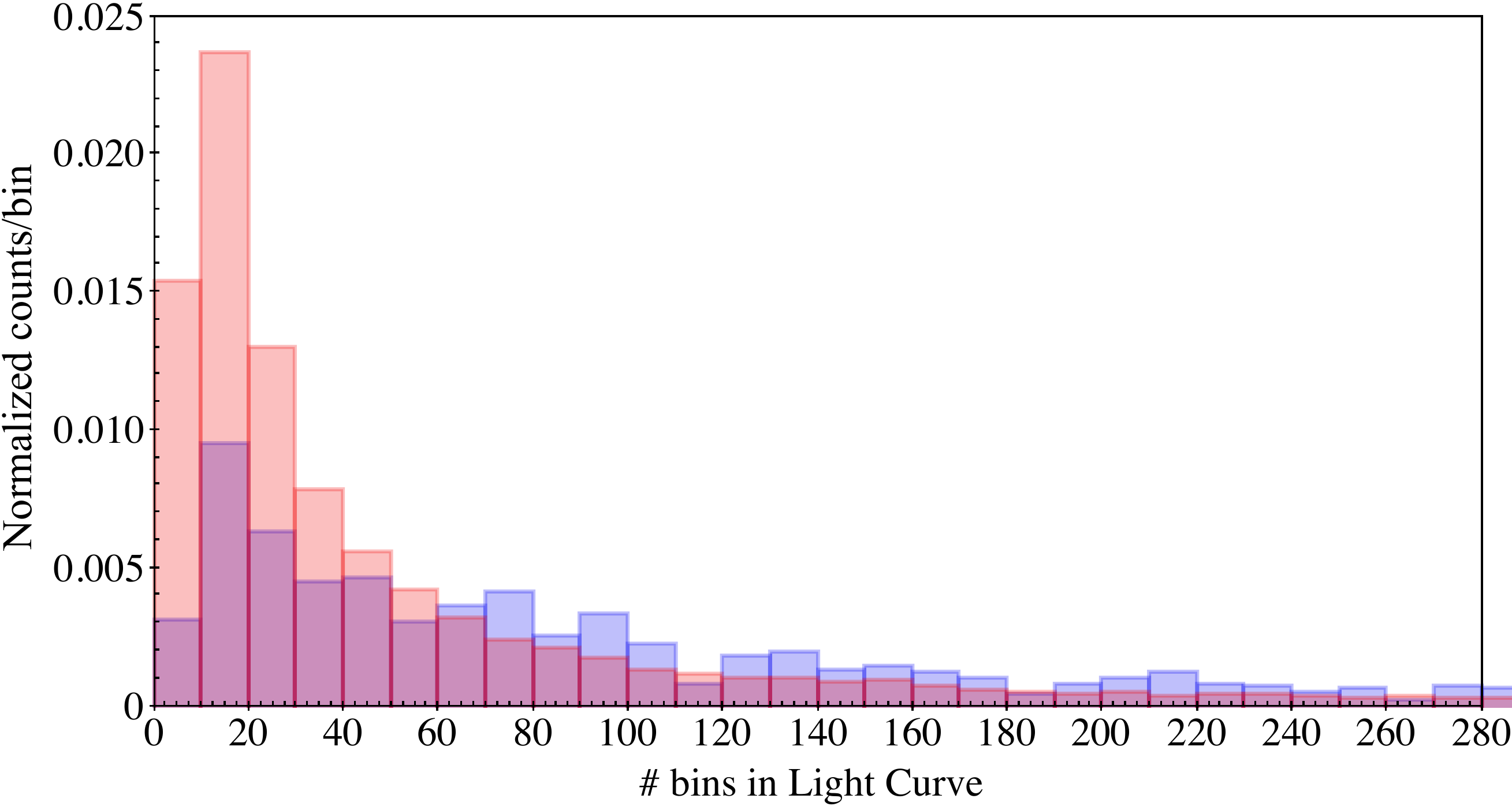}
\caption{Same quantities as Fig.~\ref{fig:ub500flarevsconst} for uniform bin light curves that have at least 25 counts per bin. Their behaviour is similar to that of 500s bin light curves (see Fig.~\ref{fig:ub500flarevsconst}), but far fewer light curves that are compatible with a constant contain a significant flare. 
} 
\label{fig:uboptflarevsconst}
\end{figure*}

\subsection{Analysis in the frequency domain}\label{sect:FFT}
For each source, we produced a representation of the time series in the frequency domain by applying the Fast Fourier Transform
(FFT) algorithm to the uniformly binned light curve with 10 s time bin. All light curves were zero-padded up to $T\sim160$\,ks before applying the FFT. This is longer than any observation while giving a number of bins that is a power of 2, which yields a faster FFT computing time. In this way, all the FFTs have the same format, the same sampling time, and the same size. This artificial windowing alters the FFT properties. Moreover, the light curve might have gaps due to gaps in the GTIs, which also introduces a windowing effect.

\subsection{Standard light curves}
We also produced light curves with uniform time binning (500 s and optimal) by following standard data analysis prescriptions using the SAS software. Data preparation and source selection were performed as described in Sect.~\ref{sect:dataprep} and ~\ref{sect:selection}. Source events were extracted from the same regions as described in Sect.~\ref{sect:sourceregion}. 
Background was sampled locally: For each source, background events were extracted from an annulus surrounding the source region. This is the same approach as was used in 2XMM \citep{Watson2009}, however, while in the 2XMM case the size of the annulus is the same for all sources, we decided here to implement a different approach in which the inner and outer radii are related to each source count rate. We set the inner radius to be 20\% larger than the source region radius and the outer radius to be the maximum between 40'' and twice the source region radius. This has the advantage of sampling the background as close to the source as possible, keeping in any case the source leakage in the background annulus at a low level. The specific values for the radii were set after extensive testing on a set of 200 exposures. 

We used the SAS tool \texttt{evselect} to generate light curves for source and background, and we combined them with the tool \texttt{epiclccorr} into a background-subtracted light curve, which also corrects  for a number of effects such as vignetting, bad pixels, chip gaps, quantum efficiency, and GTIs. These light curves are released in our archive for comparison purposes only. 

\subsection{Characterisation of variability}
Our STV analysis encompasses a large number of tests for variability and a set of measurements to extract synthetic information from uniform bin and Bayesian blocks light curves, power spectra, and hardness ratios. All results are stored in the EXTraS archive and are also included in the headers of the files themselves.

We fit a series of analytical models of the source rate evolution to each light curve (both with uniform time bin and Bayesian blocks). Every single light curve was tested against a constant and a linear model, including all the light curves extracted in the three energy sub-bands. Full band light curves were also tested  against more advanced models: a quadratic function, an exponential decay, and local features such as flares and eclipses in addition to a constant. For each model we extracted the best-fit value for each parameter and its associated $1\sigma$ error, the $\chi^2$ value, the number of degrees of freedom, and the tail probability for the model.

We provide a number of other variability indices to characterise the light curves. These include the weighted average of the count rate with its uncertainty; the weighted standard deviation, skewness, and kurtosis of the distribution of the count rates; the relative variance given by the ratio between the variance and the average count rate; the relative excess variance with its uncertainty; the correlation coefficients between the source and background light curves; the amplitude of count rate excursion given by (max(rate)- min(rate))/2; the median absolute deviation; and the maximum relative offset from the median given by max(|rate-median|)/median.

Our characterisation of short-term variability did
not take advantage of tools such as autoregressive models. This is a promising extension beyond EXTraS. 

Other synthetic parameters were extracted by analysing the cumulative distribution of the count rate. These include the
 fraction of time spent more than 1, 3, and 5 $\sigma$ below and above the average count rate; the fraction of time spent within 10\%\  of the median count rate; the width of the range of rates in which the source spends 90\%  of its time; the fraction of such a range in which the source spends 20, 35, 50, 65 and 80\%\  of its time; and the asymmetry of the count rate distribution in which the source spends 20, 35, 50, 65 and 80\%\  of its time.

We provide a number of variability indices that are specific to Bayesian block light curves. These include the number of blocks; the {\em \textup{fragmentariness}}, that is, the number of blocks per ks of observation; the {\em \textup{steadiness}}, defined as $\sum (rate^2 / rate_{err}^4)$ per ks; the minimum time for doubling and halving the count rate in the light curve (derived from the maximum positive and negative slope between any two blocks); and the maximum negative and positive deviation of the rate from the weighted average in sigma units.

We characterise the spectral variability of XMM sources with two separate approaches. On the one hand, we fit simple models (constant and linear) to the hardness ratios light curves, and on the other hand, we provide a basic characterisation of the light curves produced in each sub-band, including computation of excess variance, weighted average, weighted standard deviation (and their corresponding uncertainties), median, and median absolute deviation.

We characterise the power spectra of each source by fitting a constant+power-law model and a constant + Lorentzian model. The results of these fits are stored in the archive, in the catalogue, and in the header of the files themselves.

\subsection{Products}
The output of the STV analysis of EXTraS consists of (i) a catalogue that lists all results of the variability characterisation for all detections included in our investigation. The catalogue is available as a fits file and is also included in the EXTraS database. It can be fully searched with an online web form (see Section ~\ref{sect:resources}). A light version of this catalogue, that is, stripped down to the most important quantities, is also available. (ii) A set of FITS and ASCII files for each source, for each exposure, instrument, and energy band: light curves in the 0.2-12 keV energy range with uniform time bin of 500 s, 5000 s, and optimal (see Sect.\ref{sect:uniformbin}); light curves in the 0.2-12 keV energy range  with adaptive binning, based on the Bayesian block approach, 
with sensitive ($3\sigma$ separation) and robust ($4\sigma$ separation) segmentation of neighbouring blocks; cumulative distribution for uniformly binned (500 s) and Bayesian blocks (sensitive and robust) light curves; background light curve in the 0.2-12 keV energy range with adaptive time binning; light curves in the 0.2-1 keV, 1-2 keV, and 2-12 keV energy ranges with uniform time bins (500 s and optimal); light curves in the 0.2-1 keV, 1-2 keV, and 2-12 keV energy ranges  with adaptive binning (both sensitive and robust segmentation); hardness ratio light curves (1-2 keV versus 0.1-1 keV, 2-12 keV versus 1-2 keV, and 2-12 keV versus 0.2-1 keV) with uniform time bins (500 s); power spectrum of the source variability (0.2-12 keV); and source and background region files.

\begin{table}[htb]
\begin{minipage}[ht]{\columnwidth}
\normalsize
\caption{Basic facts about the EXTraS STV analysis.}
\label{tab:stvoverview}
\small \centering \renewcommand{\footnoterule}{} \tabcolsep 0mm
\begin{tabular}{
l @{\extracolsep{2mm}} c} \hline \hline

\hline 
\# Selected Observations  &  7,007 \\ 
\# Selected Exposures &  19,962\\
\# Selected Detections & 418,387\\
\# Unique sources &   297,351\\
\# Detections with Uniform Bin light curves (500 s) &  327,104 (pn) \\
 &  225,888 (MOS1) \\
 &  247,274 (MOS2) \\
\# Detections with Bayesian Blocks light curves  & 320,142 (pn)  \\
 & 221,236 (MOS1) \\
  & 242,056 (MOS2) \\
\# Detections with Hardness ratios &  154,870 (2 bands) \\
 & 144,293 (3 bands) \\
\hline

\end{tabular}
\end{minipage}
\normalsize
\end{table}

\begin{figure}[ht]
\centering 
\includegraphics[width=9cm]{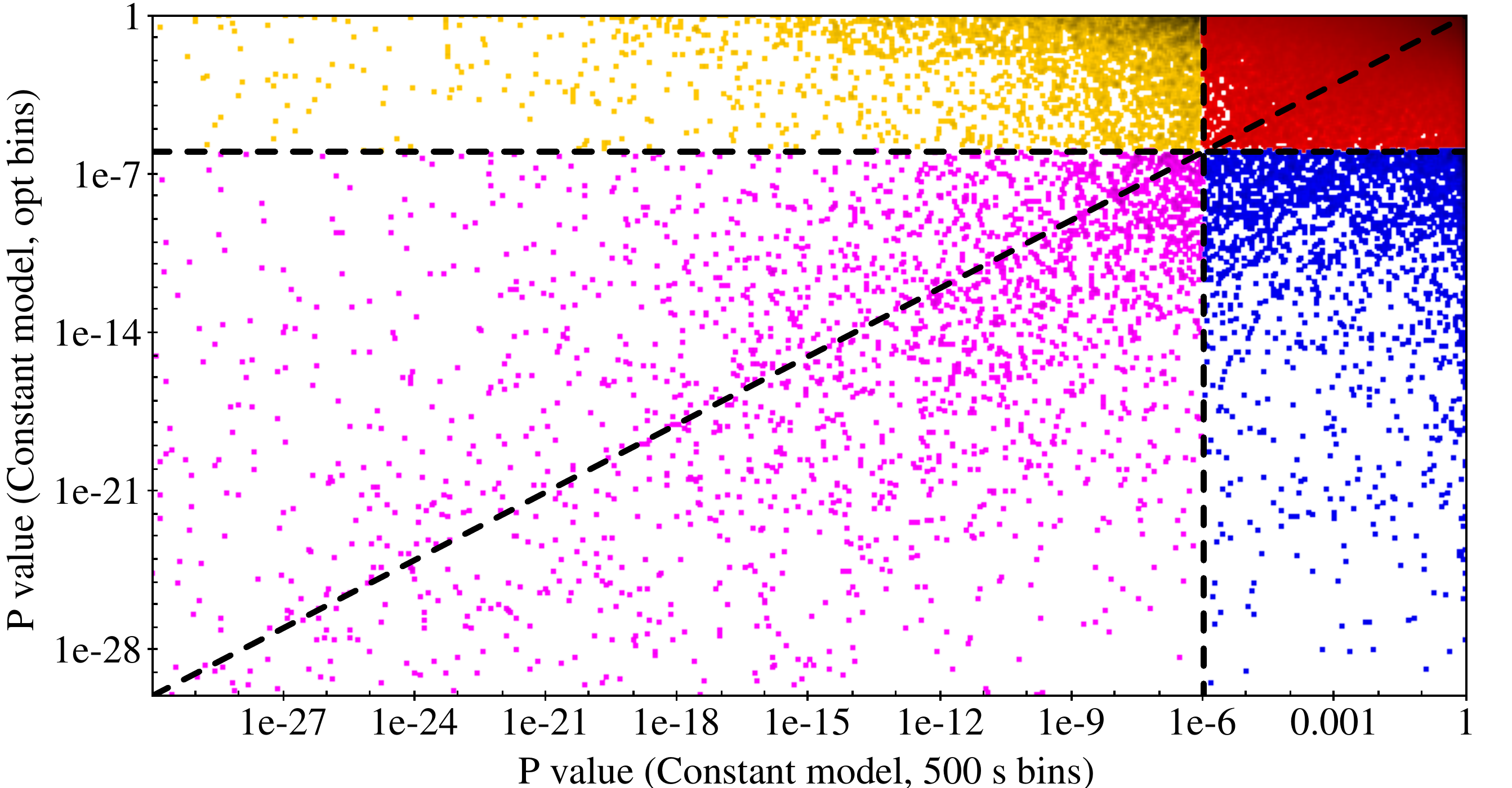}
\caption{ Distribution of the p-value for a constant model as applied to optimal bin and 500s bin light curves for each source. The threshold lines at p-value=$10^{-6}$ 
divide the plot into four regions that are colour-coded as in Fig.~\ref{fig:ub500vsubopt2}.  
Yellow points 
(2,789) in the top left corner correspond to sources whose optimal bin light curves are compatible with a constant, while 500s light curves are not. 
Red points (787,023) in the top right corner correspond to sources whose light curves are compatible with a constant in both cases. 
Magenta points 
(2,324) in the lower left corner correspond to sources whose light curves are not compatible with a constant in either case. 
Blue points 
(2,257) in the lower right corner correspond to sources whose 500s bin light curves are compatible with a constant, while optimal light curves are not. 
See also Fig.~\ref{fig:ub500vsubopt2}.
} 
\label{fig:ub500vsubopt}
\end{figure}

\subsection{STV database and its properties}
We provide here a very concise statistical analysis of results and products of the STV analysis to help understand their meaning, reliability, and usage. An overview of the basic properties of the STV analysis is given in Table~\ref{tab:stvoverview}.

\begin{figure*}[ht]
\centering 
\includegraphics[height=4.5cm]{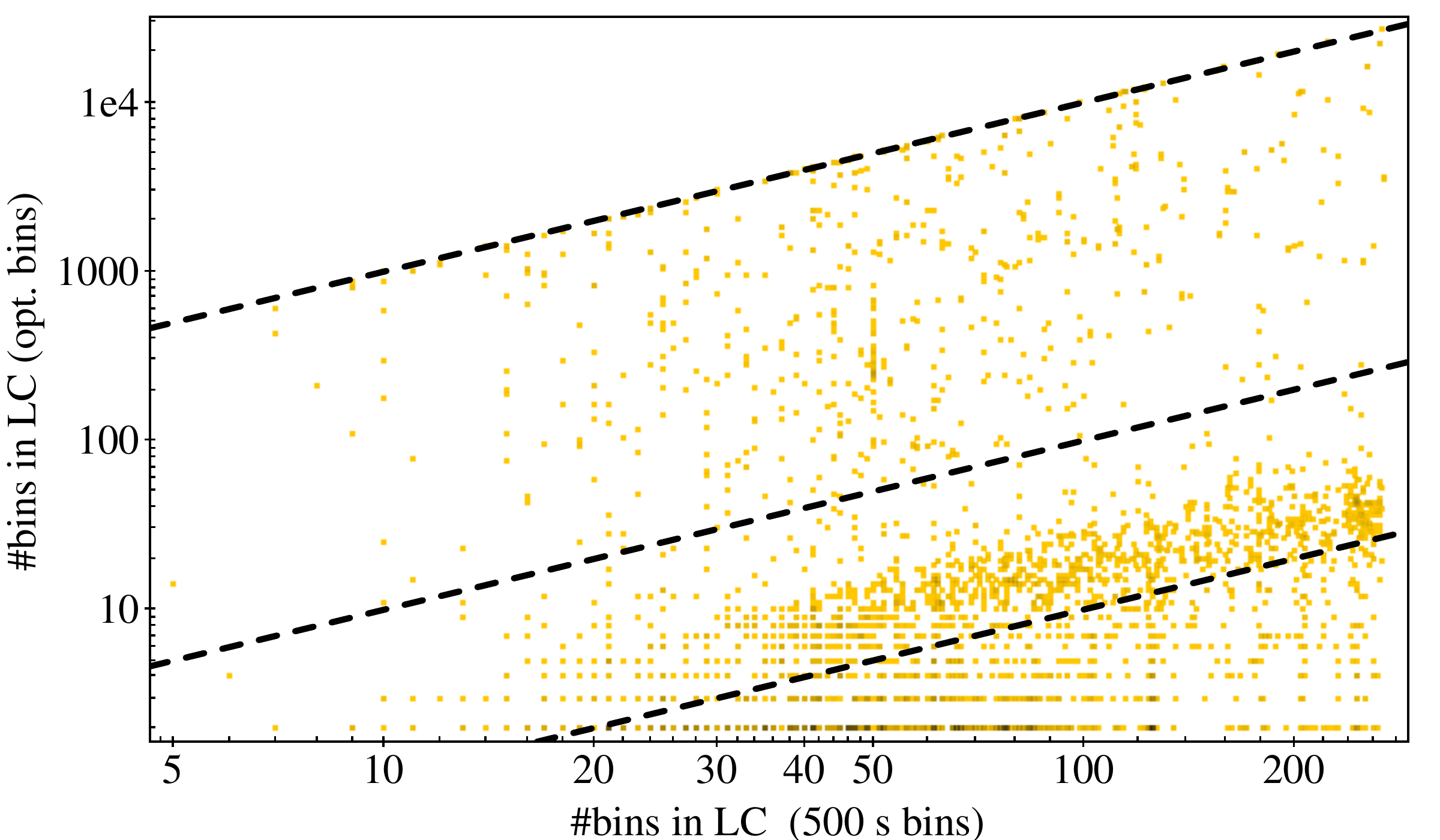}
\includegraphics[height=4.5cm]{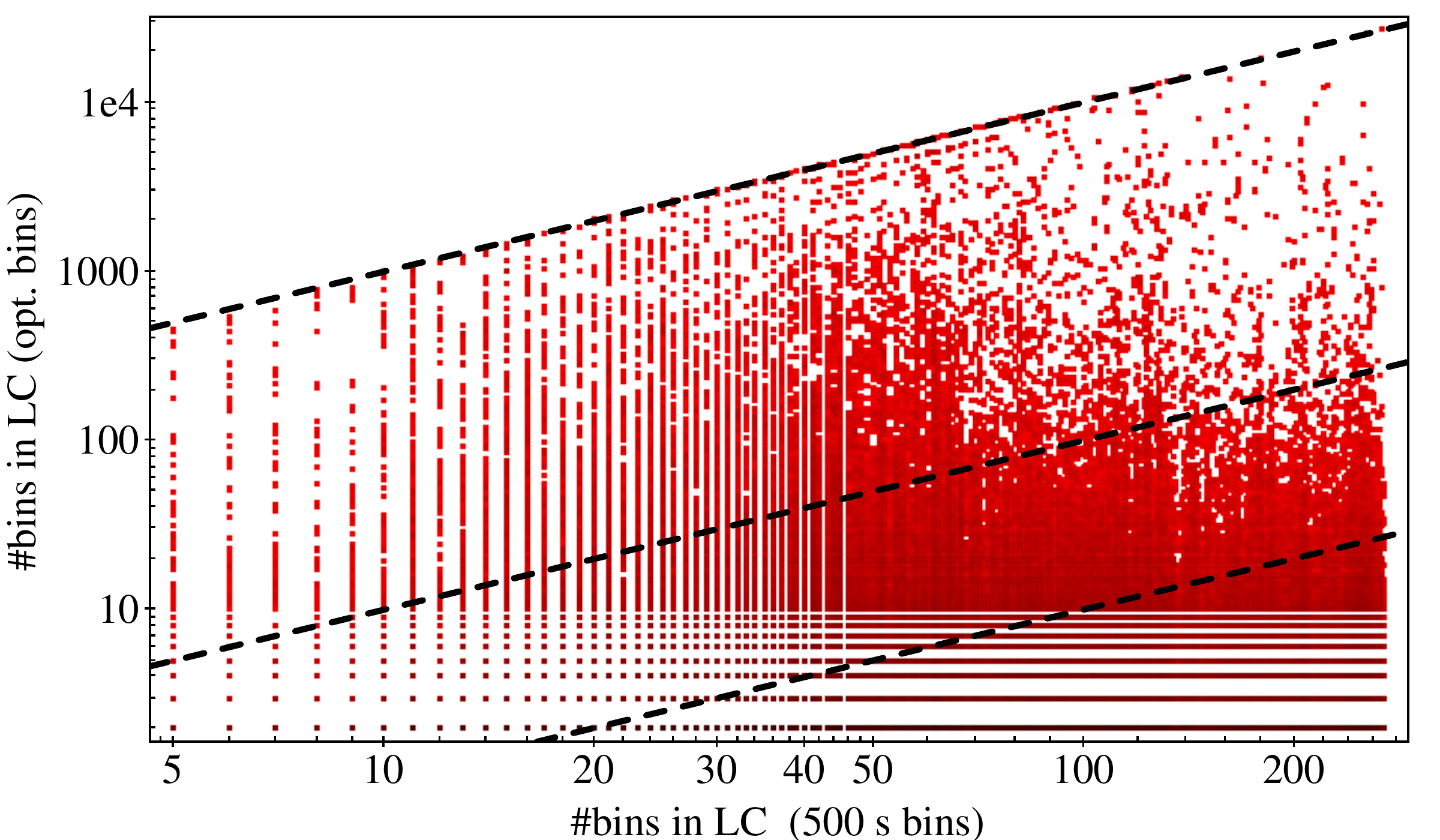}
\includegraphics[height=4.5cm]{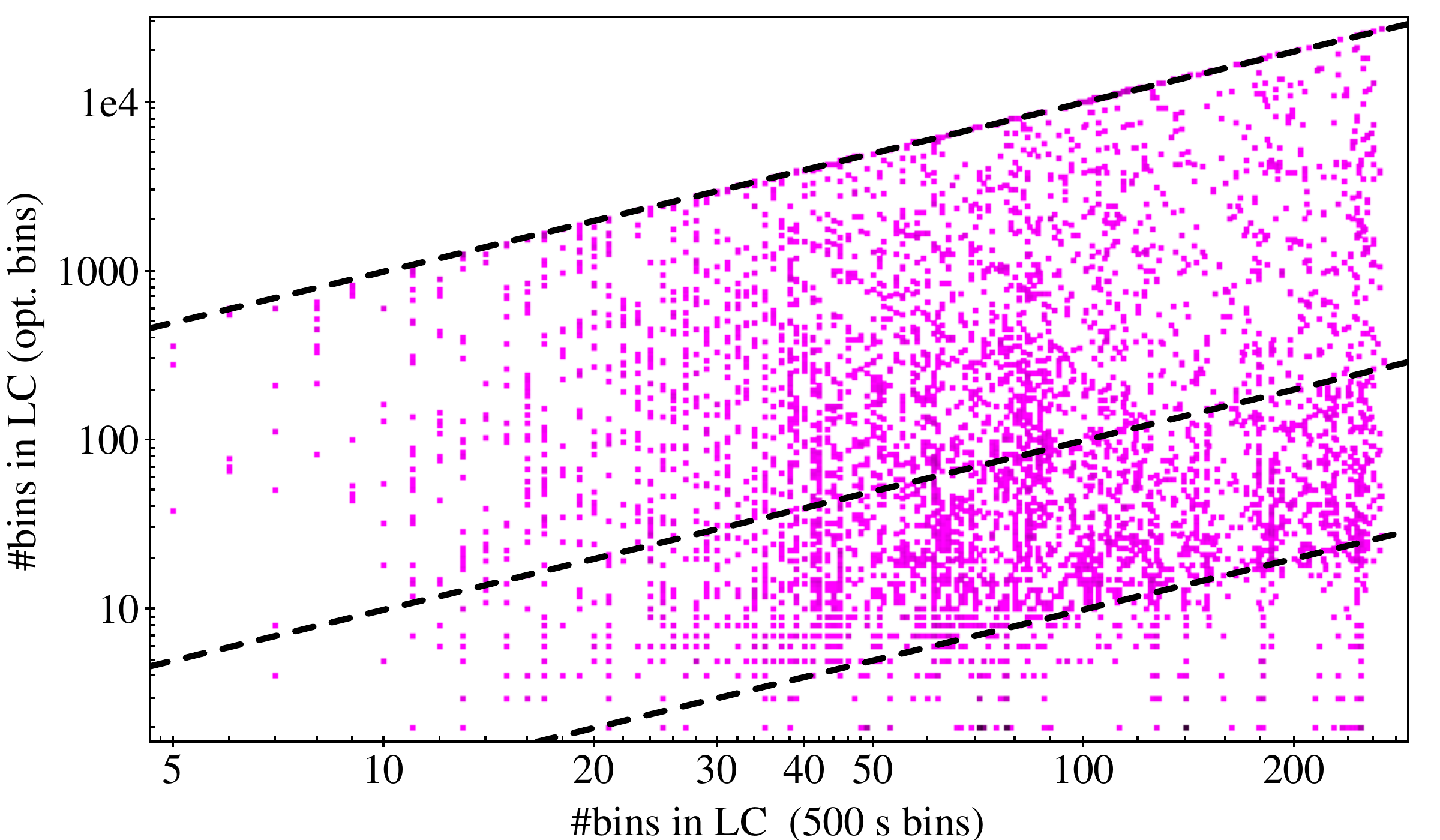}
\includegraphics[height=4.5cm]{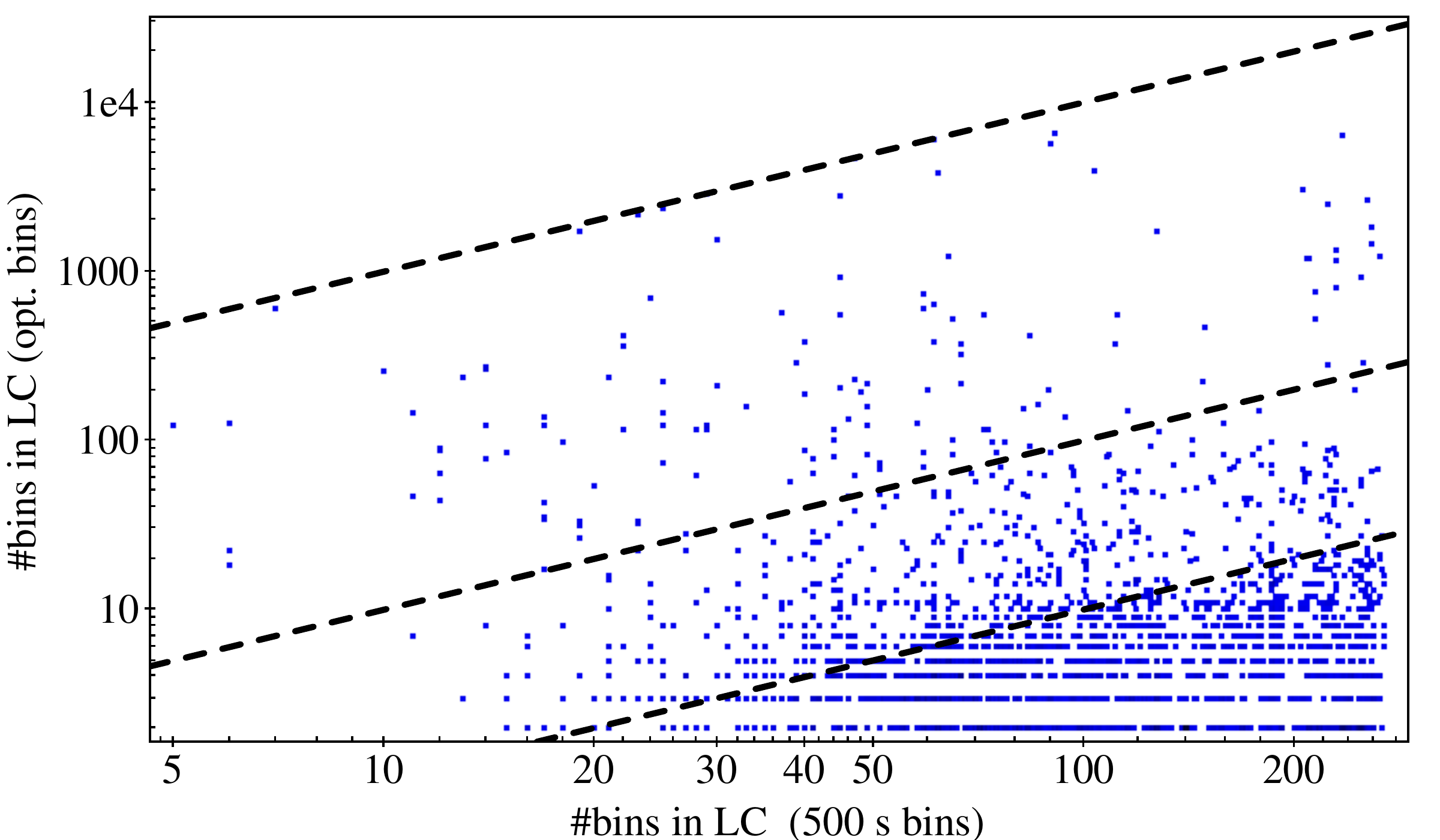}
\caption{Comparison of the number of bins in 500s bin light curves and in optimal bin size light curves, colour-coded as in Fig.~\ref{fig:ub500vsubopt}. In each panel, the top dashed line indicates the minimum allowed time bin of 5s; the middle dashed  line indicates sources for which the optimal bin size is exactly 500s, and the bottom  dashed line does the same for 5ks. The number of bins for the 500s bin light curve is an indication of the observation length, while lines parallel to the three lines already drawn indicate sources with a similar rate.} 
\label{fig:ub500vsubopt2}
\end{figure*}

In Figure~\ref{fig:stvvs3xmm} we show the distribution in count rate of detections included in EXTraS/STV analysis together with the distribution in count rate of detections displaying variability. These were selected according to different markers for variability. In the case of light curves with uniform time bins, we used the p-value associated with a constant model. This is defined as the probability of observing rates as scattered or more scattered than those observed when the model holds, and the source is not variable. We selected   $p<10^{-5}$  here, yielding 7,279 and 5,650 candidate variable sources in the 500 s light curves and in the optimal bin light curves, respectively (9,265 candidates when the condition is fulfilled in at least one light curve, either with 500 s bins or with optimised bins). In the case of Bayesian block light curves, a number of blocks greater than 1 is an obvious marker for variability. This yields 7,379 and 14,939 candidate variable sources in the robust and sensitive Bayesian block segmentations, respectively. In Figure~\ref{fig:stvvs3xmm} we include all detections with either $p<10^{-5}$ in uniform bin light curves or more than one block in the sensitive Bayesian block segmentation, totalling 18,529 candidate variable detections. In the same figure, we also show the distribution in count rate of 3XMM-DR4 sources (and variable sources, according to the 3XMM variability flag), and the same information for the recent 4XMM-DR9 catalogue, restricted to the DR4 dataset.

The expected fraction of false positives in selecting candidate variable sources can be estimated as follows: Based on the number of light curves with 500 s bins and with optimal bins in our archive (800,266 and 797,697, respectively), the assumed threshold $p<10^{-5}$ yields about eight spurious candidate variable detections for each sample ($\sim0.1\%$ of the candidate variable sources) due to statistical fluctuations. In the case of Bayesian block light curves, as discussed in Sect. 3.16, through extensive simulations of sources with a constant count rate,  we estimated  the number of spurious blocks we expect for each detection due to statistical fluctuations based on the number of cells in the initial grid. This yielded 10 ($\sim0.1\%$) and 1,320 ($\sim8.8\%$) detections with more than one expected block in the samples with robust and sensitive  segmentation, respectively.

A basic question regarding the description of variability is whether it is possible to characterise the variability further if a light curve is consistent with a constant model. We expect that a local feature such as a flare might sometimes be detected with good confidence, even if the global fit of the light curve to a constant is acceptable. 
This is confirmed by Fig.~\ref{fig:ub500flarevsconst}, where it is clear that a significant flare in uniformly binned light curves (500 s bins) can be missed by the fit to a constant model when the light curves have a large number of bins. Statistical fluctuations can lower the global test statistic against a constant for all points far from the flare, and the contribution of the few discrepant points to the global test statistic is negligible. In  optimally binned light curves (see fig.~\ref{fig:uboptflarevsconst}), far fewer cases that are compatible with a constant contain a significant flare. This might be due to the lower number of bins in the optimal bin size light curves with respect to those with 500s bins.

We now compare some variability indicators between uniform bin light curves obtained with 500 s binning and with optimal binning. In the first type of light curves we expect many more frequent departures from the Gaussian regime. 
The p-value for a constant model applied to 500 s bin light curves may give an incorrect indication in a fraction of cases for faint sources. For sources reaching the Gauss approximation for bin sizes larger than 500 s, the characterisation based on optimally binned light curves is more robust. This is shown in Figure~\ref{fig:ub500vsubopt} and \ref{fig:ub500vsubopt2}, where we adopt a threshold at p-value=$10^{-6}$, implying less than one false positive when we try to reject the constant model. Sources whose optimal bin light curves are compatible with a constant while 500s light curves are not 
(yellow points in Figure ~\ref{fig:ub500vsubopt} and ~\ref{fig:ub500vsubopt2}) 
are mostly  faint and reach the Gaussian approximation for bin sizes of 2 to 4 ks, 
therefore 500s light curves are mostly not appropriate. 
Sources whose light curves are compatible with a constant in both cases (red points) have a distribution of bin sizes that essentially mimics the distribution of the entire source population, with faint sources (with fewer optimal bins) dominating. Sources whose light curves are not compatible with a constant in either case (magenta points) are uniformly scattered in the plot of Fig.~\ref{fig:ub500vsubopt2}, indicating that it is easier to detect variability in bright sources. Sources whose 500s bin light curves are compatible with a constant while optimal light curves are not (blue points) have an optimal bin size that is concentrated beyond 5ks, indicating that the 500s binning is not adequate.

\begin{figure}[ht]
\centering 
\includegraphics[width=9cm]{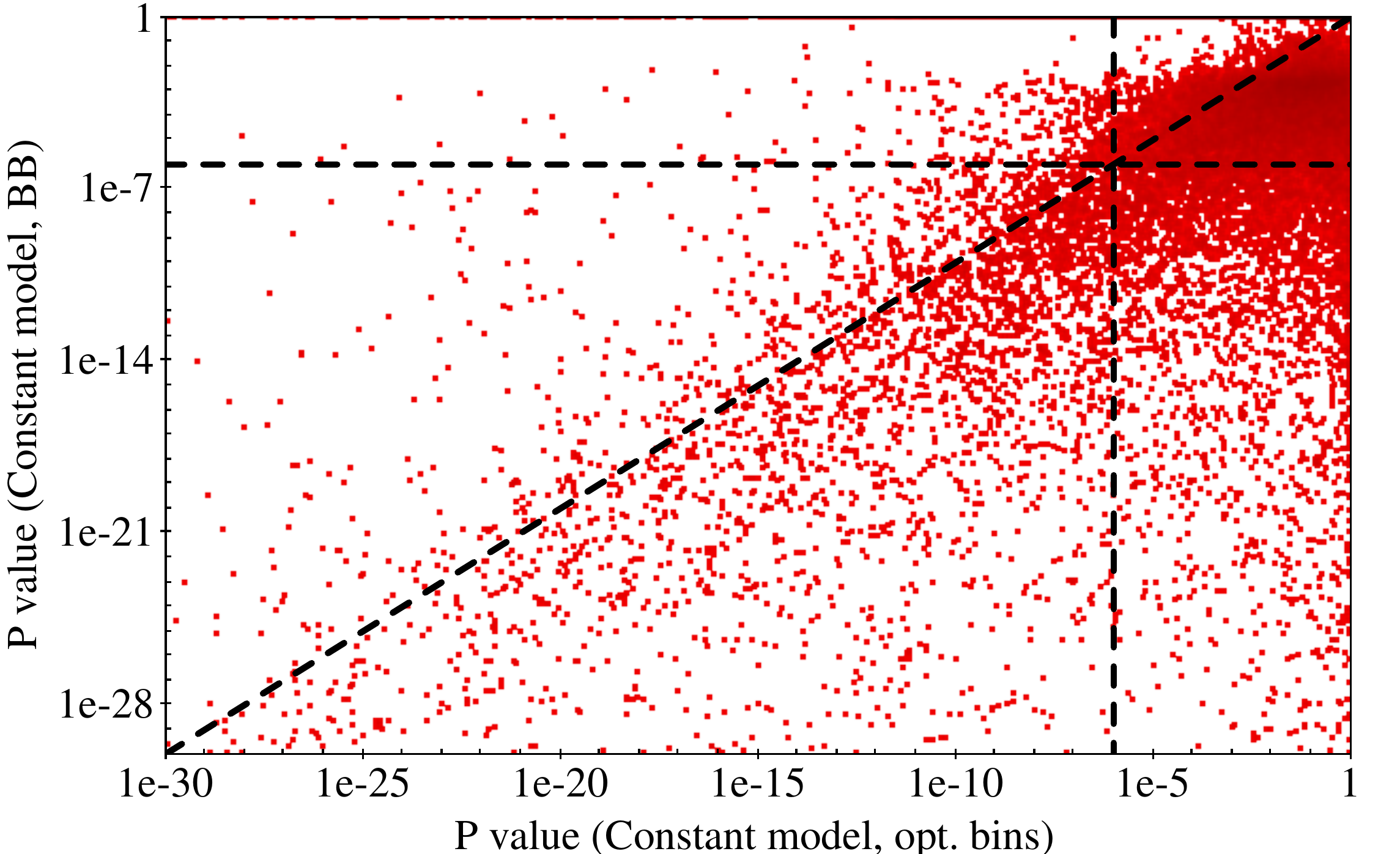}
\caption{Comparison of the p-value associated with a constant model for uniform bin light curves that have at least 25 counts per bin and the same p-value for Bayesian blocks. The top right region contains 786,262 sources whose light curves are both compatible with a constant model.
The bottom left region contains 6,272 sources whose light curves are both incompatible with a constant model. The top left region includes 1,501 sources that are variable according to their uniform bin light curve, but are compatible with a constant in their Bayesian block light curve. 
The bottom right region includes 3,582 sources that are variable according to the Bayesian block analysis, but are compatible with a constant in the uniform bin light curve.
} 
\label{fig:uboptvsbb}
\end{figure}

\begin{figure}[ht]
\centering 
\includegraphics[width=9cm]{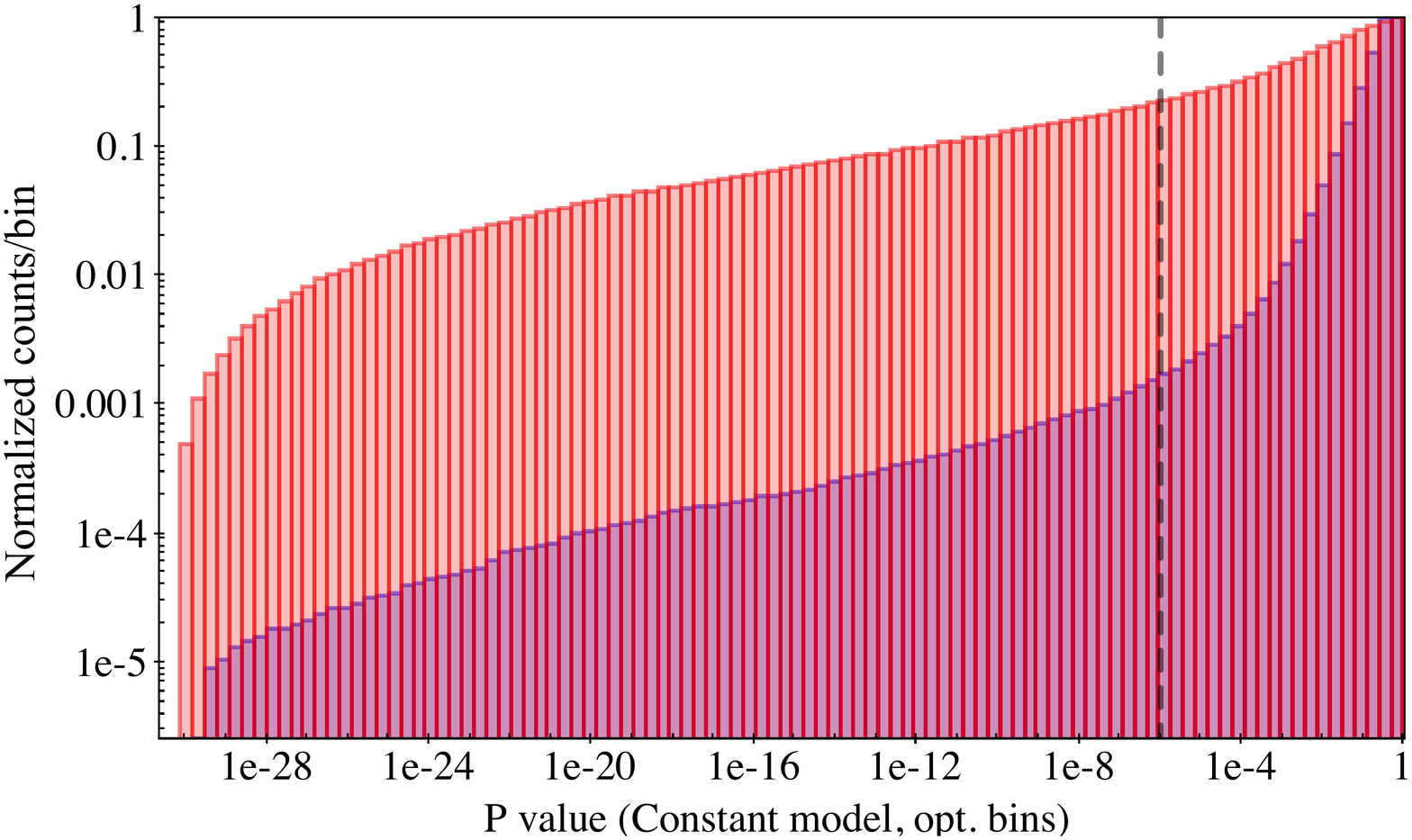}
\caption{
Histogram of the p-value associated with a constant model for uniform bin light curves with optimal bin size (see also Fig.~\ref{fig:uboptvsbb}). 
Sources  that are described by a single Bayesian block are plotted in purple. Histograms are normalized to the peak. 
}
\label{fig:uboptvsbb2}
\end{figure}

We also compared some variability indicators between Bayesian blocks and uniform bin light curves with optimal binning.
Figure~\ref{fig:uboptvsbb} shows that uniformly binned light curves are often less effective in spotting localised short features such as flares or eclipses than Bayesian blocks. Figure ~\ref{fig:uboptvsbb2} shows that a consistent fraction of sources described by a single Bayesian block have an associated uniform bin representation that is highly variable. This may be due to several reasons: If the initial grid for the Bayesian block segmentation has only one cell, variability cannot be tested, while optimally binned light curves always have at least two blocks; a bug in the script that generates the Bayesian block light curves misrepresents all light curves that should have as many blocks as initial cells as a single block (see Sect.~\ref{sect:wp2bugs}). 

\subsection{Some usage examples}
In this section, we provide some usage examples to illustrate the science capabilities of EXTraS STV products. First, we give a short account of two investigations that were recently published by our team: the discovery of an X-ray superflare from an L dwarf star \citep{deluca2020}, and the study of the properties of flares from supergiant fast X-ray transients \citep{sidoli2019}. These two cases are selected as a clear demonstration of EXTraS STV potentialities for the search for and characterisation of peculiar objects and rare events, and for the study of the properties of classes of sources. Further examples are for instance the discovery of X-ray flaring activity from a young pre-main-sequence star \citep{pizzocaro16}, the discovery of a puzzling flaring X-ray source in the Galactic globular cluster NGC6540 \citep{mereghetti18}, and the study of the X-ray activity-rotation connection in cool stars \citep{pizzocaro19}. In the last part of this section, we examine a science project published before the release of EXTraS catalogues: the study of the X-ray variability in a large sample of stars by \citet{pye2015}. We also compare its results to those quickly derived from our database (comparison with the recent 4XMM-DR9 catalogue is also shown).


\begin{figure}[ht]
\centering 
\includegraphics[height=9cm, angle=-90]{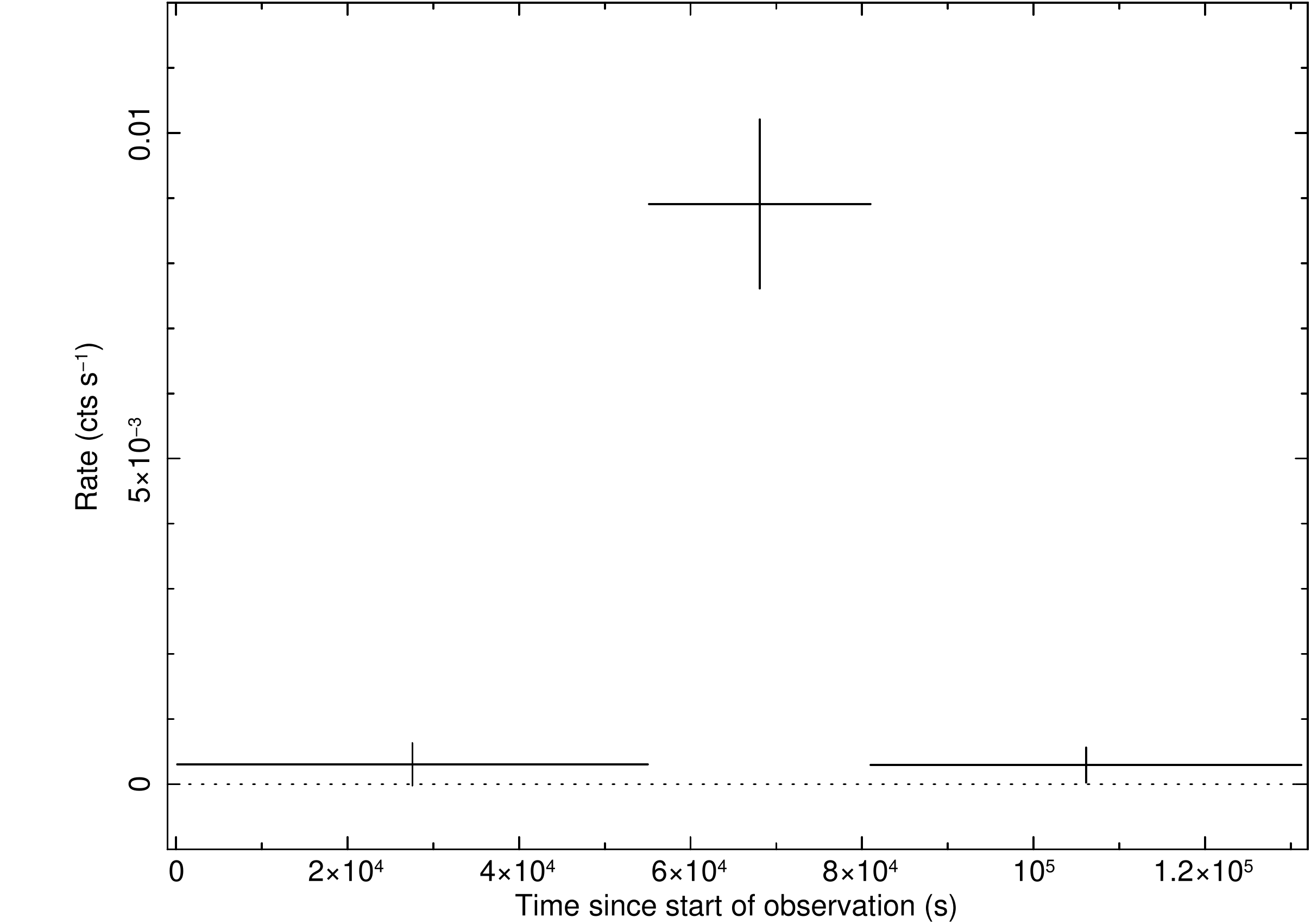}
\caption{EXTraS Bayesian block representation of the light curve of the source 3XMM J033158.9-273925 (pn camera, Obs.Id. 0555780101), 
the first detection of flaring activity from an ultra-cool dwarf star of spectral class L. See text.} 
\label{fig:0331_bb}
\end{figure}

\begin{figure*}[th]
\centering 
\includegraphics[height=6cm]{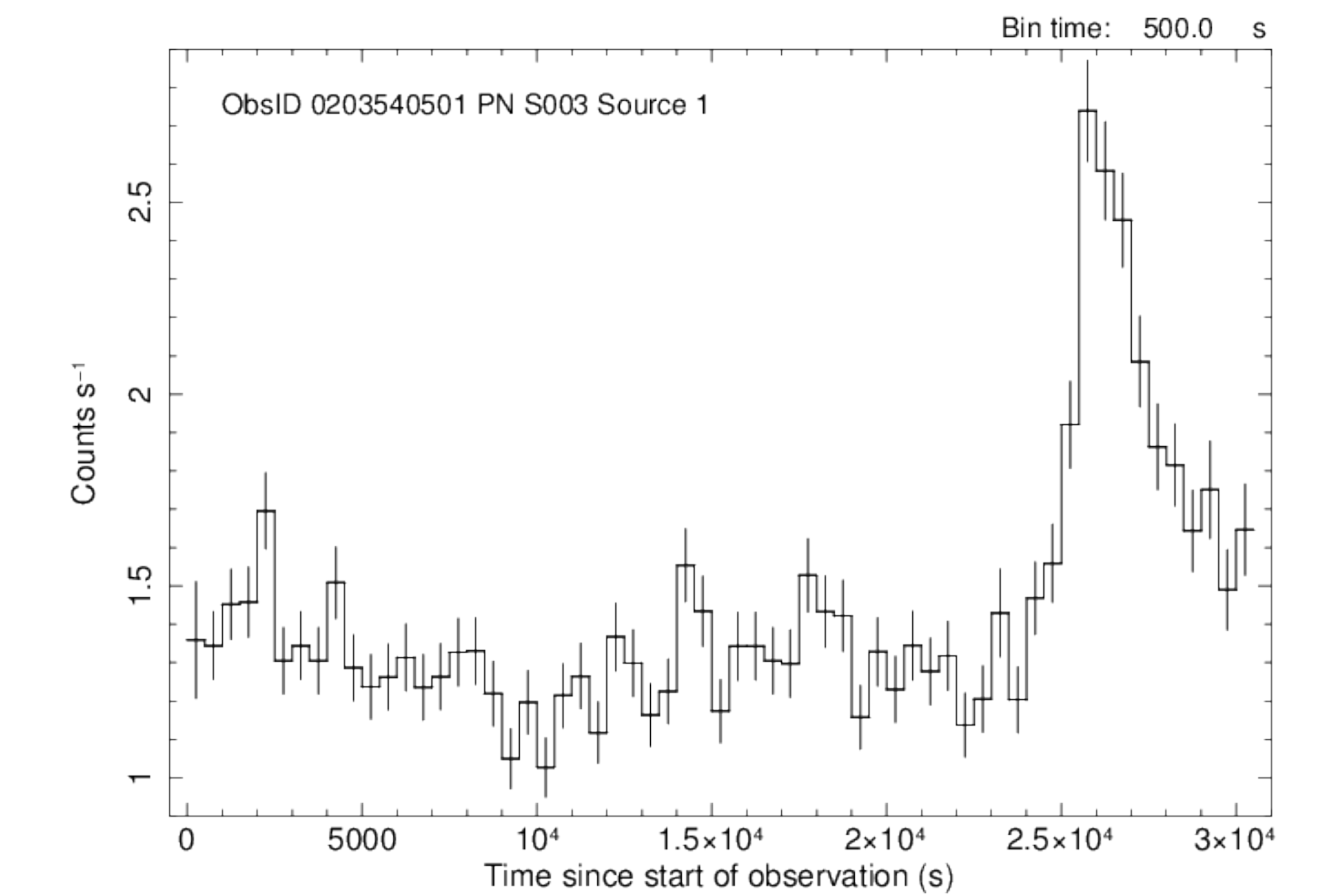} 
\includegraphics[height=6cm]{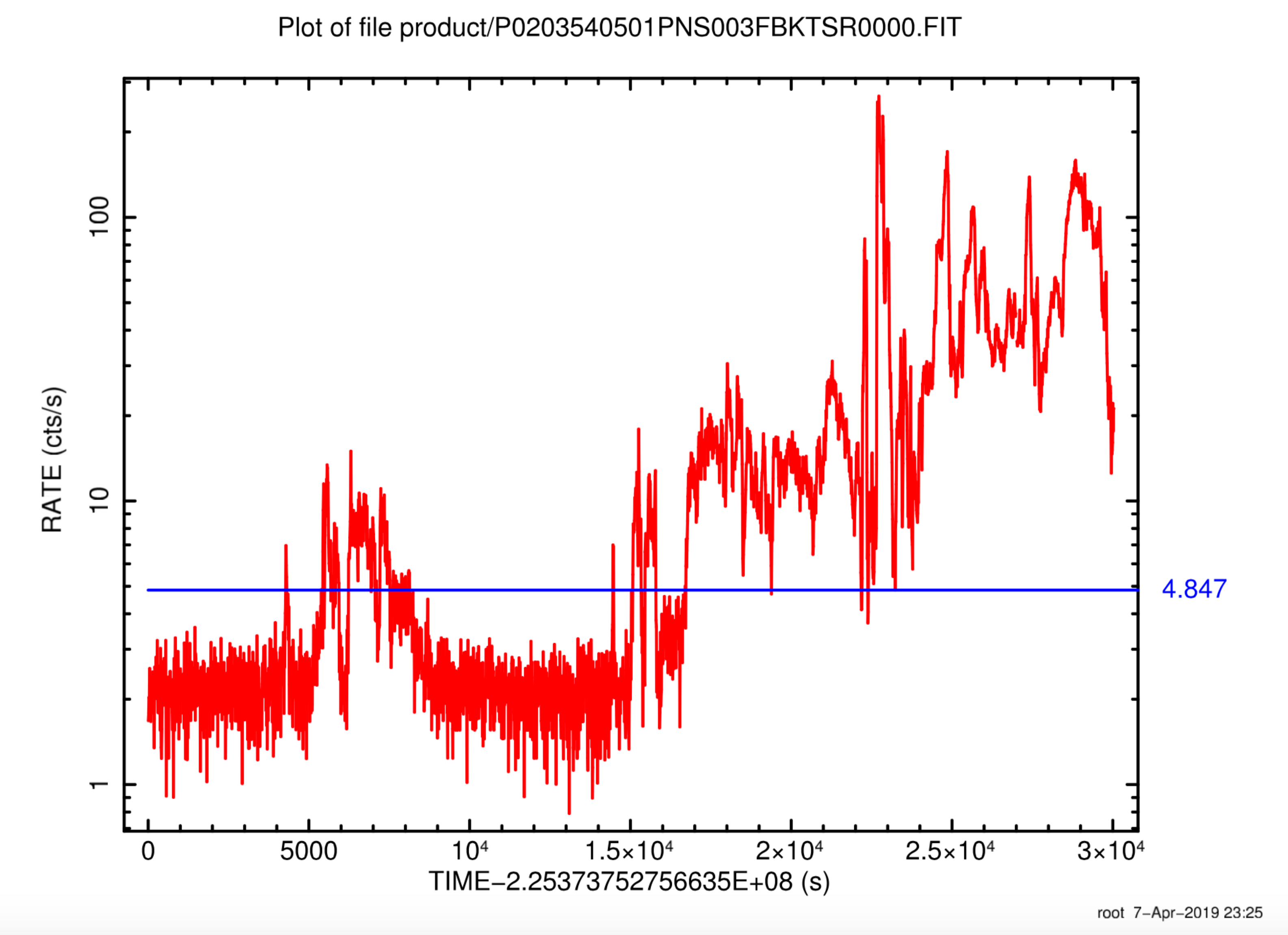} 
\caption{{\it Left panel:} EXTraS/STV light curve of the star HD 283810 with uniform time binning (500 s). A large flare is seen in the second half of the observation. The tail probability of a fit with a constant model is $<\,10^{-25}$. The flare is missed by the 4XMM variability analysis because it occurred outside of the GTI (see right panel). {\it Right panel:} Background light curve of the same observation, taken from the 4XMM products. All observing times with a background rate exceeding the threshold marked by the blue line are rejected. The time interval of the flare (see left panel) is therefore excluded.} 
\label{fig:wp2vs4xmm1}

\end{figure*}

\subsubsection{Discovery of an X-ray superflare from an L dwarf.} 
Magnetic activity in stars at the low-mass end of the main sequence is poorly understood. We cross-correlated the EXTraS STV catalogue with the catalogue of L- and T-class ultra-cool dwarfs by \citet{carnerorosell2019}. This selected 3XMM J033158.9-273925 (hereafter J0331-27) as a very interesting case.  J0331-27 matches the position of an L0 candidate object within 1$\farcs$1. Inspection of EXTraS light curves clearly shows an X-ray flare (e.g. the Bayesian block light curve is shown in Fig.~\ref{fig:0331_bb}). 
The source is located in the FoV of the Chandra Deep Field South survey, and a large multi-wavelength dataset is available. As discussed by \citet{deluca2020}, analysis of these data showed (i) the spectral type to be L1. This is only the second L dwarf detected in X-rays after a previous four-photon detection of the binary system Kelu-1 \citep{audard2007}. (ii) The flare energetics is in the regime of {\em \textup{super-flares}}, showing that strong magnetic reconnection events and the ensuing plasma heating are still present even in objects with photospheres as cool as $\sim 2100$\,K. (iii) The flare energy number distribution is inconsistent with the canonical power law $dN/dE \sim E^{-2}$, suggesting that magnetic energy release in J0331-27,  and possibly in all L dwarfs,  takes place predominantly in the form of giant flares. 

\subsubsection{Statistical properties of flares from supergiant fast X-ray transients.} 
The sub-class of high-mass X-ray binaries called supergiant fast X-ray transients \citep[see][]{sidoli2017}
shows flaring activity in their entire dynamic range of luminosities, even outside outbursts. We used EXTraS STV products to investigate the properties of these X-ray flares in a sample of nine supergiant fast x-ray transients \citep{sidoli2019}. Adopting the Bayesian block decomposition of the EPIC X-ray light curves, we selected 144 X-ray flares covering a wide range of luminosities ($10^{32}-10^{36}$ erg s$^{-1}$), from quiescence to outbursts. The Bayesian block light curves also allowed us to measure in a model-independent way the flare rise and decay time, the flare duration, and the time separation between adjacent flares. We also measured the luminosity at the peak of flares, the average accretion rate, and overall emitted energy. The observed properties of flares from supergiant fast X-ray transients are 
in qualitative agreement with the expectations of the subsonic settling accretion regime model \citep[see e.g.][]{shakura12}, where the development of flares is related to the onset of Rayleigh-Taylor instabilities in the hot quasi-spherical shell of plasma accumulated at the boundary of the neutron star magnetosphere, resulting in unstable accretion of the entire shell \citep[see][for details]{sidoli2019}.

\begin{figure}[ht]
\centering \includegraphics[width=9cm]{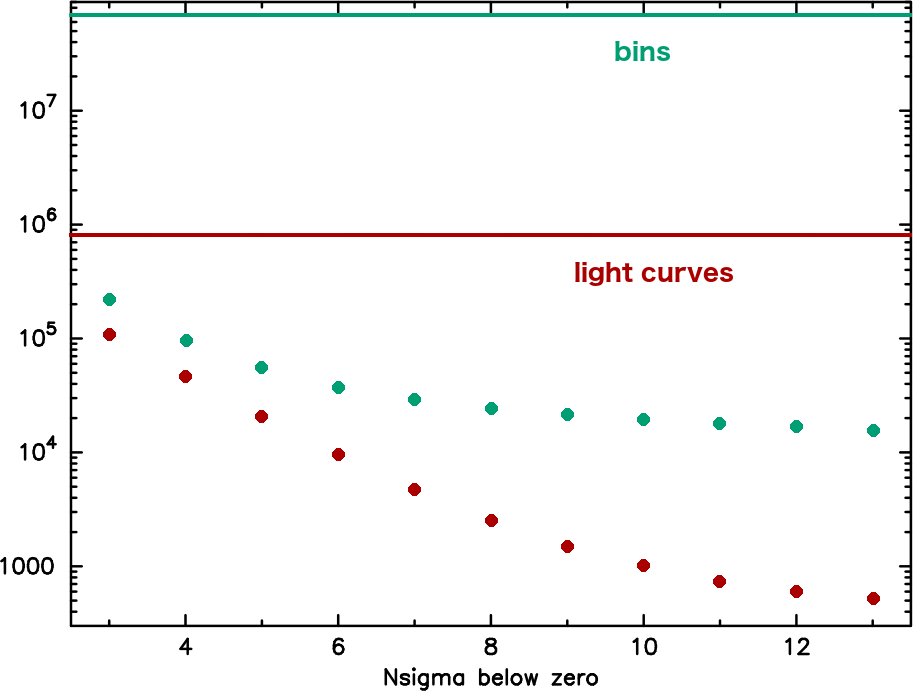}
\caption{Number of light curves with at least one negative bin by more than N $\sigma$ and the overall number of bins below this threshold. The horizontal green line indicates the overall number of bins ($8\times10^7$) in 500s bin light curves, and the green dots show their number at least N$\sigma$ below 0. Similarly, we show the light curves that contain these points in brown. The vertical separation between green and brown dots indicates the average number of negative bins per light curve. The number of bins per light curve increases as the threshold decreases.
}
\label{fig:nbins1}
\end{figure}

\begin{figure}[ht]
\centering 
\includegraphics[width=9cm]{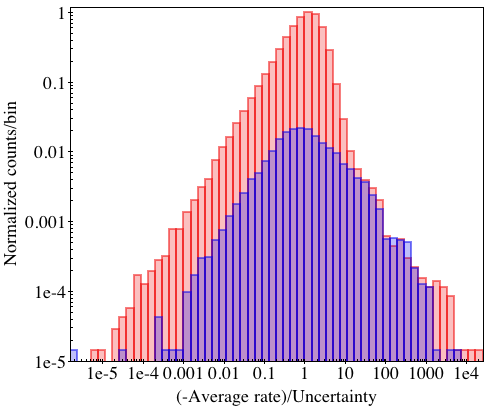}
\caption{
Histogram of the distribution of the number of standard deviations below zero of all negative average rates for light curves with a negative average rate. 
The histogram is normalised to the peak.
Red bars: light curves with uniform bins of 500s. 
Blue bars: 
optimally binned light curves. 
}
\label{fig:nbins2}
\end{figure}

\subsubsection{Variability in a large sample of cool stars}
We focused on the sample of 2,357 X-ray emitting cool stars selected by \citet{pye2015}  by matching the Hipparcos-Tycho-2 catalogue with the 2XMM catalogue. 
In order to study X-ray variability and flares, \citet{pye2015}  considered all light curves flagged as variable by the 2XMM catalogue. This yielded 118 light curves. Of these, 22 had spurious variability, which is related to different issues in the analysis. This left 96 actually variable light curves, called the cool variable sample (CVS). As a further step, \citet{pye2015} visually inspected all of the remaining 815 light curves that were not marked as variable in 2XMM and found that 12 of them displayed apparent variability. These were called the cool low variable sample (CLVS). 

As a simple exercise, we selected all EXTraS/STV products generated for the same sample of stars using the same data and searched for variable light curves.
First, we cross-matched the coordinates of all 2,357 stars with the EXTraS/STV catalogue by adopting a correlation radius of 15$''$. Second, we selected all matches related to the same observation ID in the two catalogues. We also matched the resulting dataset with the 4XMM catalogue by using the DR4DETID identifier. This exercise yielded 2,880 detections with EXTraS light curves of 2,039 stars
(duplications related to light curves at the single-exposure and camera level in EXTraS were not included in this count),
 including 91 light curves (out of 96) from the CVS and all 12 sources from the CLVS of \citet{pye2015}.
We searched for variability in the resulting sample using the EXTraS/STV output. As a simple selection criterion, we selected detections whose EXTraS Bayesian block light curves had more than one block in the sensitive representation and a tail probability $<10^{-3}$ for a constant model fit to uniform time bin light curves with either 500 s or optimised time bins. This allowed us to select 217 detections with variable light curves
(as above, possible duplications related to light curves at the single-exposure and camera level in EXTraS were not included in the count).
These include 86 light curves from the CVS and 11 from the CLVS and 120 additional variable light curves that are not mentioned by \citet{pye2015}. We visually inspected the EXTraS/STV results for these 120 detections and identified 108 light curves with genuine variability and 12 likely artefacts related to incorrect background subtraction during intense soft proton flares or to spurious Bayesian blocks due to statistical fluctuations in very bright sources (see Sect.~\ref{sect:wp2bugs}). 

\begin{figure*}[ht]
\centering 
\includegraphics[height=6cm]{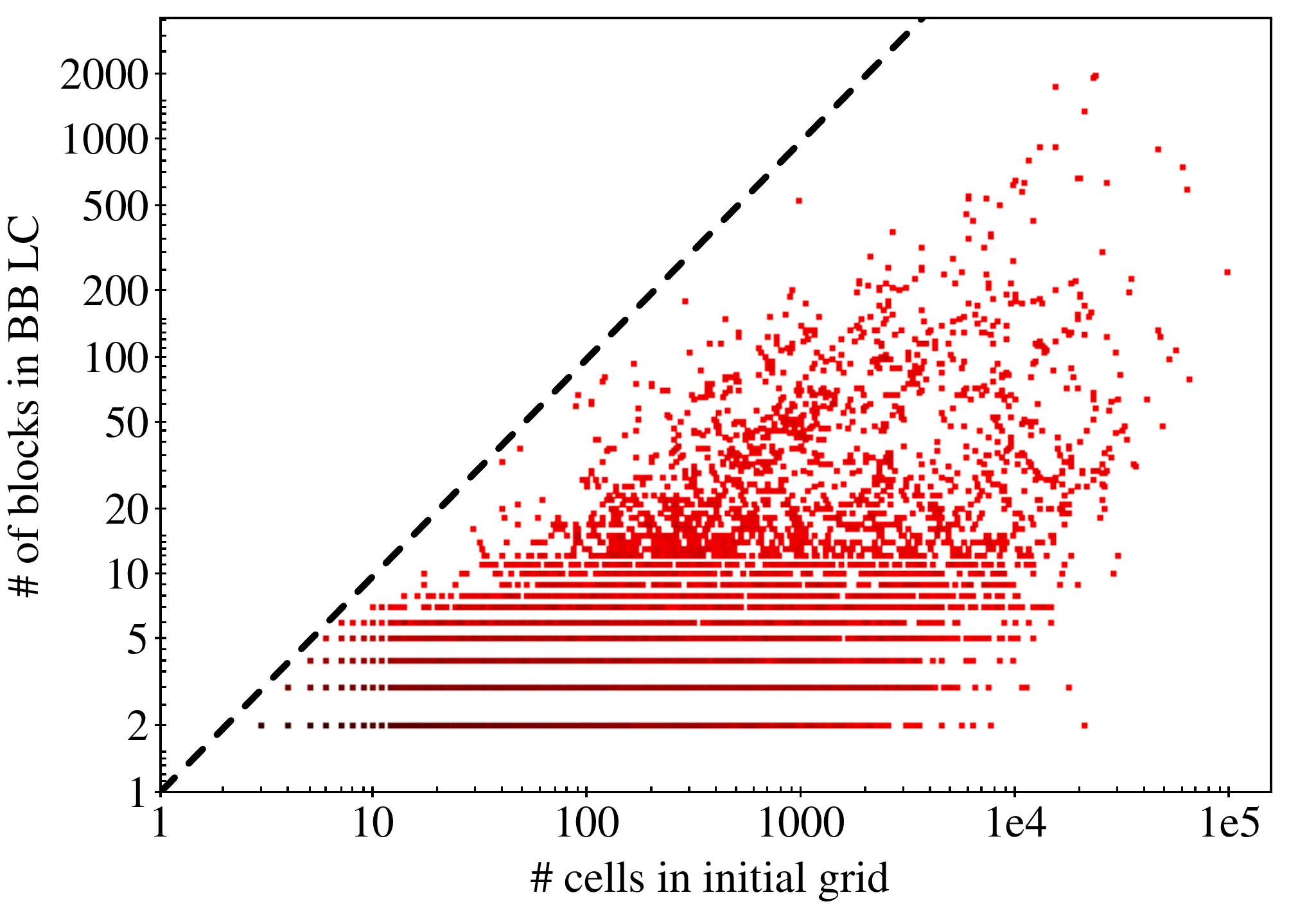}
\includegraphics[height=6cm]{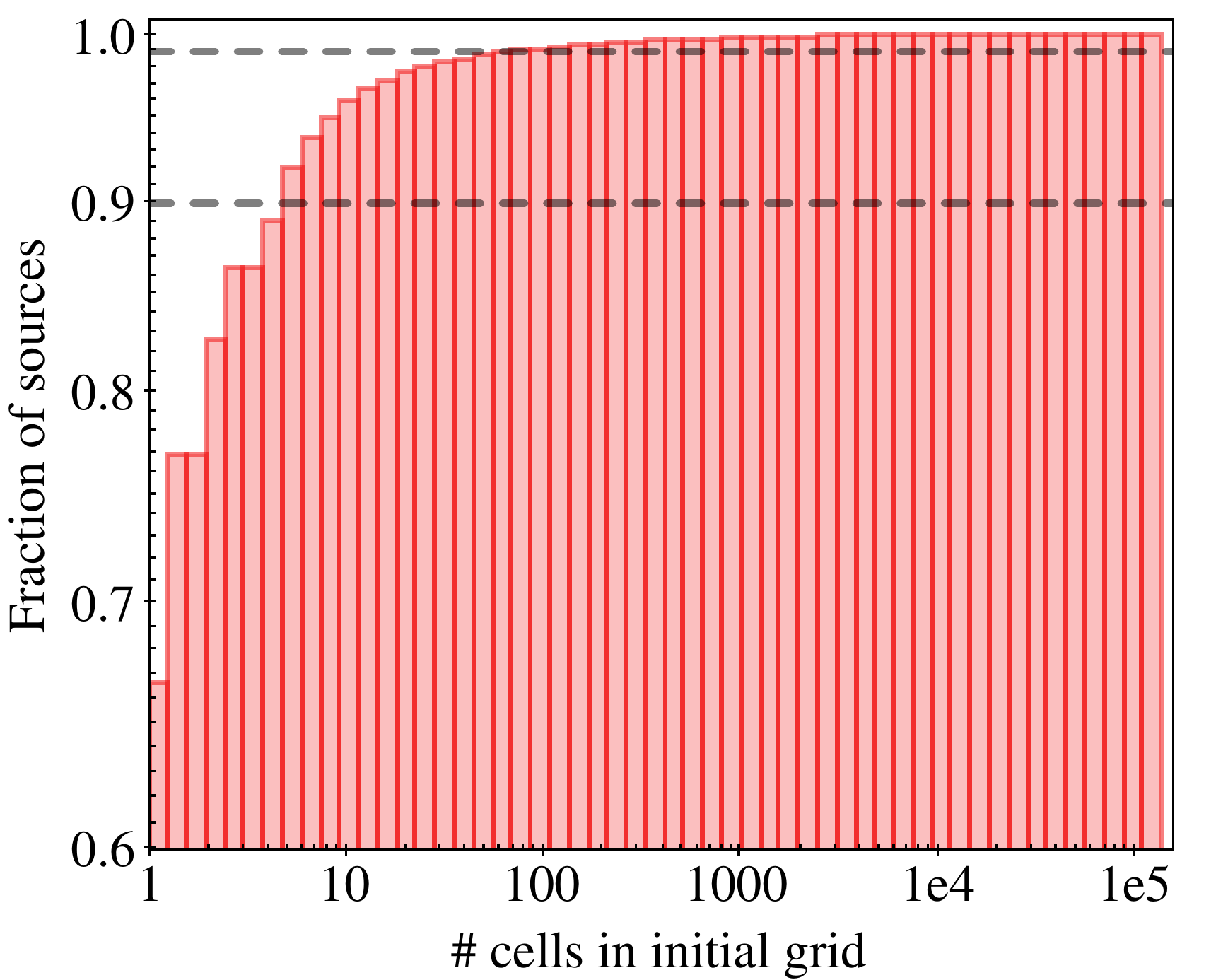}
\caption{Left panel: Distribution of the number of cells in the initial grid and its effect on the Bayesian block algorithm (see Sect.~\ref{sect:bb}). The dashed  line is an upper limit to the number of blocks that can never exceed the number of initial cells. A bug in the code, described in Sect.~\ref{sect:wp2bugs}, prevents us from reaching  this line. The distribution in the left panel clearly shows that this bug has no effect on light curves that start from a grid with $> 10$ cells. Right panel: Cumulative distribution of the number of initial cells. We show the 0.9 and 0.99 values with solid lines. Currently, about 77\% of the sources are not characterised by Bayesian blocks because their initial grid has fewer than three cells. }
\label{fig:bbcells}
\end{figure*}

As a further test, we repeated the same exercise using results from the recent 4XMM-DR9 catalogue. First, variable sources were selected based on the variability flag that is included in the catalogue (set to true if the tail probability of the source is constant $<1\times10^{-5}$). This yields 109 variable light curves, including 82 light curves from the CVS and 5 from the CLVS, and 22 additional light curves that are not mentioned by \citet{pye2015}. By visual inspection, we identified 16 actually variable light curves and 6 likely artefacts among the latter. We also tried to assume a different less conservative threshold for variability (tail probability of a constant fit $<1\times10^{-3}$). This allowed us to select 155 light curves, including 86 from the CVS, 8 from the CLVS, and 43 additional light curves. Of these, 33 feature apparent variability, and 10 are likely artefacts. Fig.~\ref{fig:wp2vs4xmm1} shows the case of the bright star HD 283810, target of Obs.Id. 0203540501, that was included by \citet{pye2015} in their CLVS sample. It is selected by EXTraS/STV as a variable source, but is not flagged as variable by 4XMM. The left panel shows the EXTraS/STV light curve of the source (pn camera, exposure S003). A large flare is apparent in the second half of the observation. The flare was missed by the 4XMM variability analysis because it occurred outside of the GTI. The observation is affected by proton flares: the right panel shows the 4XMM background light curve and the GTI adopted in the catalogue analysis. The resulting tail probability of a constant model in the EXTraS/STV catalogue is $<\,10^{-25}$ (pn camera, light curve with 500 s bins), while in 4XMM (where the flare was excluded) it is  $\sim3\times10^{-3}$. 

Thus, using the EXTraS/STV catalogue, the number of variable light curves identified by \citet{pye2015} can be extended by a factor of two using the same set of observations with relatively little effort. While an astrophysical characterisation of the variable stars that were not studied by \citet{pye2015} is beyond the scope of this work, the figures reported above show that the EXTraS/STV catalogue allows us to perform a very sensitive but robust search for variability in any sample of sources.

\subsection{Known problems and caveats}\label{sect:wp2bugs}
In some light curves, one or more bins can assume negative values. In principle, this is expected because we often subtract the background component in a low-count regime. However, the number of negative bins is larger than expected from simple fluctuations. Figure~\ref{fig:nbins1} and ~\ref{fig:nbins2} show that there are two kinds of problems. 
Figure~\ref{fig:nbins1} shows that most light curves that have bins at least 3$\sigma$ below 0 have only some such bins on average. Light curves that have bins at least 10$\sigma$ below 0 have dozens of such bins. 
Figure~\ref{fig:nbins2} shows that the distributions of the number of standard deviations below zero of all negative average rates for light curves with uniform 500s bins and for optimally binned light curves have a clearly different shape. The bulk of the points within 10$\sigma$ from 0 in the former is not apparent in the latter; the points beyond 10$\sigma$ are instead very similar.
In many cases, light curves have a few negative bins whose errors are underestimated due to the failing Gaussian assumption. Optimally binned light curves are indeed less affected by this issue. In a few cases entire light curves have a baseline much below 0. This is due to problems in background modelling. This interpretation was confirmed by visual inspection of problematic cases, which we found flawed by rare issues in the background characterisation such as bright extended sources in the vicinity of the point source under study. 
These entries are flagged as bad (QUALITY\_FLAG) in the light version of our catalogue. 


We describe below some caveats about the usage of a fraction of Bayesian block light curves. By construction, the number of blocks in the Bayesian block representation depends on the initial segmentation in cells (see Section~\ref{sect:bb}). An initial grid of a few cells results in a light curve with at most several blocks. This is the most common situation: 67\% of the sources start with an initial grid made of a single cell, 90\% start from 5 or fewer cells, and only 1\% of the sources start from a grid with more than 50 cells (see Fig.~\ref{fig:bbcells}). The user should therefore check the number of initial cells through the column BB\_LC\_NCELLS  in the light version of the catalogue. We also warn users that all light curves that should have as many blocks as initial cells are misrepresented as a single block because of a bug in the script that generates the Bayesian block light curves. As shown in Fig.\ref{fig:bbcells}, this bug has no effect on light curves starting from an initial grid with $> 10$ cells.  

Visual inspection of Bayesian block light curves
highlighted recurring narrow spurious features, both flare-like and
eclipse-like, that are made of pairs of closely separated false blocks. The occurrence of this feature is strictly correlated to the number
of expected false blocks. The geometric prior in Bayesian blocks gauges
between the sensitivity to weak real features in the light curve and
the robustness against false features. We simulated constant sources in addition to real observations, spanning a wide range of parameters,
to estimate the number of false blocks in each single light curve.
This number essentially depends on the number of cells in the initial
segmentation that generated the light curve. We included these estimates
for sensitive (BB\_LC3\_NFALSE) and robust (BB\_LC4\_NFALSE)
segmentations.

Finally, we include the description of some caveats affecting our results for few peculiar cases.
As reported in Section \ref{sect:dataprep}, and following 3XMM selections, we excluded time periods with attitude change $>$3'. These changes can lead to an incorrect coordinate conversion within the SAS tools. During these occurrences, our event selection, based on celestial coordinates, could fail to extract events around the selected source but extract events from a shifted region, thus resulting in spurious variability. We have provided the user with a column in the light version of the catalogue, ATT\_FLAG. This reports the maximum attitude variation during the observation in arcseconds.\\
Because background maps are produced with a 1 arcmin smoothing, sources falling at the edge of bright extended sources can be affected by an incorrect background subtraction. Very bright flares ($>$ few counts s$^{-1}$) could be underestimated because we did not consider the loss of counts due to pile-up effects. Moreover, because we did not treat OOT events, light curves of sources falling on OOT trails can be contaminated.\\
Photons from extremely bright optical sources can excite a significant number of electrons in the X-ray CCDs and can be (falsely) recognised as (X-ray) events (this phenomenon is known as optical loading). Sources contaminated by optical loading do not follow the expected PSF. Therefore our modelling over- or under-estimates their count rate. The 3XMM catalogue does not flag such sources.

%
%

\section{Search for pulsations}\label{sect:pulsators}

\subsection{Aims and scope}
\begin{figure}[ht]
\hspace{-4mm}
\includegraphics[angle=-90,width=9.3cm]{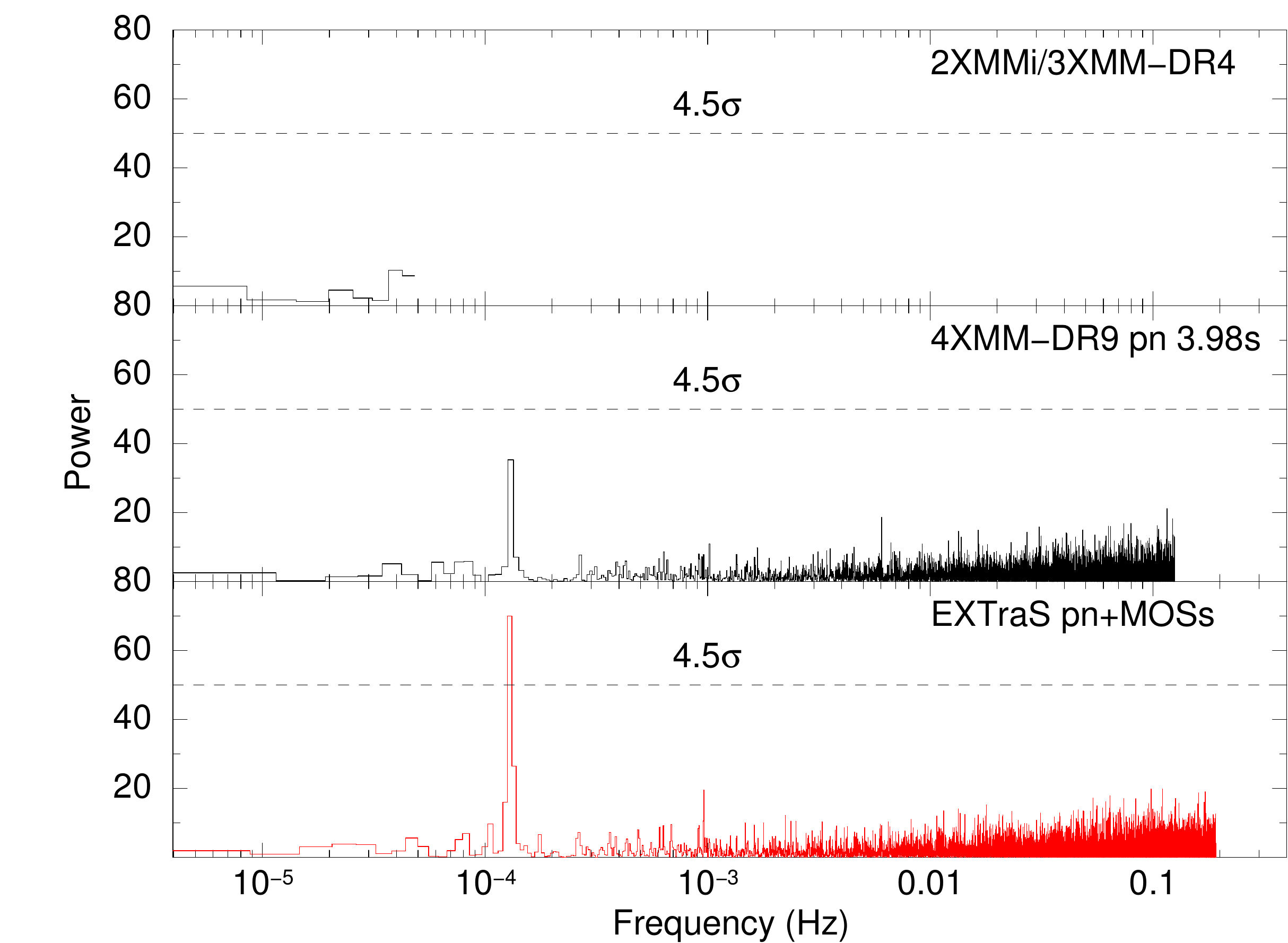}
\caption{EPIC power spectra of one of the new faint ($\sim1.3\times10^{-14}$ ergs\,s$^{-1}$\,cm$^{-2}$ 1-10\,keV flux) X-ray pulsators discovered within the EXTraS project with a period of about 128.5 minutes, namely 3XMM J221900.5+722508 (3XMM id. 233383, 4XMM id. 204025302010035). 
Upper panel: based on 2XMMi/3XMM-DR4 binned light curves (in the specific case, the 3XMM-DR4 time bin is 11.\,020s); mid panel: based on 4XMM-DR9 pn camera binned light curve (bin time 3.98 s); lower panel: from EXTraS catalogue (pn+MOS data, counting 300 photons).
}

\label{fig:unbinned}
\end{figure}

The main goal is to search for signals in all the 3XMM detection time series with more than 50 counts in a systematic and automatic fashion. In particular, our search is optimised for coherent signals, that is, signals that are characterised by only one characteristic variability timescale, as opposed to quasi-periodic oscillations (QPOs), for example, where an interval of characteristic variability timescales is present. For more than 500,000 time series and about one million timing analyses, we searched for coherent signals in a period range spanning from $\sim$150\,ms (in the majority of cases) up to the highest value allowed by the length of each specific time series (observation). In particular, we worked directly on photon arrival times rather than on binned light curves in order to optimise the signal search sensitivity that is strongly dependent on the number of counts and the binning time, among other things.  
This is particular relevant for sources with relatively poor statistics and/or faint signals. In this respect, we note that the corresponding time series of these sources provided by the 3XMM-DR4 products are often heavily rebinned, which hampers a sensitive search for signals; see the upper panel of Fig.\,\ref{fig:unbinned} as an example where the signal at about 1.3$\times$10$^{-4}$\,Hz is not even sampled in the power spectrum of the 3XMM-DR4 light curve. 
While we drafted this paper, the 4XMM-DR9 catalogue and products were released. From the point of view of the analysis discussed in this section, no major changes with respect to 3XMM-DR4 are registered. In particular, although a different rebinning is considered in 4XMM-DR9, including  a pn light curve that is rebinned to few seconds (in most cases 1.46s; see the central panel and caption of Fig.\,\ref{fig:unbinned}), no search for signal is foreseen or performed. 

Our analysis was carried out on both the pn and the two MOS detectors individually for about 50\% of the total FFTs that were carried out during the search  in order to keep the original time resolution and to rely upon unbinned data, and merging all the available data  from the three different instruments. Correspondingly, depending on the observational mode and the position of the source in the CCDs, it is possible to have one, two, or three EPIC time series for each detection  (in a small fraction of cases more than three time series for a detection are available) in order to maximise the statistics and therefore the signal search sensitivity. Again, this is rather important for faint sources and/or faint signals (see the central and lower panel of Fig.\,\ref{fig:unbinned} as an example where the power spectra of the pn only and pn plus MOS time-series is shown, respectively).


\subsection{Data preparation, filtering, and source event selection}\label{sect:dataprep4}

As in the search for short-term aperiodic variability in the previous section, we used the PPS event files. For the pn data we used the same reprocessed data as described in Section \ref{sect:dataprep}  in order to cope with the counting-mode bug  that causes incorrect time tagging of events and strongly affects the search for periodic signals and that is not filtered for  time intervals affected by high particle background.  Good events were selected using the PPS extraction flag parameters, which for the pn camera exclude events near the CCDs borders. By using the events flag \#XMMEA\_2 in the SAS tool \texttt{evselect}, we selected sources near borders. For these sources we performed an additional extraction with the standard \#XMMEA\_EP \& PATTERN<=4 filters, as described in the official SAS threads\footnote{ https://www.cosmos.esa.int/web/xmm-newton/sas-thread-timing}. 
For each source we used the GTIs, the time intervals during which cameras are properly working and looking at the target field, which correspond to the CCD in which the source is observed. Sources that fall near the borders may have photons in two or more CCDs. In these cases we used the GTI of the CCD that contained most of the photons.
When more than one exposure in a single observation was available, the events of single sources were merged with the SAS tool \texttt{merge} if the time resolution used in each exposure was the same. Otherwise, the event files of different exposures were analysed separately. 
After extraction and before the search for signal, we shifted the time of arrival (ToA) of each event to the Solar System barycenter reference frame using the SAS task \texttt{barycen} and the relevant {\tt *ROS.ASC} PPS file. About 99.7\% of the event lists were successfully corrected. When the file was missing (and a correction was not possible), the timing analysis was carried out while a warning flag was set and added to the final database for future checks. For the correction of each events file we used the central coordinates of the corresponding circular extraction region.

\subsection{Search for a coherent signal}
\label{sec:psearch}
The pn and MOS event files extracted and prepared as described above were then ingested into the signal search pipeline. This pipeline is structured as described below.  
    
{\bf Step 1.} It sets the file groups (in order of decreasing time resolution) for the file events of each source in order to cope with the different observational modes and sub-modes of each EPIC detector. 
Sub-modes affect the time resolution of events, which is an important parameter for the timing analysis: the better the resolution, the higher the frequency  range in which we can search for signals. For a given source, the pn and MOSs can have different time resolution, and different sources within the MOSs can have different time resolutions depending on the CCD in which they lie. For each source, a decision tree has therefore been implemented, starting from the instrument and/or science mode event file with the highest time resolution and subsequently adding all the other instrument and/or science mode event files (whenever available) with lower resolution. This approach  optimises the signal search capability in different frequency intervals based on the specific sub-modes of each single source. Correspondingly, for each source in a given observation, we can have from one to several groups of event files where the signal search is applied. \\
When the ratio of the length of the observation and the sampling time is higher than 2 million, the analysis is split into two modes: The first mode is aimed at keeping the maximum Fourier resolution, $1/T$, and rebinning the original sampling time such to have only one interval with two million or fewer time bins. The second mode keeps the original time resolution and cuts the observation into two or more time intervals, each one with 2 million time bins.  
    
    
{\bf Step 2.} It carries out the signal search with the validated detection algorithm for all the groups with more than 50 counts. Different algorithms were taken into account, such as the $Z^2_N$ \citep{z2n}, the Rayleigh periodogram, and the FFT. For several reasons, an FFT was considered the best solution for the specific task or project. The choice was driven by the CPU-time consumption and frequency resolution among other things, independent of the signal frequency itself (1/T rather than P$^2$/2T, where T is the observation length). The adopted FFT includes a logarithmic smoothing algorithm in order to evaluate the spectrum continuum plus a detection algorithm that derives the main signal parameters from the Fourier transform properties (such as period, pulsed fraction, and statistical significance). The smoothing module is needed in order to cope with non-Poissonian power spectrum noise components, which might be present as a consequence of source intrinsic aperiodic variability or background radiation flares (\citealt{israel96}; note that the different length of pn and MOSs time series  also introduces spurious aperiodic variability in power spectra). Correspondingly, a local (frequency-dependent) power threshold level for candidate signals is computed. For the project, we set a 3.5$\sigma$ detection threshold assuming a number of trials equal to the number of FFT frequencies in each power spectral density (PSD; see also Sec.\,\ref{issueswp3}). The inferred main signal parameters are the period 1/$\nu_j$ (where $\nu_j$ is the j-th Fourier frequency), the pulsed fraction (defined as the semi-amplitude of the sinusoid divided by the source average count rate), and the probability of being a noise fluctuation. For the latter quantity, for which we cannot apply the properties of the $\chi^2$ statistics, we refer to \cite{israel96}. Upper limits to the pulsed fraction are inferred at the 3$\sigma$ level if no significant peak is found in the PSD. For the source groups with fewer than 50 counts a FFT is computed but no search is attempted due to the poor statistics.

{\bf Step 3.} It further inspects the candidate signals by means of a Rayleigh periodogram and inferring the pulse shape, pulsed fraction, and period. With the aim of confirming the goodness and source-origin of each detected signal, a number of follow-up analyses were carried out. These include a search for spurious signals with similar frequency in the background event file (one for each CCD and after removing the extraction regions of all detected sources in it), a Rayleigh periodogram with a slight overestimation (about a factor of 10) of the period Fourier resolution ($P^2/2T$), and a light curve folded on the detected period. The latter algorithm does not have an automatic routine to decide whether the detected peak is intrinsic to the source (true) or spurious (false). The decision is left to the archive user and to a more accurate analysis (see also Sec.\,\ref{issueswp3}).
    
\begin{figure*}[ht]
\centering 
\includegraphics[angle=-90,width=8.5cm]{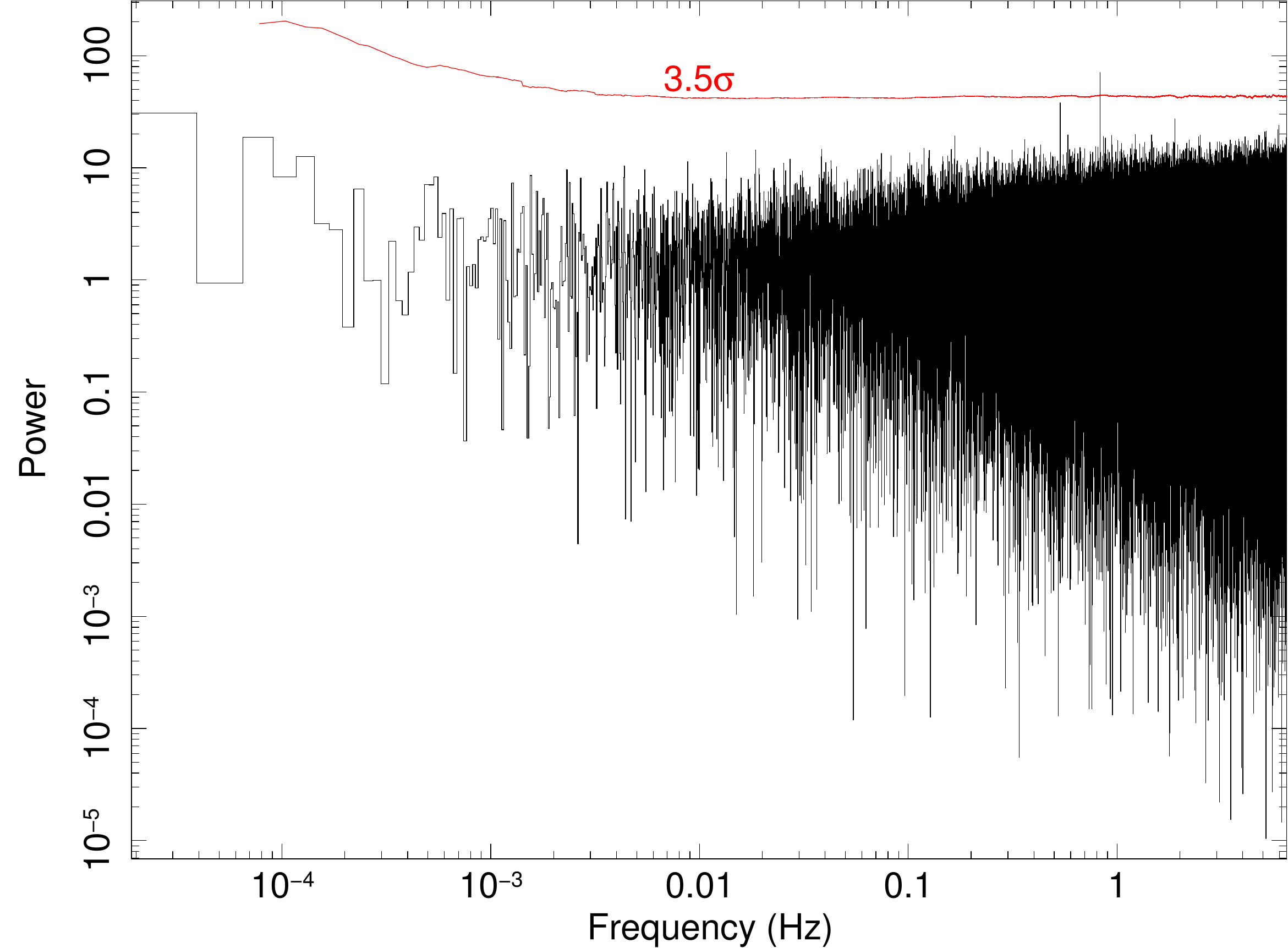} 
\hspace{5mm} 
\includegraphics[angle=-90,width=9.1cm]{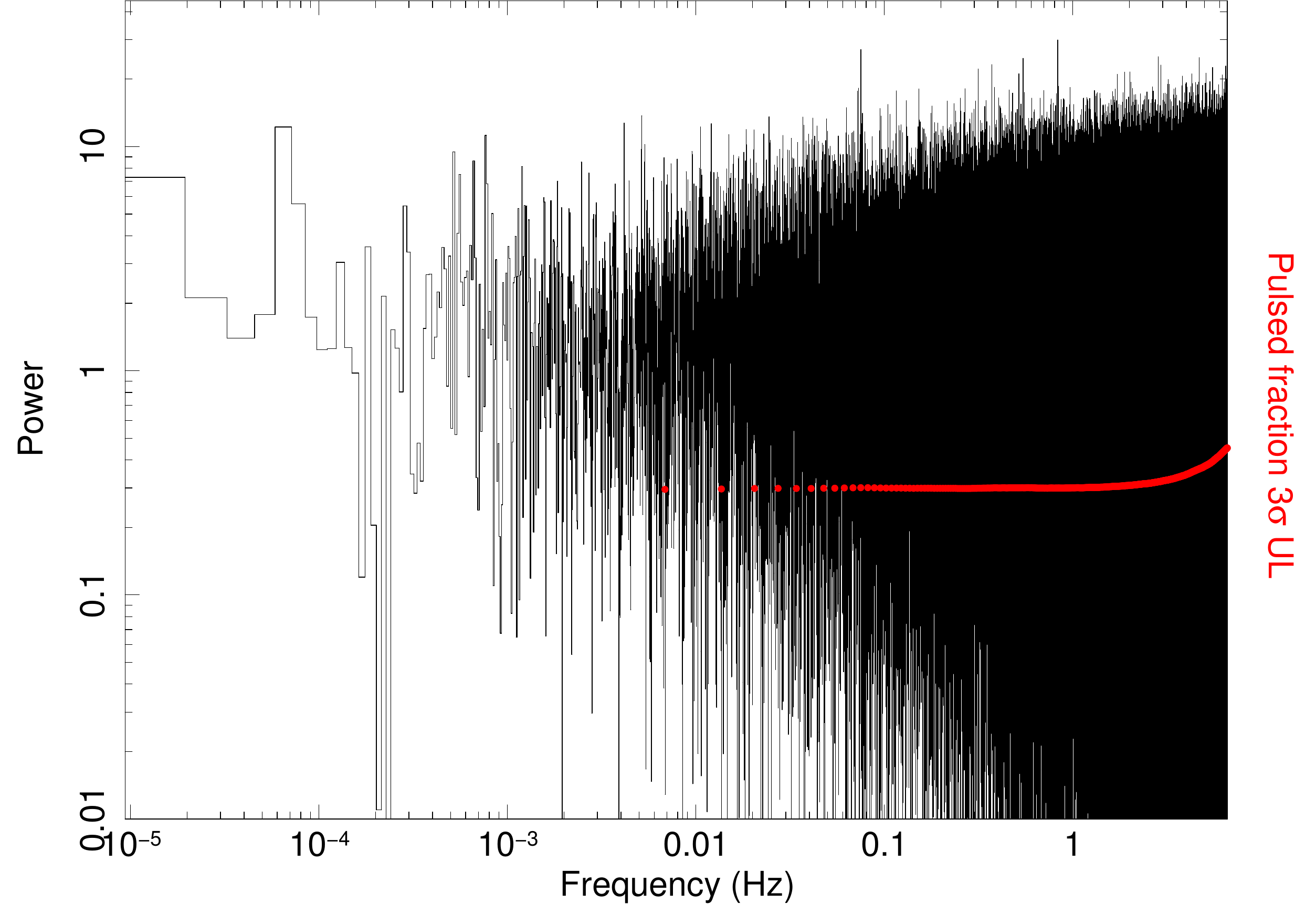} 
\caption{Left panel: EPIC pn power spectrum of 3XMM J004301.4+413016 (Obs. Id. 0650560301) unbinned events, the first accreting NS found in the nearby galaxy M31 (\citealt{eib16,zb17}; see also Section\,\ref{sec:M31}). One peak is above the 3.5$\sigma$ detection threshold (solid red line) and corresponds to the 1.2s period. Right panel: EPIC pn power spectrum of the same source for an earlier pointing (Obs. Id. 0112570101) during which the signal is not detected: the 3$\sigma$ upper limits (red dots) are obtained. Upper limit units are in \%/100 (e.g. 0.1 stands for 10\%). The increasing values towards high frequencies are due to the $x/\sin{x}$  term in the relation between the signal amplitude (pulsed fraction) and FFT powers (\citealt{leahy83}; see also section \ref{issueswp3}).} 
\label{fig:dps}
\end{figure*}

{\bf Step 4.} It creates the database and products.  We divided the information contained in the catalogue into four categories: 1) Observation (OBSID) parameters, 2) single source (SRC) information, 3) parameters of the periodic signals search, and 4) peak parameters. For the last item, different information is stored in the catalogue depending on whether a peak is found above the statistical detection threshold. If no signal is found, the analysis efficiency was recorded, measured in terms of the percentage of Fourier frequency with upper limit values below 100\%, together with the highest value of the ratio of the detection threshold and the powers in the FFT. Furthermore, for each group of event files the following plots were generated (in gif format): the light curve, the power spectrum with the 3.5$\sigma$ detection threshold if at least a peak is found, or the 3$\sigma$ upper limits in the case of negative detection (see Fig.\,\ref{fig:dps}). The Rayleigh periodogram carried out around the detected signal and the folded light curve were also stored.

\subsection{Products}

The products of our search for periodic signals consist of (i) a PSD per detection, (ii) a discrete periodic search (DPS) with the signal detection threshold, (iii) a DPS with the pulsed fraction (PF) upper limits (if no significant signal is found), (iv) a folded light curve (for each signal found), and (v) a catalogue that lists all results of the search for periodic variability, the parameters used for the search, in particular, the smoothing width of the DPS, the probability of the power spectra being chi-squared distributed, the time resolution of the search, its Nyquist frequency, the quantity of frequencies analysed, and the quantity of PSD used in the analysis. The most relevant information about the analysed source is listed there as well: its unique identifier, its celestial coordinates, its International Astronomical Union name, the OBSID of the specific detection, the quantity of significant peaks in its DPS or the efficiency of the PF upper limits (if no peaks were detected), a flag indicating whether the event ToA was shifted to the Solar System barycenter, the instrument that made the current detection, the best period found, its amplitude, the probability of the signal being noise, and if there were no peaks detected, then the ratio of the detection threshold and the power of the highest peak found; and if a signal was detected, its associated power and its Fourier frequency. The catalogue also contains information about the observation in which the source was observed: its exposure time, the quantity of events in the source region, the CCD in which the source is located, a flag indicating if there is a CCD border near the source (see Section \ref{sect:dataprep4}), the observation pointed object, and a link to the Simbad database\footnote{\url{http://simbad.u-strasbg.fr/simbad/}} for the source position. A detailed description and full list of all the catalogue columns and the catalogue itself are included in the EXTraS database. This is fully retrievable via an online web form (see Section ~\ref{sec:archive}).

\subsection{Statistical properties}
\begin{figure}[ht]
\centering 
\includegraphics[angle=0,width=9cm]{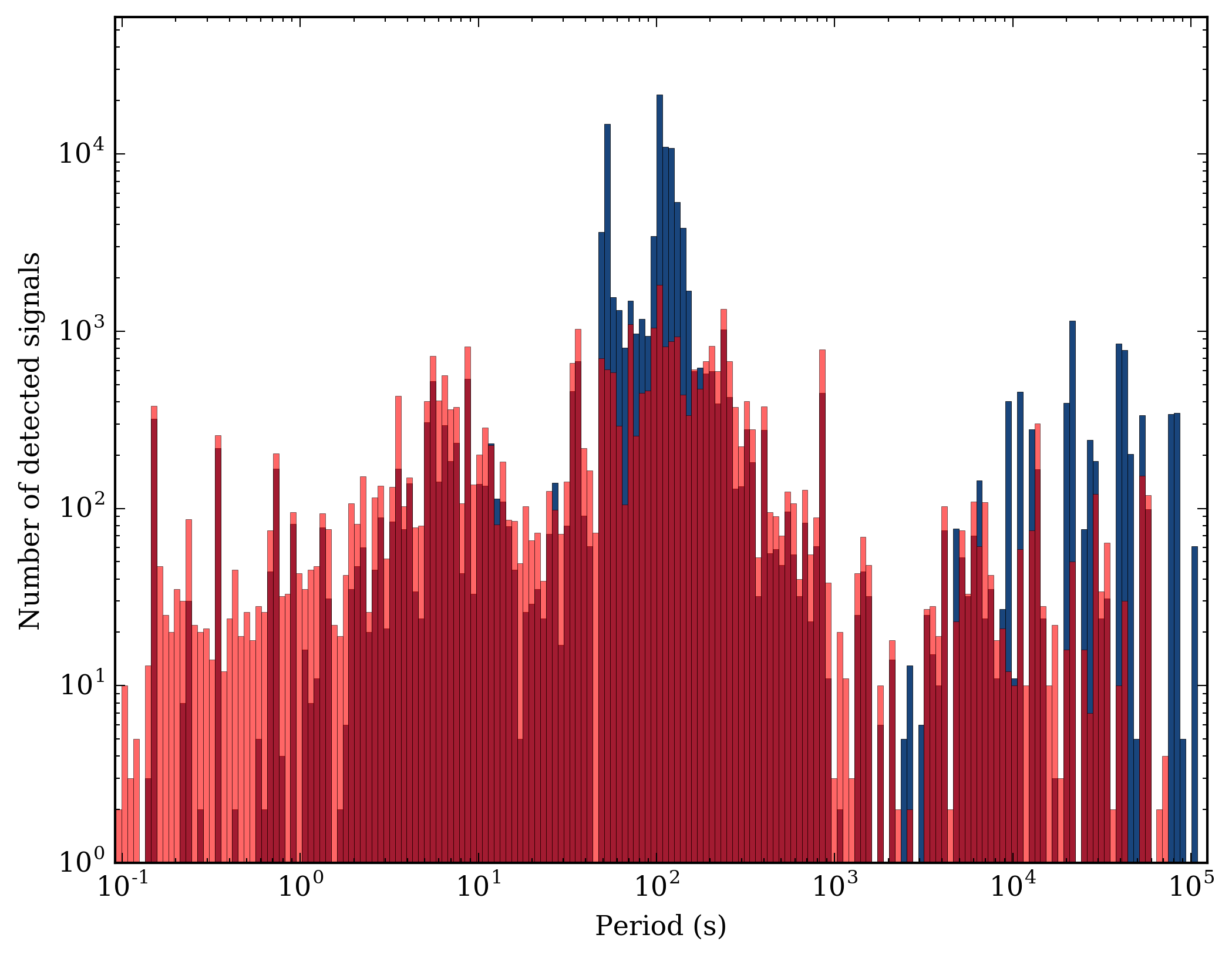} 
\centering  
\caption{Comparison of period distributions between recorded peaks (above the 3.5$\sigma$ threshold) over observations affected by counting-mode switches (blue bars) and a cleaned sample of observations that is not (or almost not) affected by the counting-mode switch (red bars).}
\label{fig:statWP3a}
\end{figure}

In order to validate the reliability of the catalogue parameters, a number of statistical checks were made and the results are briefly outlined below.
As a first step, the distribution of all signals found within the whole 3XMM-DR4 dataset by the pipeline was extracted. This is shown in Fig.\,\ref{fig:statWP3a} (blue region). The main feature is the relatively large and  high (in terms of number) peak at $\sim$100s, which is mainly composed of spurious signals that are due to the counting-mode switch that still affects the results. This is also emphasised by superimposing the distribution of periods detected in observations that are not affected (or are slightly affected) by  counting-mode switches (red region). 
The spurious signals are present in the pn and pn plus MOS time series  FFTs (see also Section\,\ref{issueswp3}). The second less evident feature in the signal distribution is for long periods above about 5.\,000-10.\,000 seconds. These candidate signals are partly due to spurious detections due to intrinsic aperiodic variability of the sources (affecting the low-frequency part of the FFTs) and partly due to time intervals that are affected by high particle background (often related to the counting-mode switch). However, after inspecting a large sample of these signals, we 
found
that there are also genuine signals, although the spurious peaks constitute the majority. 
The third distribution peak is in the range of 5-15 seconds and is  dominated by XMM observations of known rapid pulsators (mainly magnetars). In all cases, a visual inspection is strongly recommended. 

\begin{figure}[ht]
\centering 
\includegraphics[angle=0,width=9cm]{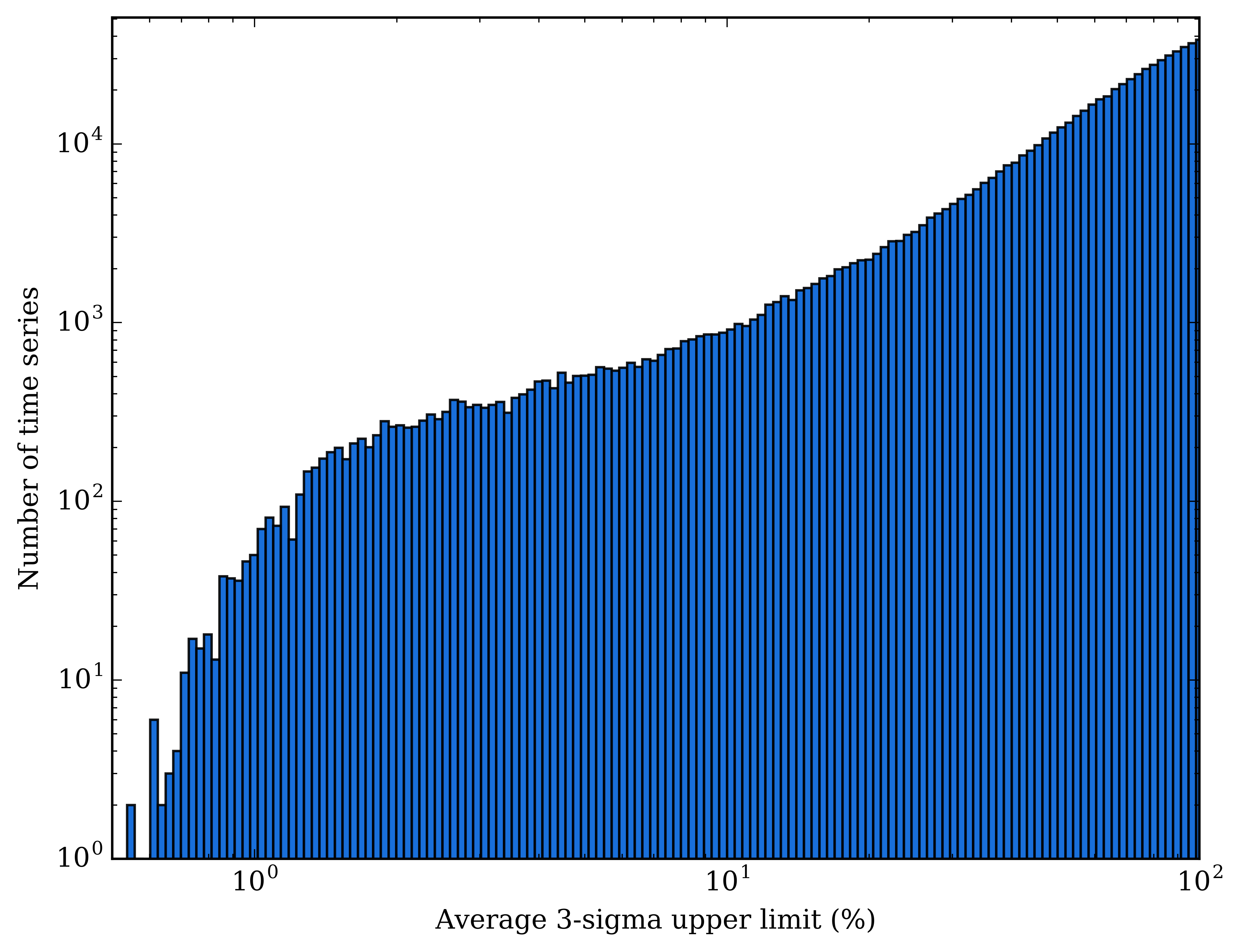} 
\caption{Distribution of the average 3$\sigma$ upper limits to the pulsed fraction (defined as the semi-amplitude of the sinusoid divided by the source average count rate) as inferred for the single-interval FFTs where no signals have been detected.}
\label{fig:statWP3b}
\end{figure}

For all the FFTs (in the maximum Fourier resolution mode) for which no signal was detected, the timing analysis capability of obtaining meaningful constraints on the average values (above all the Fourier frequencies) of the 3$\sigma$ upper limits were derived and stored in the catalogue. The distribution of the  average values of the 3$\sigma$ upper limits is shown in the bottom left panel of Fig.\,\ref{fig:statWP3b}. About 38\% of all the FFTs have PF upper limits (ULs) below 100\%.
In the best (a few) cases, ULs close to 1\% are obtained (for the pulsed fraction, we adopted the definition of  the semi-amplitude of the sinusoid divided by the source average count rate). 

\begin{figure}[t]
\centering 
\includegraphics[angle=0,width=9cm]{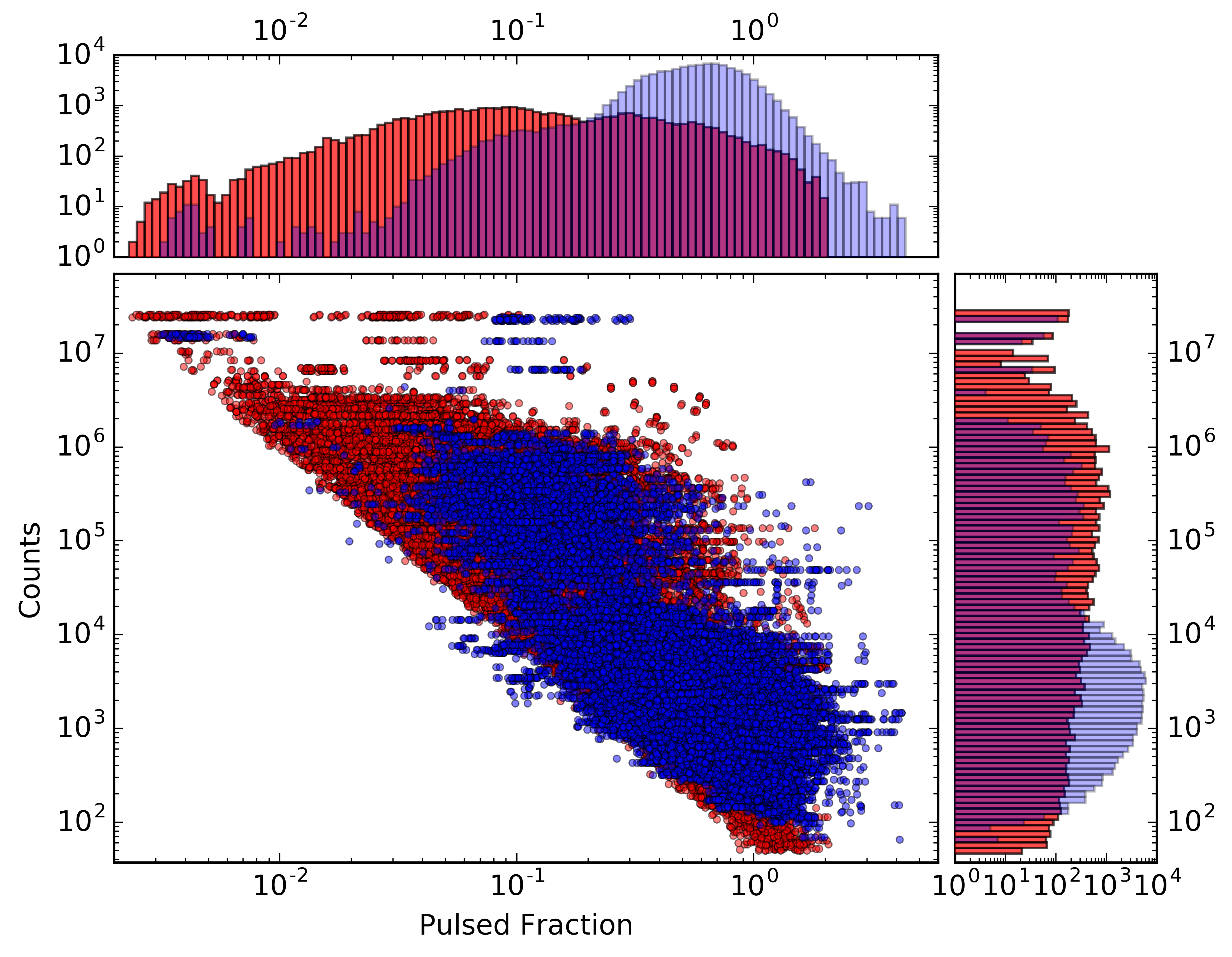} 
\caption{Distribution of all detected period pulsed fractions (defined as the semi-amplitude of the sinusoid divided by the source average count rate) vs. source counts. The blue dots correspond to observations that are affected by counting-mode switches. The red dots correspond to periods detected in observations that are not (or almost not) affected by the counting-mode switch. Histograms of the PFs and counts are presented at the top and left, respectively, for the two datasets.}
\label{fig:statWP3c}
\end{figure}

Finally, the above defined pulsed fractions of detected signals were plotted with respect to the total counts (of the corresponding time series) for periods in observations with none or moderate counting-mode switches (red dots and histograms in the bottom right panel of Fig.\,\ref{fig:statWP3c}) and for periods in observations that are highly affected by the problem (blue dots and histograms). 
The comparison of the two samples clearly shows that for time series with decreasing statistics, the spurious signals have large pulsed fractions.
Although these findings confirm that the majority of signal detection in the $\sim$ 20-200 seconds range is from spurious signals, genuine signals in the same period interval cannot be excluded. Correspondingly, we decided to keep all the detections in the catalogue. We did not reject any period range. More in general, these findings further strengthen the need of carefully inspecting the signal(s) in the catalogue that one might be interested in.

A final check concerns the capability of the smoothing algorithm to recover a white-noise FFT from a noisy FFT in which additional non-Poissonian noise components are present. This control was conducted by means of a Kolmogorov-Smirnoff (K-S) test in which the original FFT was locally (for each Fourier frequency) normalised to the obtained smoothed power spectrum continuum and multiplied by 2N (where N is the number of averaged FFTs) and finally compared with the statistical properties of a pure white-noise FFT (see \citealt{israel96} for extensive simulations and checks).  In this framework, K-S numbers of the statistics close to 1 mark a good agreement, that is, the capability of the smoothing algorithm to model the power spectrum continuum well. 
As expected, the greatest part of the smoothed FFTs has K-S probabilities close to one. The number decreases for decreasing probability values. We note that the FFTs in which signals are detected are less cleaned on average. the smoothing algorithm is less efficient in modelling the peaks themselves (as expected). Correspondingly, it is worth emphasising that a low K-S value in the catalogue for an FFT with a signal does not necessarily mark a failure of the smoothing algorithm. We emphasise that the K-S is mainly used as a control test to confirm whether there are substantial issues in running the algorithm, and it does not affect the solidity of the results in any way.

\subsection{The case of M31}
\label{sec:M31}
The EXTraS catalogue for periodic variability has a huge potential for new discoveries, as demonstrated by our detection of pulsations in the ultraluminous X-ray sources (ULXs) NGC 7793 P13 \citep{israel17a} and NGC 5907 ULX-1. The latter is the most extreme accreting pulsar ever observed \citep{israel17b}. We do not describe these findings in this section, but show and briefly discuss the case of the nearby Galaxy M31. With tens of XMM archival observations, its relatively large number of X-ray sources, and properties similar to those of the Milky Way, M31 is one of the best regions of the sky in which to search for X-ray pulsations. In particular, despite the similarities with the Milky Way and the extensive monitoring of M31 since the Einstein mission, no accreting X-ray pulsar was found before the beginning of the EXTraS project. Within the project,  85 pointings of M31 have been analysed, and signals were searched for among 14438 detections (with more than 50 counts) and 36584 FFTs. About 860 peaks above the 3.5$\sigma$  limit have been found for 498 sources. When objects with known periodicities (mainly orbital periods) were removed, only two sources showed convincing signals in their power spectra. These objects are 3XMM J004301.4+413017 (hereafter J004301) with a period of about 1.2\,s and 3XMM J004222.9+411535 (hereafter J004222) with a period of about 464\,s. No periodicity is reported for them in the literature.  In Table\,\ref{table:M31} we list the main parameters of the EXTraS catalogue for the filtered signals detected by the pipeline in M31 (see also section \ref{issueswp3} for a comment on the low-probability values of J004222). The ``Prob.'' column reports the probability of the detected peak to belong to the power estimate noise distribution. Values above 1 are often due to the presence of strong non-Poissonian noise components lying below the peak(s).  Below we report the main characteristics of the latter two sources and the corresponding signals.

\begin{table*}
\caption{M31 EXTraS signals}             
\label{table:M31}      
\centering                          
\begin{tabular}{l l l l l l l l l}        
\hline\hline                 
ObsId & Inst. & N$_{\gamma}$ & Peaks & Period & Frequency & PF & Prob. & Power  \\    
& & \# & \#  & (s) & (Hz) & ($\times$100 \%) & & \\
\hline             
\multicolumn{9}{c}{3XMM J004301.4+413017}\\
\hline
0650560301 &EPXPN & 1286 & 1 & 1.2038 & 0.8307 & 0.44 & 3$\times10^{-8}$ &       60.07 \\        
0505720301 &EPXPN & 870 & 1 & 1.2036 & 0.8309 & 0.42 & 3$\times10^{-5}$ & 46.45 \\
\hline                                   
\multicolumn{9}{c}{3XMM J004222.9+411535}\\
\hline
0600660501 &EPXPN       &13463  &1      &457.9091       &0.00218 &0.16 &87.63   &70.91  \\
\hfill ---  &EPXPNM1M2  &22843  &1      &463.0261 &0.00216 &0.51 &1.17  &149.61 \\
0650560601 &EPXPN       &10623  &2 &469.0776 &0.00213 &0.15 &0.30       &82.35  \\
\hfill ---   &EPXPN &10623 &2   &463.4261       &0.00216        &0.18   &31.74  &124.56 \\
\hfill ---      &EPXPNM1M2 &18042       &3 &468.1143 &0.00214 &0.42 &0.39       &133.64 \\
\hfill --- &    EPXPNM1M2 &18042 &3 &463.0261   &0.00216        &0.48 &4.08     &176.06 \\
\hfill ---      &EPXPNM1M2      &18042  &3      &232.7782       &0.00430        &0.34   &4.30 &86.55\\
\hline
\end{tabular}
\end{table*}

\begin{figure*}[ht]
\centering \includegraphics[angle=0,width=17cm]{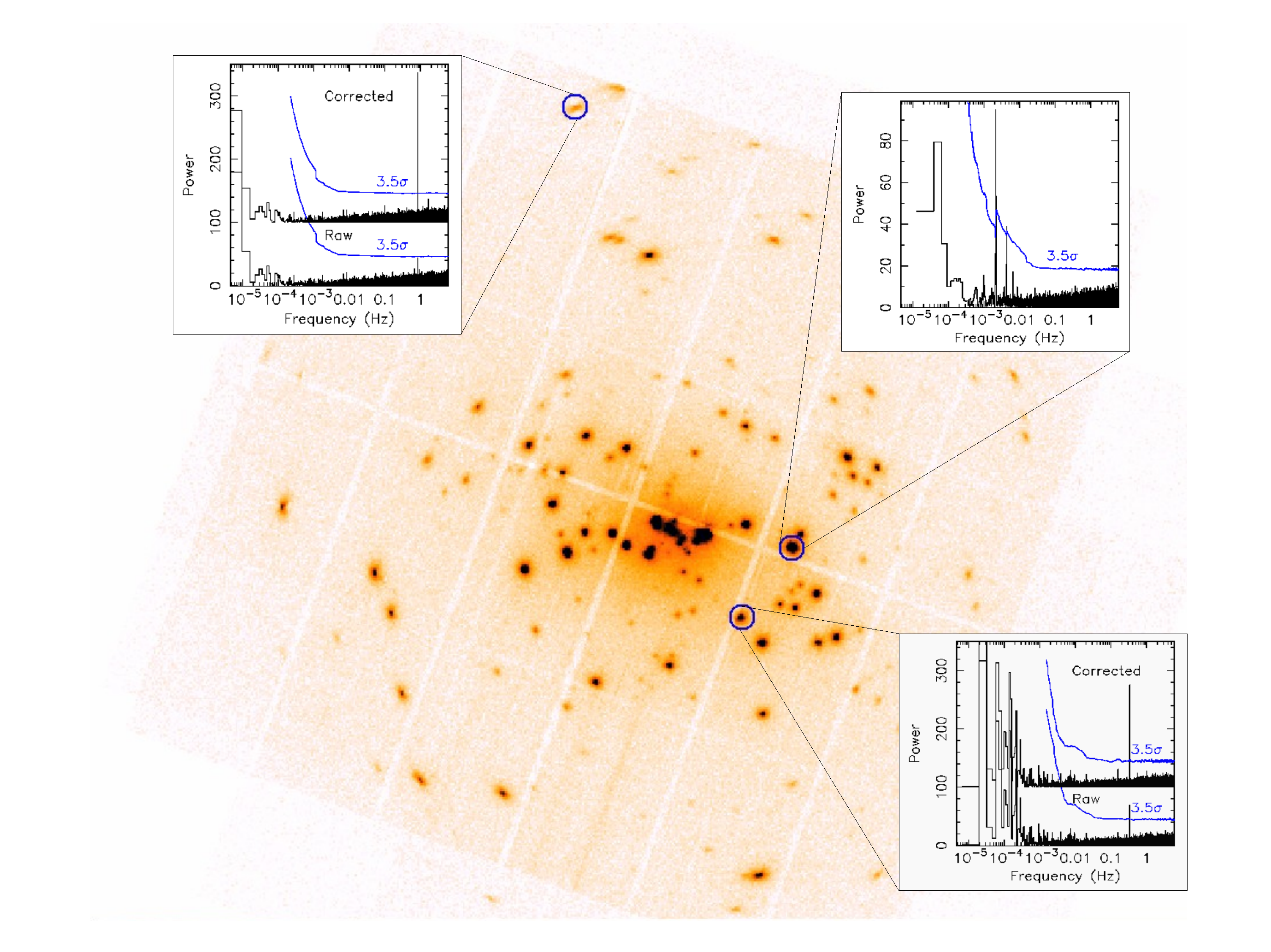}
\caption{XMM pn plus MOSs cleaned image of ObsId  0112570101 (64ks effective exposure time) for the galaxy M31. The insets show the {\rc pn} power spectra (solid black lines) together with the local 3.5$\sigma$ detection threshold (solid blue lines) obtained from  the  unbinned event lists of the three pulsating sources discovered during the EXTraS project and discussed in Section\,\ref{sec:M31}: J004301 (ObsId 050572030), J004222 (ObsId 0600660401, 0600660501, and 0650560601 together) and J004232 (ObsId 0764030301) from top to bottom. In the cases of J004301 and J004232, the discovered X-ray pulsars revolve around a companion star. In the corresponding insets we show the power spectra with (corrected) and without (raw) the best inferred orbital corrections (shifted by 100 in power on the y-axis for clarity). }
\label{fig:M31}
\end{figure*}

{\bf 3XMM J004301.4+413017:} The 1.2\,s period signal from this source testifies to the spin of the first accreting X-ray pulsar ever discovered in M31 \citep{eib16}. The 1.2\,s coherent signal is affected by the Doppler motion of the neutron star around its companion in a  1.27-day orbit. Seven further detections of the 1.2\,s signal were obtained after correcting for the orbital motion in the event files of all the M31 pointings where J004301 was detected (see also Fig.\,\ref{fig:M31} and the upper left inset). This  allowed the timing parameters to be sampled as a function of time over a baseline of about 11 years. The nature of the binary system is still unclear and ranges from an intermediate-mass X-ray binary similar to Her X-1 in our Galaxy to a peculiar low-mass X-ray binary such as 4U1822--37 or 4U1626--67 to the slowest spinning neutron star in a globular cluster (\citealt{eib16,zb17}). Regardless of its real nature, J004301 represents a milestone in the study of extra-galactic X-ray pulsars.

\begin{figure}[ht]
\includegraphics[angle=-90,width=9cm]{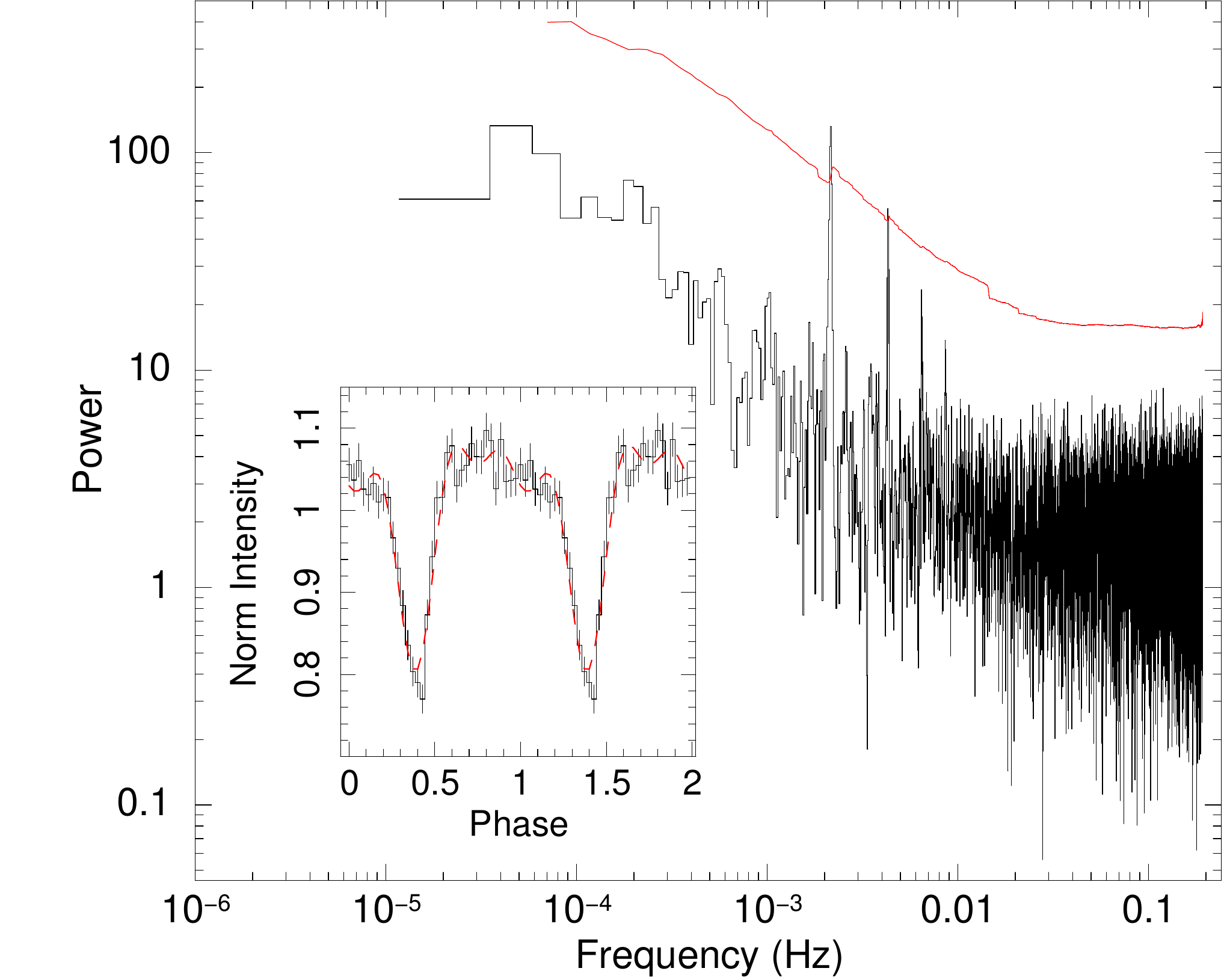}
\caption{Cumulative power spectrum of pn plus MOSs data for observations 0600660401, 0600660501, and 0650560601 of J004222. Superimposed, we show the local 3.5$\sigma$ detection threshold (solid red line): both the 464\,s fundamental and first$^{}$ harmonic peaks are detected at a $\sim$6$\sigma$ confidence level. The light curve folded to the best period is shown in the inset for the same data.}
\label{fig:J004222}
\end{figure}

{\bf 3XMM J004222.9+411535:} The nature of this source is unclear. In particular, it might be  a super-Eddington accreting neutron star or black hole at the distance of M31 or a foreground closer  cataclysmic variable. Correspondingly, the  $\sim464$\,s signal might be ascribed to the spin of an X-ray pulsar or the orbital period of a compact low-mass X-ray binary, if at the distance of M31, or to the spin of an accreting white dwarf (likely an intermediate polar) if within the Milky Way (see Fig.\,\ref{fig:M31} and upper right panel of Fig.\,\ref{fig:J004222}). A more detailed analysis of the whole sample of XMM data for J004222 has revealed that the modulation is detected at high confidence level in three observations over a baseline of one year, during which the flux was significantly higher than the remaining pointings and with virtually no change in the period over the same time interval. All these findings together disfavour the super-Eddington accreting X-ray pulsar  scenario. \\    

During the last year of the EXTraS project, we followed two distinct data analysis approaches. We started analysing the new archival XMM observations of M31 that are not included in the 3XMM-DR4 release, and we developed more sophisticated timing analysis pipelines aimed at taking the possible presence of a strong period first derivative of the putative signal and/or an orbital motion of the compact object (causing the signal) around its companion star into account. During this process, we detected a new signal from the same source, namely 3XMM J004232.1+411314 (hereafter J004232) by using the EXTraS pipeline on new archival data and by applying the newly developed pipelines to old data. Although this result is not included in the EXTraS catalogue, we briefly comment on it because it is a natural evolution of the project.  

{\bf 3XMM J004232.1+411314:} Pulsations at about 3\,s have been detected from this bright hard X-ray source located at 3.7' from the bulge of M31, and it is known to show dips with a likely orbital period of about 4.01\,hr  \citep{guill18,marelli17}. By correcting the archival data for the unknown orbital parameters, we  detected the 3\,s signals from nine datasets over a baseline of 16 years (see also Fig.\,\ref{fig:M31}). J004232 is another milestone in the study of extragalactic X-ray pulsars. It is the first low-mass X-ray binary hosting a young magnetised neutron star (rotating at or close to its equilibrium period $P_{eq}$) outside the Milky Way, a rare evolutionary path for a binary system. Alternatively, it might be a mildly magnetised NS (rotating close to $P_{eq}$; \citealt{guill18}). 

\subsection{Known problems }
\label{issueswp3}
As for any adopted approach or algorithm, a number of assumptions and/or approximations potentially affect the signal search capability in this case as well. We discuss these below.

{\bf Background:} FFTs have been obtained without taking into account the background component, mainly because a relatively constant count rate background level, even when it represents a significant fraction of  the periodic source count rate, does not affect the powers of the noise and of the signal. The statistical significance of the signal is not affected either. What is affected is the pulsed fraction of the signal, which must be evaluated in a different way by the catalogue user. The situation is different for a highly variable background (proton flares), which introduces non-Poissonian components in the power spectra. Nonetheless, the detection algorithm we used takes into account any additional noise components (regardless of their instrumental or source-intrinsic origin). On the other hand, the advantage of considering the whole observation length is reflected in a higher Fourier resolution (important in the search for coherent signals). Correspondingly, the pulsed fractions stored in the catalogue represent a lower limit and need to be carefully inferred by the catalogue user by subtracting the corresponding background level.  

{\bf Number of trials:} In principle, the number of trial periods that should be considered to infer the probability of each candidate signal to be a noise fluctuation is the total number of Fourier frequencies in all the FFTs carried out in the whole project. However, this precept cannot be applied for two main reasons. One is that the total number of searched sources and Fourier frequencies are unknown until the end of the project (the search for coherent signals in the XMM archive is an ongoing project, and  the total number of final trials is therefore still unknown). More importantly, the second reason is that a good number of sources has been observed more than once with Chandra and/or XMM-Newton (or with other missions). We therefore preferred to select the candidate signals based solely on the statistical properties of each individual time series so as to leave open the possibility to later confirm the recurrence of the same signal within the project or confirm based on data from other missions (Swift, NuSTAR, Suzaku, ASCA, etc.). This recipe has been adopted for a similar project on the Chandra archive, namely CATS@BAR in \cite{israel16}, and it proved to be rather efficient, with about 10 signals confirmed by further Chandra pointings carried out during the 20-year interval of the project, and about 20 signals confirmed by archival data from other missions and/or follow-up observations.  
Furthermore, in the case of XMM, the goodness of a candidate signal detected in the time series of one EPIC detector can also be verified by means of the other cameras within the same observation.

{\bf Spurious Signals:} As already discussed above (see Section\, \ref{sect:dataprep4}), independently of the other work packages of the EXTraS project, we found several spurious signals in those observations during which EPIC cameras switched to the so-called counting mode. Correspondingly, in order to minimise the spurious signal and maximise the signal detection capability of the pipelines in the affected  period interval (mainly in the $\sim20-200$ second interval), we relied upon reprocessed data provided within the project. This solved the problem for the greatest majority of the time series. Still, a significant fraction of spurious detections occurred within the same period interval for reprocessed data due to the timing properties (distribution and/or length) of the GTIs introduced to correct for the counting-mode switches. Nonetheless, these spurious signals are relatively easy to spot: First, in most cases they present a very wide profile in the PSD (similar to a QPO component), and second, they are often present in other sources of the given observation. Generally, we urge users to carefully inspect the corresponding PSDs of the signal under investigation (verifying if similar signals have been detected in PSDs of other sources of the pointing).

{\bf Low-frequency signal sensitivity:} The FFT capability of recovering the signal power, and therefore the intrinsic signal detection efficiency, is known to be maximum towards the first Fourier frequencies due to the $(x^2/\sin^2{x})$ term in the pulsed fraction formula (see eq.\,10 and Fig.\,3 in \citealt{leahy83}; $x=\pi j /N,$ where N is the number of time bins and $j$ the $j$-th Fourier frequency). However, the logarithmic smoothing algorithm adopted here (in order to cope with the low-frequency noise) is such that there is a significant decrease in signal detection sensitivity in the same frequency interval if low-frequency noise components are present in the FFT.  Correspondingly, it is very likely that many low-frequency signals have not been detected by the algorithm, even though they can easily be spotted by a visual inspection. Moreover, the algorithm provides underestimated probability values for detected signals at low frequencies, in particular when low-frequency noise components are present. Correspondingly, no probability filter has been applied to the detected signals and all the detections have been stored in the catalogue. Therefore low-probability values in the catalogue for detected signals at low frequencies do not necessarily imply a weak or spurious detection: A visual inspection is strongly suggested in this case as well.   

Finally, we note that different new versions of the catalogue and products have been obtained with the aim to mitigate some of the aspects and limits we have presented in this section. In particular, we mitigated the counting mode effects and increased the capability of detecting low-frequency signals (see also Fig.\,\ref{fig:statWP3a}). Relying upon a personal effort basis, we will try in the future to update and upgrade the EXTraS database.  

%
%

\section{Search for new transients}\label{sec:transients}

\subsection{Aims and scope}\label{TransAims}
The goal is to find new X-ray transient sources, that is, sources that can be detected in a short time interval, but not by a time-integrated analysis of the whole XMM observation. 
This happens to strongly variable sources that are too dim to emerge from the background of a long observation or that are bright enough only during periods of high particle background that is removed by the standard analysis. Such sources are thus not listed in the XMM serendipitous source catalogue.
We implemented a detection algorithm that (i) applies existing source detection tools to time-resolved images, and (ii) compares the positions of the detected sources with those of the source list included in the PPS products of the full observation.
To identify the time intervals containing the flare candidates, we applied a Bayesian block analysis \citep{Scargle2013} of time variability in different regions of the EPIC detectors.
The search for short X-ray transients was performed by systematically running our software pipeline on all the XMM-Newton observations from which the 3XMM-DR5 catalogue \citep{Rosen2016} was derived.
After this step, we carefully selected the high-confidence transients through visual screening of the pipeline products in order to distinguish astrophysical transients from spurious transient candidates and to study them in detail.
Our search can be divided into different steps that we describe below.
\begin{enumerate}[i]
\item 
Data cleaning and preparation. The aim is to select, filter, and format the data of the observations for the analysis.

\item 
Time interval construction and source detection to be applied to these intervals. Depending on the selected options, the time intervals can have a fixed or variable (optimised through a Bayesian block analysis) duration, and the detection algorithm can also operate with different parameters with respect to the full observation.

\item 
Position matching to identify transient sources. The aim is to compare the source lists obtained in the full observation (list present in the PPS products) and in each time interval in order to identify new sources, which we define as transients.

\item 
Position matching to compare results in different instruments, bands, and catalogues. They are needed to identify the presence of the same transient or variable object in different lists of candidates obtained in the same observation with different options.
\end{enumerate}

\subsection{Overview of the pipeline}
The software tools run by our pipeline are combinations of  C-shell scripts, C++, and Python programs and already existing FTOOLS (HEASOFT version 6.15.1), and SAS (version 14.0) tasks.
The datasets selected for our analysis include observations of different durations, operating modes (full frame, extended full frame and large window for the pn, full frame and, partially, any other mode for the two MOSs), and targets (young star clusters, nearby galaxies, and extragalactic fields, including multiple visits of the same objects, in some cases at different off-axis positions).
Several fields are very crowded and contain regions of bright diffuse emission.

Here is a summary of the main steps of the pipeline:


{\bf Step 1.} Production and standard cleaning (PATTERN 0-4 for the pn and 0-12 for the MOS, FLAG=0 to avoid pixels close to CCD edges and dead columns) of the event file. The pn raw data (at ODF level) were reprocessed in order to correct the PPS event files for the counting-mode bug reported in Appendix~\ref{appendix:SAS_issue}. The time intervals with a high background rate were not filtered out.
In order to maintain consistency with other work-packages, we decided to barycenter the data, that is, correct the arrival time of each event for the satellite orbit, as if it were detected in the reference frame of the Solar System barycenter.


{\bf Step 2.} Source detection based on the SAS emldetect task was performed on snapshot images obtained by dividing the observations into adjacent time intervals of a fixed duration, or into variable time intervals optimised through a Bayesian block analysis.
To maximise the sensitivity, we decided to analyse the data of the combination of the three EPIC cameras, selecting the seven energy bands included in the 3XMM-DR5 catalogue \citep{Rosen2016}: 0.2–0.5 keV; 0.5–1 keV; 1–2 keV; 2–4.5 keV; 4.5–12 keV; 0.5–4.5 keV; and 0.2–12 keV.


{\bf Step 3.} Matching positions of all point-like sources (i.e. with null extension according to emldetect) detected at step 2 with the reference sources available from PPS products. When no counterpart (within a given tolerance accounting for both statistical and systematic uncertainties) is found, a ”transient” flag is set.


{\bf Step 4.} Identification of transient sources, avoiding duplications within the same observation (transient sources with consistent positions detected in different snapshots are considered the same object, and the time intervals of these snapshots are registered). A transient is defined as any source flagged as ”transient” at step 3 and with a detection likelihood above a given threshold ($\rm DET\_ML$\footnote{\url{http://xmm-tools.cosmos.esa.int/external/sas/current/doc/emldetect/node3.html}}$>$6, the standard detection threshold adopted for the XMM-Newton source catalogues).


\subsection{Time intervals with fixed duration}

As a first step, we implemented a software pipeline to perform the source detection on images of fixed time duration and compare its output with the source list of the full observation that is included in the PPS products. We define transient candidates as all the sources detected in at least one time interval, but with no counterpart in the PPS source list.
This pipeline was systematically run on all EPIC observations included in the 3XMM-DR5 catalogue \citep{Rosen2016} with time intervals of 1000 and 5000 s. This analysis required more than 45,000 computing hours, and produced a very large number of transient candidates: 104,583 and 95,410 sources for the 1 ks and 5 ks time bins, respectively. Because only point-like sources are expected to be variable, we could consider as promising transient candidates only those with $\rm EXT$=0, but their number was still very large (80,211 for 1 ks and 60,883 for 5 ks) and the manual screening of a random sample unveiled a very high fraction of false positives (mainly spurious detections close to bright and/or extended sources).

\begin{figure*}[ht]
\begin{center}
\includegraphics[scale=0.3,angle=0]{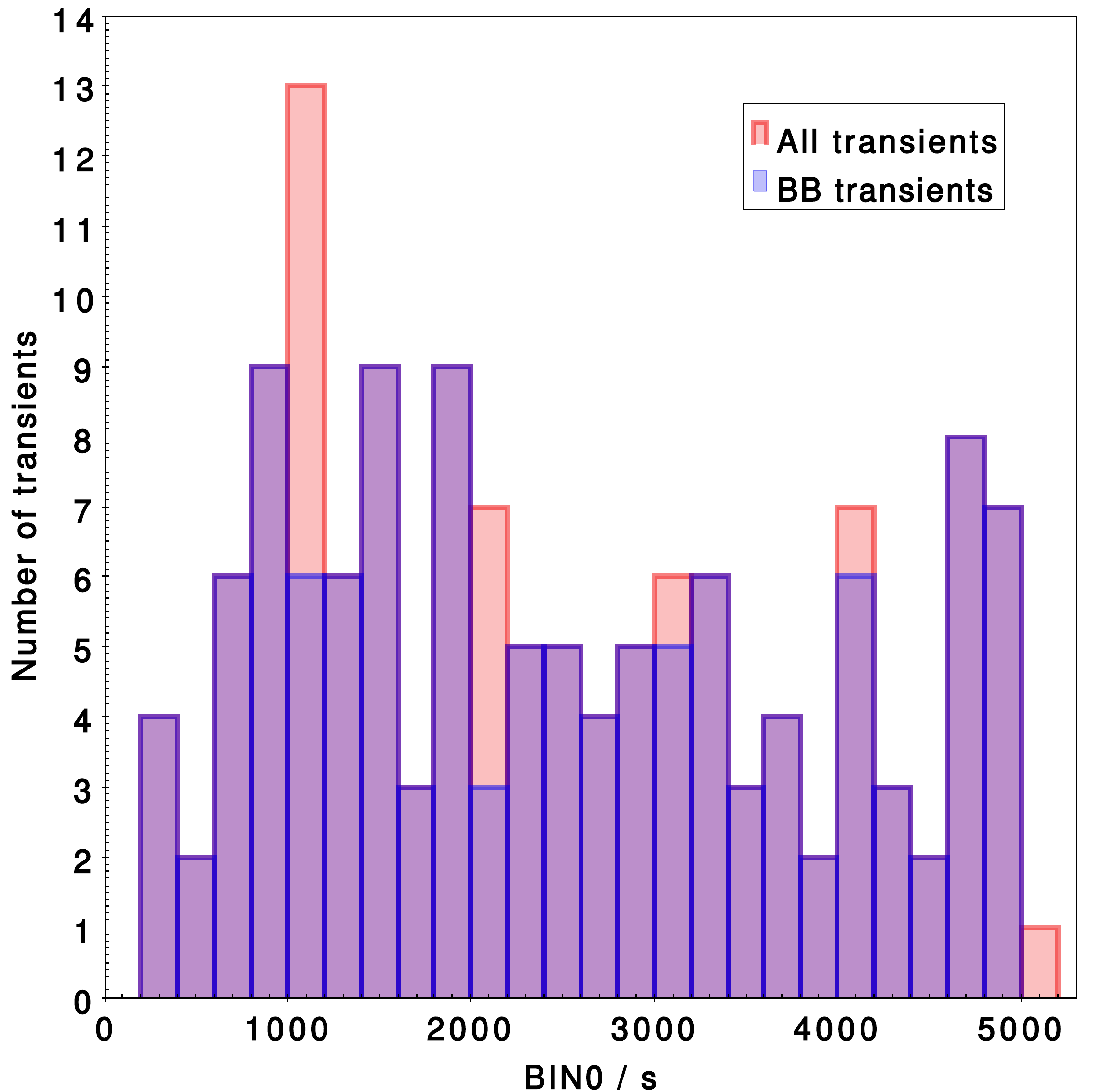} 
\includegraphics[scale=0.45,angle=0]{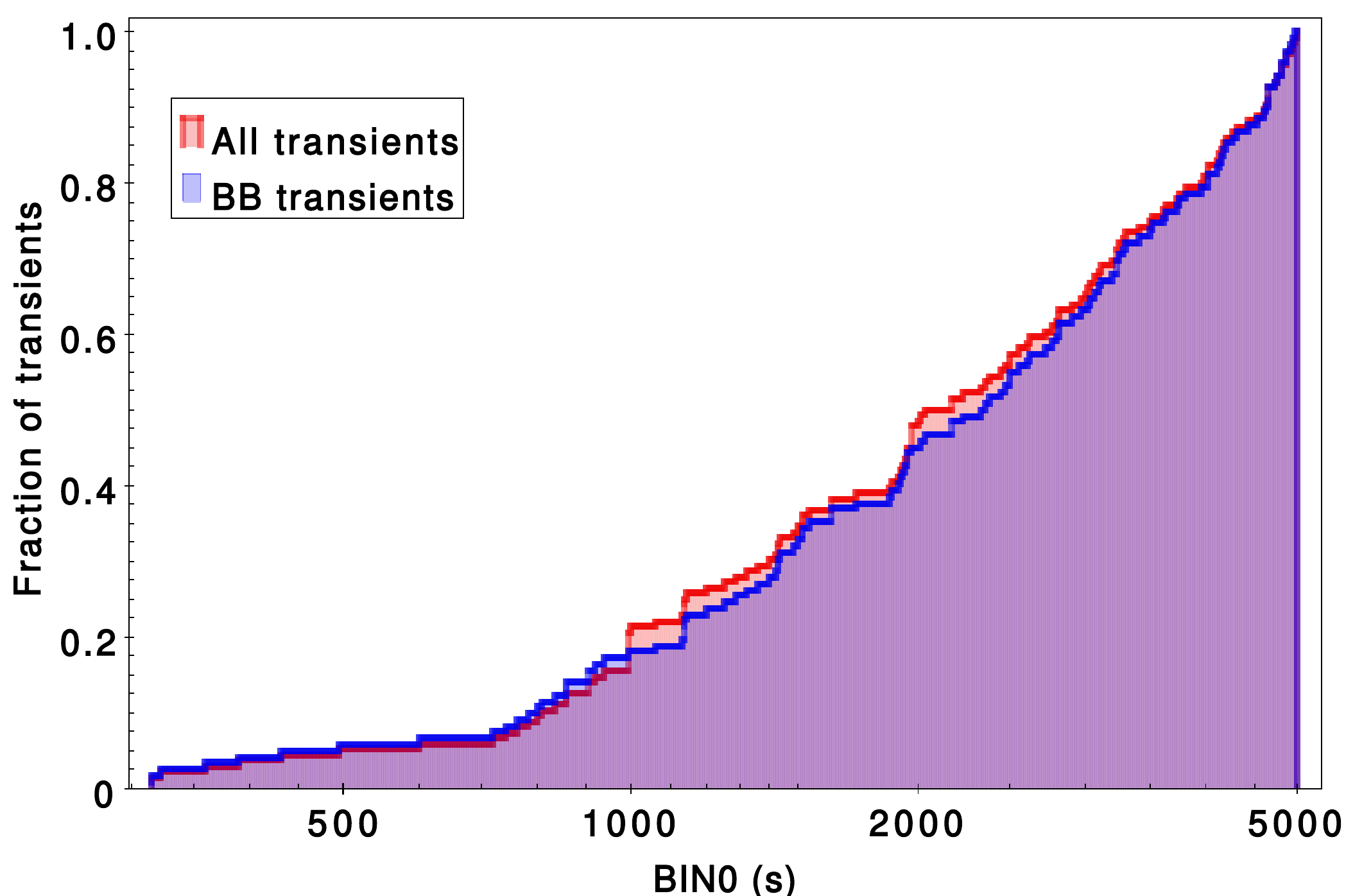} 
\end{center}
\caption{\label{fig_histoBIN0cumul} \footnotesize
Differential (left panel) and cumulative (right panel) distribution of transient durations (in seconds), defined as the length of the time interval where the source was most significantly detected. The subset of transients discovered using the Bayesian block algorithm is indicated (blue) to exclude the sources discovered with the {\it \textup{close-to-source}} algorithm, whose durations can only be integer multiples of 1000 s.} 
\end{figure*}
\begin{figure*}[ht]
\begin{center}
\includegraphics[scale=0.4,angle=0]{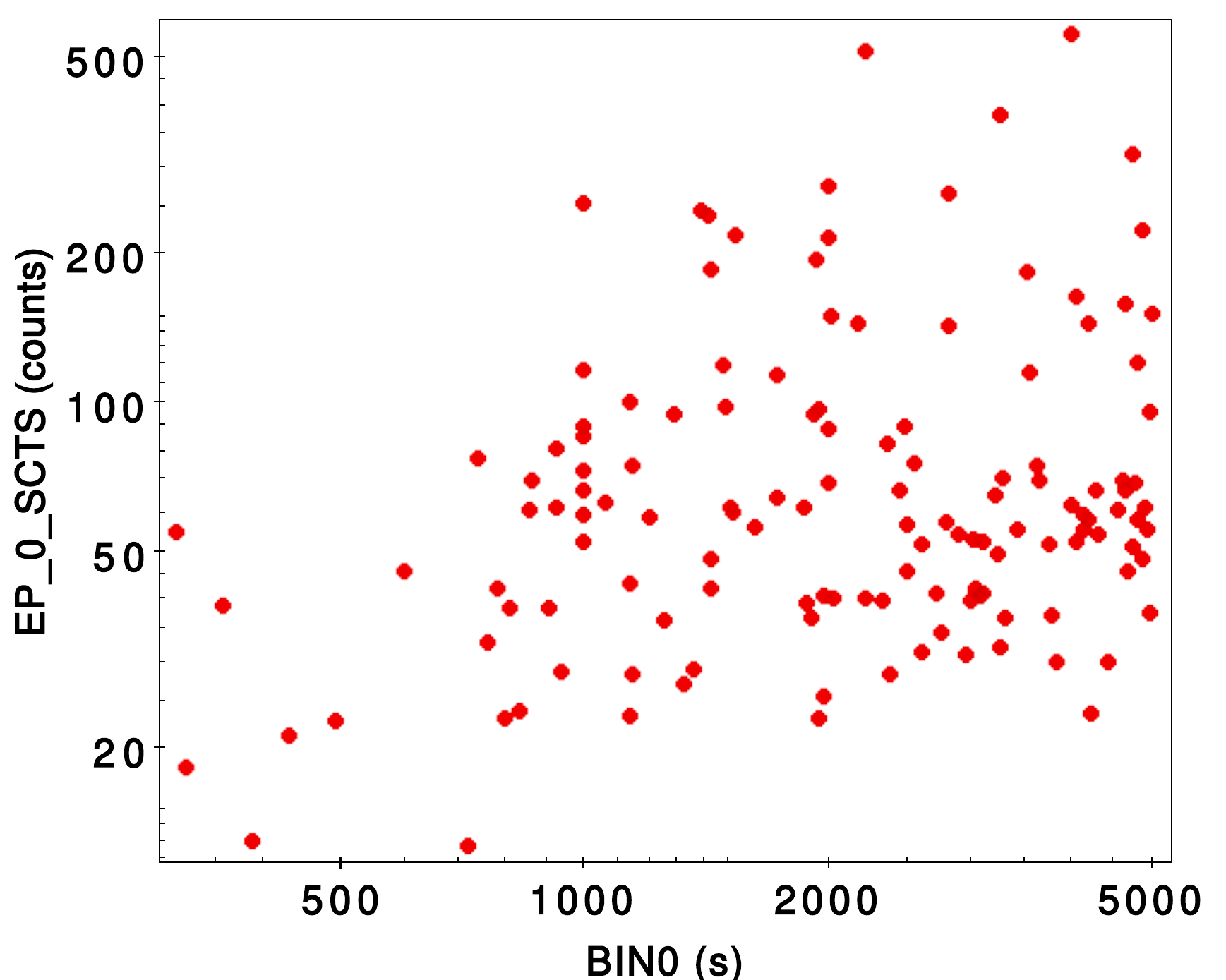} 
\includegraphics[scale=0.34,angle=0]{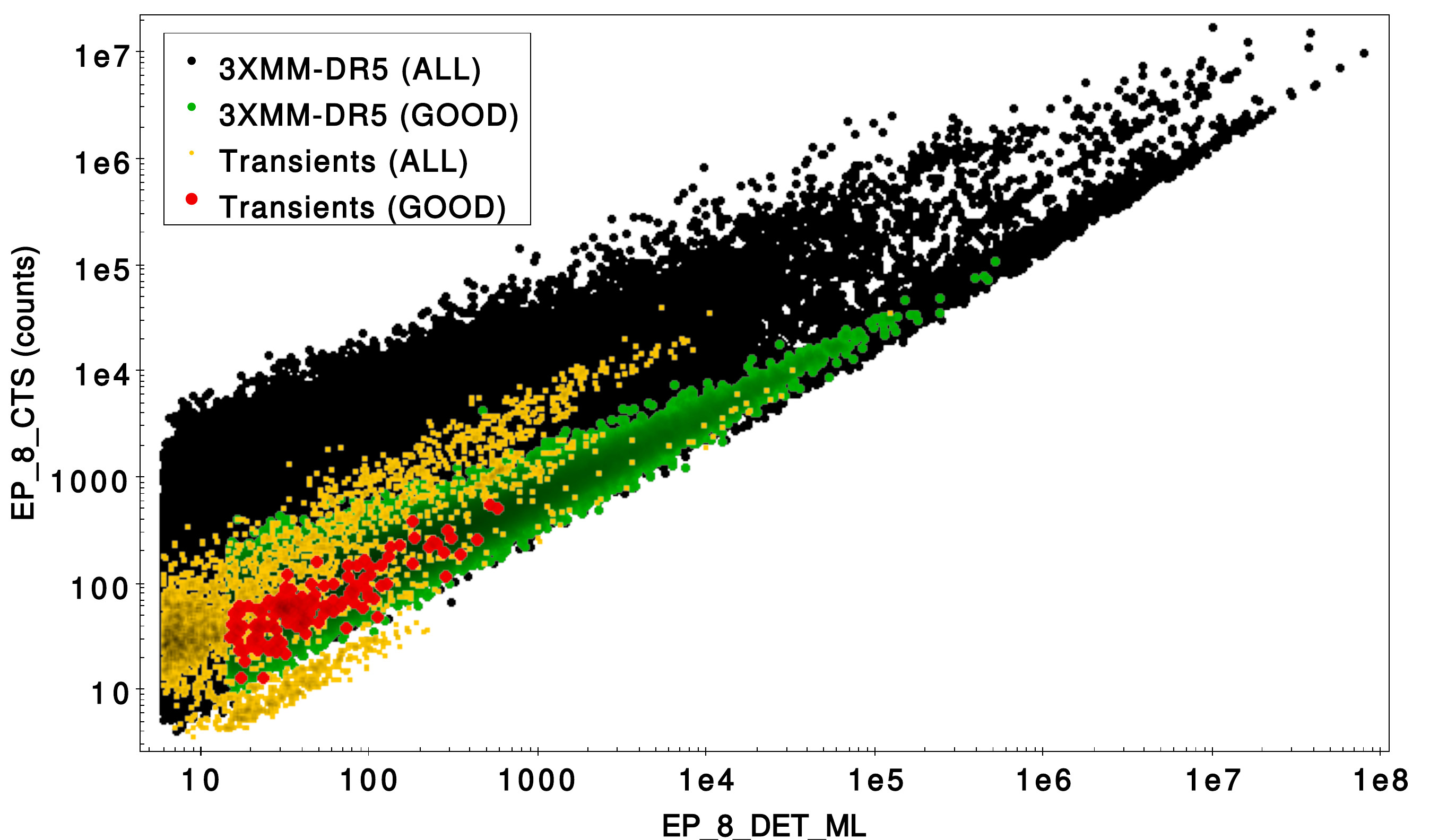} 
\end{center}
\caption{\label{fig_SCTSBIN0DETML3XMM} \footnotesize
Source counts as a function of transient durations (left panel) and detection likelihoods (right panel) in all active instruments and in the 0.2--12 keV energy band (corresponding to band 0 for the transients and band 8 in the 3XMM-DR5 catalogue). Red points indicate the 136 transient sources in the EXTraS catalogue. In the right panel, yellow points represent all transient candidates before selection and screening, black points are all 3XMM-DR5 sources, and green points their subsample with EP\_EXTENT=0, DET\_ML$>$15 and SUM\_FLAG=0.} 
\end{figure*}

\subsection{Variable time intervals}
\label{sec:newtransbb}
Instead of searching for transients by dividing the observations into many intervals of equal duration and analysing the point sources in each of these snapshots, it is more convenient to perform this analysis only for those time intervals that contain an indication for the presence of a variable source.
Determining the best interval for the analysis is indeed a key aspect for the detection of transient sources.
For this purpose, we used the Bayesian block algorithm \citep{Scargle2013}.
Bayesian block is a well-known adaptive-binning algorithm 
that finds the statistically significant count rate change-points 
by optimising the fitness function 
for a piecewise-constant representation of the light curve, starting from an event list. 
The time interval of an active transient is then identified by two count-rate change points.
Specifically, for each observation we divided the active FoV into many small partially overlapping regions, with a size comparable to the EPIC PSF (about 10,000 $30''\times30''$ regions in the current version).
For each region, we independently ran the Bayesian block algorithm, finding change points in which the count rate varied significantly.
We defined $\lambda$ as the average rate of the whole signal during a block; in the case of an ideal background with a constant count rate when there is no active transient,
$\lambda$ = $\lambda_B$
($\lambda_B$, this parameter represents the background average rate of
events during the block).
In the case of EPIC data, the background signal is variable and its count
rate is seldom constant, especially over long time
intervals, that is, $\lambda_B$ = $\lambda_B$(t).
We therefore modified the Bayesian block algorithm to remove
the effects of background variations.
We started by noting that a 
non-stationary
Poisson signal
with a time-dependent count rate $\lambda$ = $\lambda_B$(t) can be
transformed into a 
stationary
Poisson signal by introducing
the time transformation

\begin{equation}\label{integral_lamda}
  t^{\prime}(t)=\int_{0}^{t} \lambda_B(z) dz
.\end{equation}

We defined $t_i$ the $ith^{}$ time series of the arrival times of n
Poisson events detected between the start time $<$ $t_1$ and the stop time $>$ $t_n$.

When we start from the time series $t_i$ and transform it in the time
series $t^{\prime}_i$ using eq. (\ref{integral_lamda}), the count rate
of the Poisson signal is constant in the new system if
the only source of events is the background.
Any variation in count rate found in this space corresponds to a
variation with respect to the background count rate.
It is clear that the accuracy with which the variability of the
background is measured becomes fundamental to finding
these change points.
We can then run the normal Bayesian block algorithm in this transformed space.
When we found a block corresponding to a possible transient, we transformed it
back into the original time reference and obtained the optimal time
window.

Regions where no significant variability is detected vs. the local background light curve return only one block spanning the whole observation, while regions including candidate transients return more blocks.
Our modified version of Bayesian block takes time-varying exposure into account, resulting, for example, from changes in attitude or gaps or defects of the CCD, and highly variable background such as that found during proton flares in XMM-Newton data.

A proper evaluation of the background light curve is crucial, minimising the contamination from the possible variability of known sources. To this aim, the Bayesian block algorithm  was applied only to the events that are not included in sufficiently large (depending on the source intensity) 
regions surrounding the point sources detected in the full observation. 
To examine such excluded regions, where interesting transients might be located (especially in crowded X-ray fields, e.g. star-forming regions and nearby galaxies), a dedicated algorithm (dubbed the close-to-source algorithm) was developed. For each observation, it produces images integrated over a fixed time interval (1000 s in the official EXTraS pipeline run) of regions with a side of 40$''$ surrounding the sources excluded  from the analysis performed using the Bayesian block algorithm and searches for the presence of excesses with respect to known sources on a grid of fixed positions using a sliding cell. This analysis is performed on each time bin; then, all intervals where the same source was found to be active are merged.
From the time intervals identified either in the close-to-source algorithm or by the Bayesian block analysis, we selected for further analysis only those lasting less than 5 ks (the minimum duration of standard EPIC exposures) and originated by regions where the spatial distribution of the events is better described (at a $>\,5\sigma$
confidence level) by adding a point-source model rather than by adopting a simple isotropic background.

In every observation, the identification of transient source candidates is based on the comparison with the reference source list present in the PPS products. Because of statistical and systematic uncertainties in the object coordinates, it is necessary to allow a coordinate tolerance to match the positions and identify different sources as the same object. 
The tolerance is determined by two parameters, called the sigma value (the minimum number of sigmas separating the candidate transient from any catalogue source) and the systematic error (a fixed value representing the EPIC astrometric accuracy).

All the sources detected in (at least) a time interval and within the 
region from which the time interval was generated by 
the Bayesian block algorithm, but 
 that are not detected in the full observation, are defined as transient candidates. The same object can be detected as a transient 
candidate in more than one time interval: In this case, it is identified as a single source, and these time intervals are registered.

\subsection{Candidate transient sources}

The analysis was performed with the Bayesian block algorithm (with 5000 s as the maximum time interval duration), using a computer cluster made available by the University of Leicester. The total number of transient candidates is 41,881, but only 4,254 of them were detected within the triggering region of the Bayesian block algorithm.
Most false transients produced by bright 
columns and particle tracks were removed from the transient candidate list by an automatic tool, which identified the sources formed by events aligned along a straight line that is inconsistent with the instrumental PSF. The screening of the remaining candidates was mainly performed using a visualisation tool with the support of TOPCAT\footnote{See \url{http://www.star.bris.ac.uk/~mbt/topcat/}.} and on-line multi-wavelength catalogues and images. Because a careful screening was possible only for a maximum of several hundred sources, we limited this analysis to
$\sim$50\% of the transient candidates by selecting the sources detected with the highest confidence by the detection algorithm ($\rm DET\_ML$ $>$ 15 in the 0.2–12 keV band in all the active EPIC cameras).
We defined nine different quality groups that identify spurious detections (produced e.g. by out-of-time events or straylight rings of bright sources, or by a poor satellite attitude reconstruction) or encapsulate the confidence level of the transient nature of the candidate (high if a flare and a point source are visible in all active instruments, but much lower for marginally variable or confused X-ray sources). In the final transient catalogue we included only the candidates classified in the first two groups (high and good reliability) for a resulting list of 136 new X-ray sources).

\subsection{Products}
The output of the EXTraS search analysis for short X-ray transients consists of
(i) a catalogue that lists all parameters for all the 136 transient sources -- 
the catalogue is available as a FITS file and is also included in the EXTraS database;
(ii) a set of BITMAP, FITS, and ASCII files for each source (Bitmap image, all detected sources marked; Bitmap image, transient marked; EPIC background image of the interval; EPIC exposure map of the interval; EPIC image of the interval; region file, all detected sources; region file, transient).

A list of the 136 transients with their basic properties is shown in Appendix~\ref{appendix:transients} in Table~\ref{table:translist}.
The full transient source catalogue and all products are available online and can be searched via a dedicated web form\footnote{See \url{https://www88.lamp.le.ac.uk/extras/adv-query/extras_transients}.}. 


\subsection{Transient catalogue statistics}
The final catalogue of transients includes 136 X-ray sources.
Most of them (122) were discovered with the Bayesian block algorithm and 14 were discovered through the analysis of the regions close to 3XMM sources using the close-to-source algorithm with 1 ks time bins.

\subsubsection{Transient time durations}\label{TransDur}
The (differential and cumulative) distribution of the transient duration, defined as the length of the time interval where the source was most significantly detected (BIN0), is shown in Fig.~\ref{fig_histoBIN0cumul}. The transients discovered using the Bayesian block pipeline are also separately considered because the total distribution also contains the 14 sources that were discovered with the {\it \textup{close-to-source}} algorithm. These sources are associated with time intervals that are integer multiples of 1 ks. The sharp cut-off of the distribution at 5 ks is also an artefact of the analysis pipeline, which rejects all time intervals with longer durations.

We note that only a few transients are shorter than 700 s (and none are shorter than 5 minutes), as expected for a population dominated by X-ray flares of active stars (see e.g. Fig. 20 in \citealt{pye2015}). This interpretation is confirmed by the positional coincidence of a large number of these transients with relatively bright optical and near-infrared stars (see Section~\ref{gaia}).

\subsubsection{Transient counts and detection likelihood}
The left panel of Fig.~\ref{fig_SCTSBIN0DETML3XMM} displays the number of counts detected in all the active instruments in the 0.2-12 keV band (EP\_0\_SCTS) as a function of the transient duration (defined as in Section~\ref{TransDur}). A mild positive correlation is visible, which can be explained by the fact that faint transients in longer time intervals are more difficult to distinguish from the background and that transients with many counts in short time intervals are intrinsically brighter and therefore less frequent.

The expected correlation between counts and detection likelihood is clearly visible in the right panel of Fig.~\ref{fig_SCTSBIN0DETML3XMM} for different data samples. The black points are all the sources of the 3XMM-DR5 catalogue \citep{Rosen2016}, and the green points indicate only the clean (SUM\_FLAG = 0), point-like (EP\_EXTENT = 0) detections with DET\_ML$>$15. For this subsample, selected according to the same criteria as the sources in the transient catalogue, the correlation between the number of counts and the detection likelihood is even more striking than in the global sample because almost all the sources that were detected with a large number of counts but relatively low detection likelihood were either extended or displayed some anomaly in the screening performed by the Survey Science Center (SSC).

Similar considerations also apply to the comparison of the yellow points, which are all the transient candidates before the screening process, and the 136 transients included from the EXTraS catalogue (red points). A remarkable difference with respect to the 3XMM sources is the group of yellow points located below the region of clean sources: All these transient candidates, with a particularly high ratio of their detection likelihood over the number of  counts, were detected in very short time intervals (from a fraction of a second to several dozen seconds) and only in one EPIC camera, and turned out to be either bright or flickering pixels or short tracks of high-energy particles.

\begin{figure*}[ht]
\begin{center}
\includegraphics[scale=0.35,angle=0]{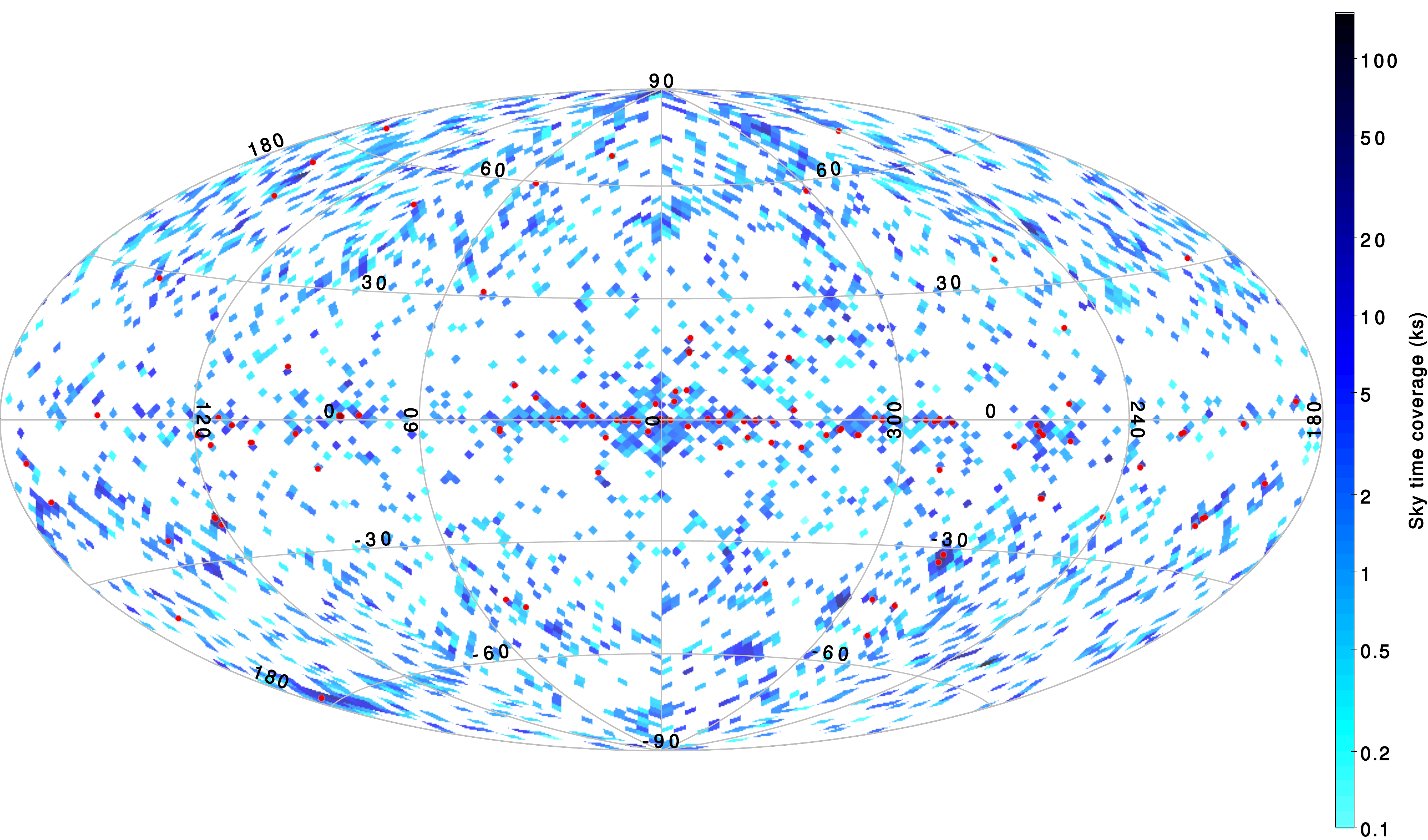}
\end{center}
\caption{\label{sky_PN} \footnotesize
Sky distribution of transients (red circles) over the map of pn sky coverage ($\sim$3.5 square degree
regions with the total exposure indicated by the colour bar) 
is clearly visible.}
\end{figure*}

\begin{figure}[ht]
\begin{center}
\includegraphics[scale=0.43,angle=0]{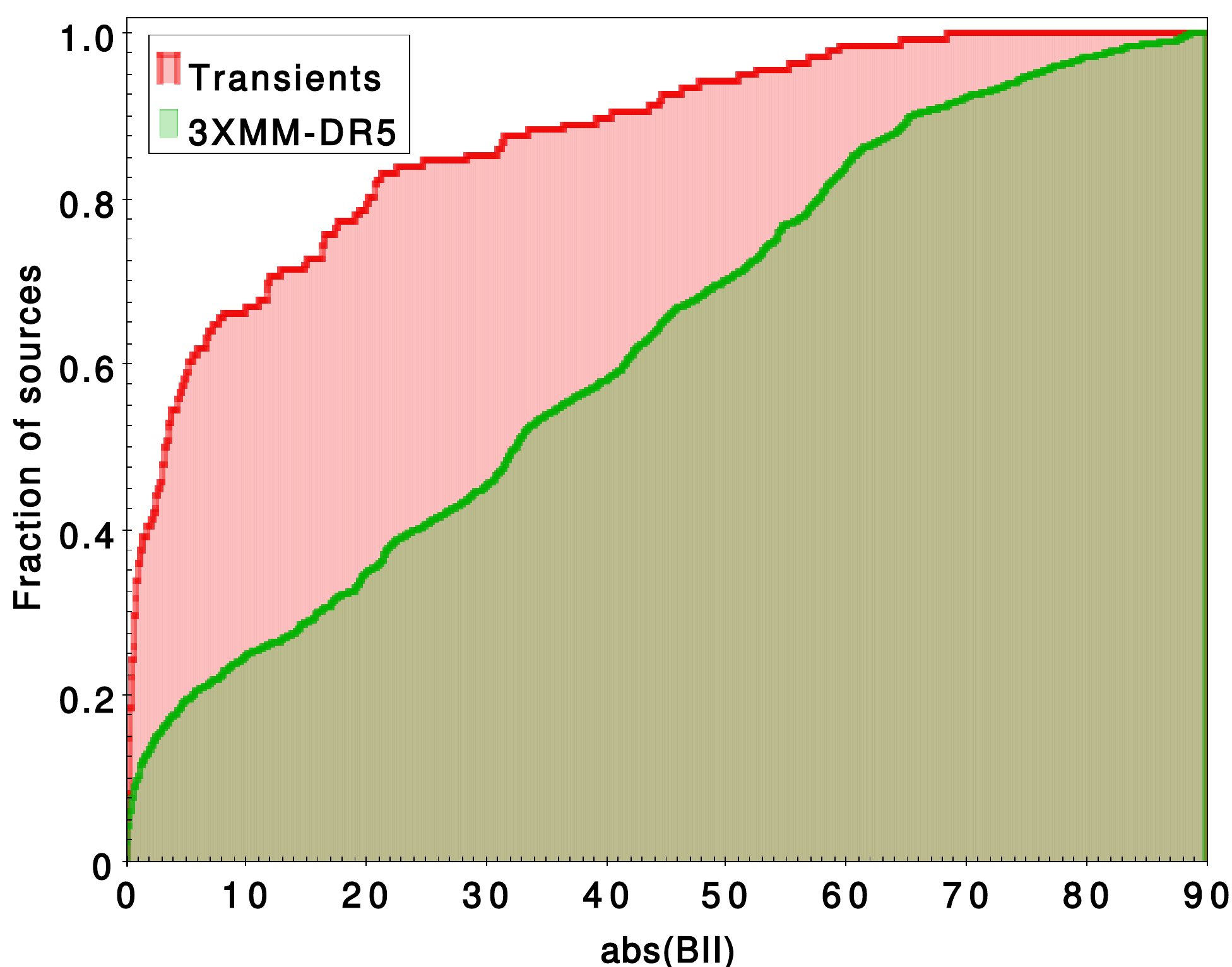}
\end{center}
\caption{\label{BhistoBcumul} \footnotesize
Cumulative distribution of the absolute value of the Galactic latitudes of the EXTraS transients (red) and of the point sources (EP\_EXTENT=0) with DET\_ML$>$15 and SUM\_FLAG=0 in the 3XMM-DR5 catalogue (green). The excess of transients at low latitudes
is clearly visible.}
\end{figure}

\subsubsection{Sky distribution of the transients}
To estimate the sky coverage of our systematic search, we produced and merged the exposure maps for each instrument (without correcting for the telescope vignetting) for all the observations that we processed with the EXTraS pipeline. The global exposure maps of the pn, MOS1, and MOS2 instruments correspond to the observation of the full sky (41,253 square degrees) for 8.7, 12.4 and 13.9 minutes, respectively.

To evaluate the sky coverage of the Bayesian block pipeline without considering the {\it \textup{close-to-source}} algorithm, which has a different sensitivity, we also created so-called cheesed exposure maps by removing the regions that were used to exclude the 3XMM sources from the Bayesian block analysis. In this case, our search for transients corresponded to an all-sky survey lasting 8.2, 11.9, and 13.3 minutes for the pn, MOS1, and MOS2, respectively.

The distribution of {\it XMM-Newton} pointings is far from being isotropic, however, and therefore we computed the same spatially averaged exposure times in $\sim$12,000 sky regions with a size of $\sim$3.5 square degrees. The corresponding map of the time coverage of the EXTraS search for transients with the pn camera (without removing the regions around 3XMM sources) is shown in Fig.~\ref{sky_PN}, together with the positions of the 136 transients in Galactic coordinates.
The sky distribution of EXTraS transients is clearly clustered on the Galactic plane, in particular, in its central part. To understand whether this effect is due to the longer exposure time dedicated by {\it XMM-Newton} to the study of Galactic objects, we compare in  Fig.~\ref{BhistoBcumul}  the cumulative distribution of (the absolute value of) the Galactic latitude of the EXTraS transients with that of the clean sample of 3XMM-DR5 sources. The striking difference between these two distributions indicates that the newly discovered transients mainly have a Galactic origin, whereas the fraction of extra-galactic objects in the 3XMM-DR5 catalogue is significantly larger. This is another confirmation that the majority of the EXTraS transients are very likely flares from relatively nearby stars (see Section~\ref{gaia}), whereas the 3XMM catalogue contains a large number of active galactic nuclei, whose X-ray variability on timescales $<$5 ks is much less prominent.

On the other hand, the position of the shortest EXTraS transient (EXMM~J023135.0--603743, with BIN0=315 s) is consistent with a galaxy at redshift $z=0.092$ and has been interpreted as the X-ray flare of a core collapse supernova \citep{novara20}. The same transient has independently been discovered in a systematic search for supernova shock break-out candidates in XMM-Newton archival data \citep{alp20}. The other 11 candidates listed in \citet{alp20} could not have been detected by the algorithm described in this section because they were either already included in the 3XMM-DR5 catalogue or occurred in more recent observations that were not covered by our analysis.

\subsubsection{Cross-match with the GAIA source catalogue}\label{gaia}

To estimate the fraction of EXTraS transients with stellar counterparts, we used TOPCAT to cross-match their positions with the \emph{Gaia} DR2 catalogue of stars with well-determined parallax \citep{gaiaDR2}. Specifically, we found that 58 out of 136 transients in the EXTraS catalogue are located within 5 arcsec from a \emph{Gaia} star with parallax/parallax\_error$>$5, which can be considered a very robust indicator of stellar nature (see e.g. \citealt{bai18}). By counting the number of these stars within 3 arcminutes from each transient, we evaluated a $<$10\% chance coincidence probability even for the most crowded fields, and therefore only a few false associations are expected.

On the other hand, the majority of the transients 
 with no clear stellar counterpart are also consistent 
with stellar flare shape and duration. They might be produced by farther and/or fainter stars, whose parallax could not be precisely measured by \emph{Gaia}.

\begin{figure}[ht]
\begin{center}
\includegraphics[scale=0.5,angle=0]{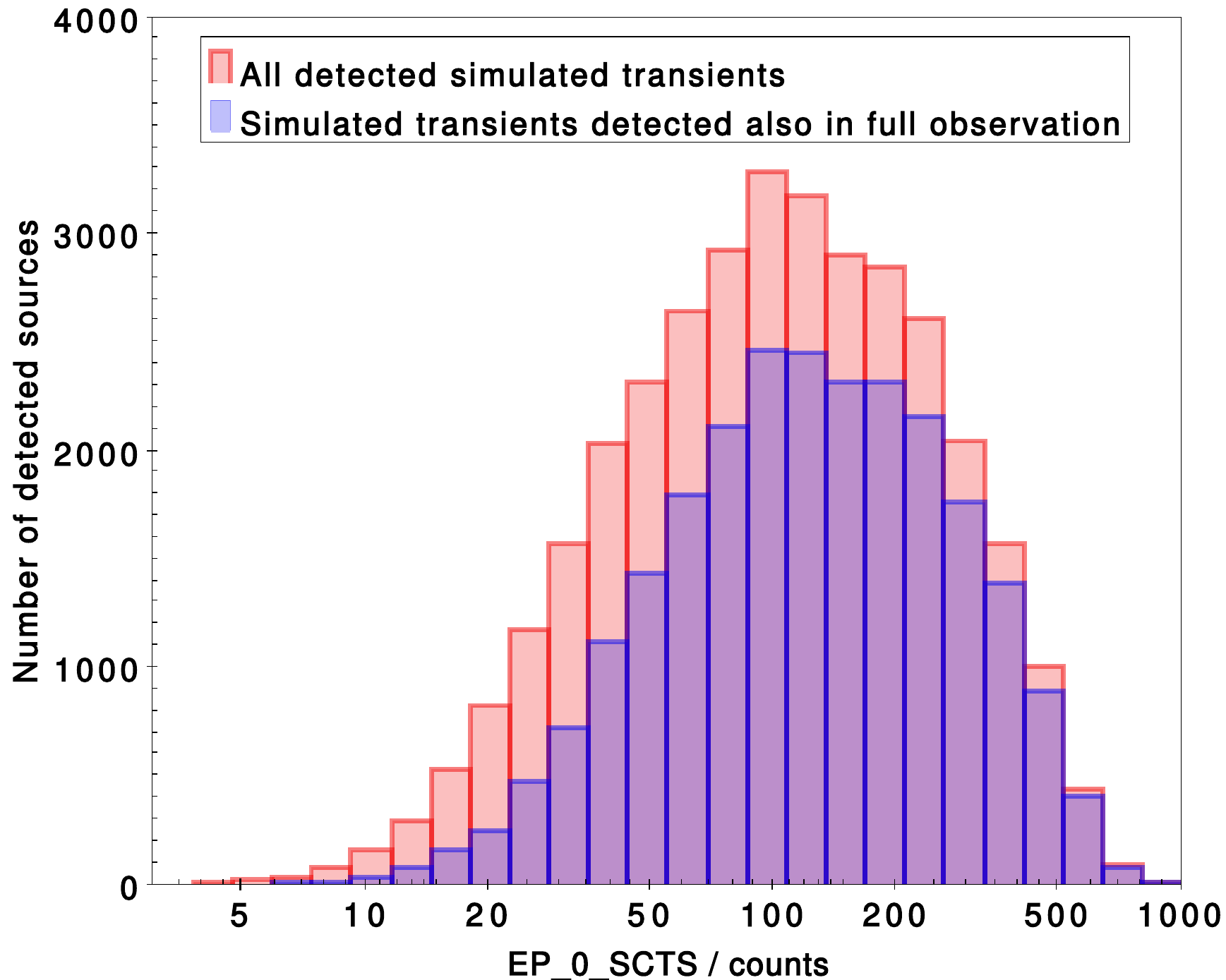} 
\end{center}
\caption{\label{SimTrans1} \footnotesize
Distribution of EPIC counts in the 0.2--12 keV energy band detected by the EXTraS pipeline for all the simulated transients (red) and for those that were also detected in the full observation (blue). 
}
\end{figure}

\begin{figure}[ht]
\begin{center}
\includegraphics[scale=0.5,angle=0]{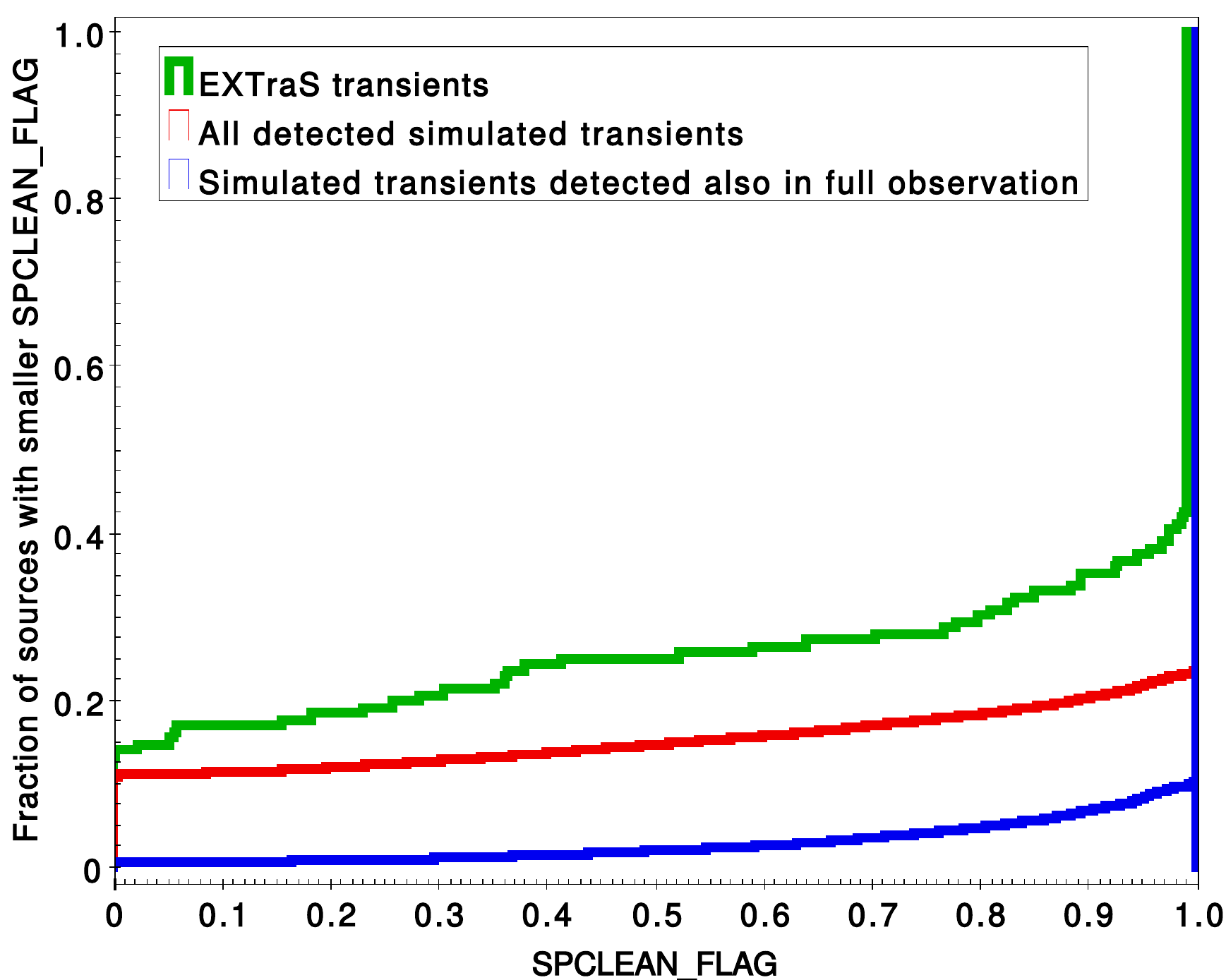} 
\end{center}
\caption{\label{SimTrans2} \footnotesize
Cumulative distribution of the SPCLEAN\_FLAG parameter, indicating the fraction of the BIN0 interval that is not affected by soft protons flares for the EXTraS transients (green), all the simulated transients detected by the EXTraS pipeline (red), and those that were also detected by the 3XMM-DR5 pipeline applied to full observations (blue).}
\end{figure}

\subsubsection{Comparison with the 4XMM-DR9 and 3XMM-DR5 serendipitous source catalogue}

After the release of the 4XMM-DR9 catalogue, we verified whether this more complete catalogue included a fraction of the 136 new transients detected by the EXTraS algorithm. Although the positions of 29 EXTraS transients are within 5 arcsec of a 4XMM-DR9 source, only 15 of them were detected during the same observation. One of them (EXMM~J162714.7--245135), detected with the close-to source algorithm, is $\sim$4 arcsec from a source that is also included in the 3XMM-DR5 catalogue, but this catalogue source and the EXTraS transient are very likely produced by two distinct bright stars in the Rho Ophiuchus open cluster, with GAIA parallaxes of 7.24$\pm$0.05 mas and 7.3$\pm$0.07 mas, respectively. 
In the remaining cases, the transient emission detected through the EXTraS analysis was missed by the standard detection procedure, which instead detected either the persistent emission or a different flaring episode from the same sky position. Both cases would not be surprising for stars that can emit multiple flares and typically have faint persistent X-ray emission that might only be detectable during relatively long XMM-Newton exposures.

We note that of the 14 EXTraS transient events that were genuinely detected by the standard analysis, only 2 are classified as variable sources in the 4XMM-DR9 catalogue. The high sensitivity of the EXTraS algorithm in detecting fast transients is also confirmed by the fact that an independent search for X-ray transients in the XMM-Newton archive, using a different approach, did not discover additional sources with respect to the 4XMM-DR9 catalogue \citep{pastormarazuela20}.

To further explore the advantages of the EXTraS search for transients in comparison with the variability study of catalogue sources, we can take advantage of the simulations performed in \citet{novara20} to evaluate the sensitivity of the EXTraS algorithm to short transients. Taking as a template the spectrum and light curve of the $\sim$5-minute X-ray flare associated with SN 2008D \citep{soderberg08}, we simulated the events of $\sim$48,000 transients with different fluxes as they would be observed by the three EPIC cameras at different off-axis angles, and added them to the event files of a randomly selected sample of the XMM-Newton observations that were used to extract the EXTraS transient catalogue. The simulations took the instrumental configuration (operating mode, filter, and good time intervals) and pointing direction (to correct the simulated spectrum according to the total Galactic absorption expected along the line of sight) of each observation into account \citep[more details of the simulation can be found in][]{novara20}.

We then applied the EXTraS transient algorithm and the same detection pipeline as was used to obtain the 3XMM-DR5 source catalogue to these data. By matching the known positions of the simulated sources with those of the sources detected by the two pipelines, we found that 34,476 of the simulated sources were detected (with DET\_ML$>$15) by the EXTraS algorithm and that the standard pipeline detected 24,296 of them in the full observation as well. As shown in 
Fig.~\ref{SimTrans1}, the fraction of sources that was detected in the full observations as well is much larger for the brightest simulated transients.

As anticipated in Section~\ref{TransAims}, the new transients in the EXTraS catalogue are missed by standard analysis because their X-ray signal is either too faint to significantly emerge from the background of the full observation or because it occurred during a time period of high background, which is filtered out by the standard detection procedure. 
The relative importance of these two effects can be evaluated by exploring the parameter SPCLEAN\_FLAG, which for each transient detected by the EXTraS algorithm is defined as the fraction of the time interval where the source was most significantly detected (BIN0) that would not be excluded by the 3XMM-DR5 pipeline, which removes high particle background time intervals. As shown in 
Fig.~\ref{SimTrans2}, $\sim$90\% of the simulated transients that were also detected in the full observations occurred during a time interval that was not affected by soft protons flares (SPCLEAN\_FLAG=1), whereas this fraction decreases to 77\% and 57\% for the total sample of the simulated sources and the EXTraS transients, respectively. On the other hand, 11\% of the simulated transients that were detected by the EXTraS algorithm and 14\% of the 136 EXTraS transients occurred in a time interval that was completely excluded by the 3XMM-DR5 pipeline (SPCLEAN\_FLAG=0).

%
%

\section{Long-term variability (LTV)}\label{sect:ltv}

\subsection{Aims and scope}\label{sect:LTVaims}

\begin{figure}[ht]
\centering \includegraphics[width=9cm]{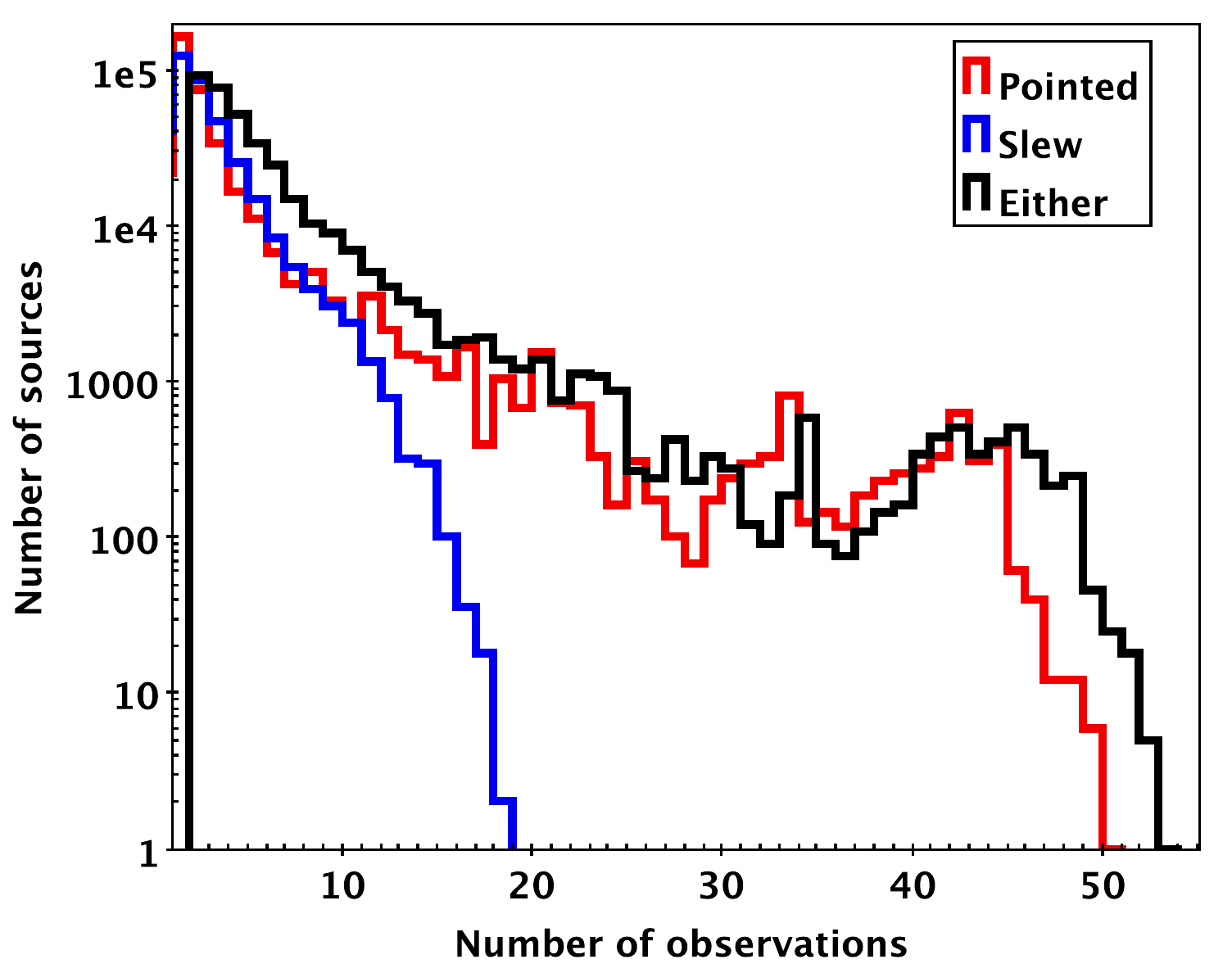}
\caption{Numbers of sources in the LTV catalogue comprising a given number of observations for pointed data only (red), slew data only (blue), and both (black). 
The latter histogram starts at 2 because sources comprising only one data point in total are not counted.} 
\label{fig:ltv_nsrcs_v_nobs}
\end{figure}

Many parts of the sky have been observed at least twice by XMM-Newton in pointed mode and/or during slews 
between pointings. The LTV component of the EXTraS project exploits X-ray photometry of a subset of sources that were
multiply observed by XMM-Newton together with upper-limit data to facilitate the 
study of potential long-term (inter-observation) variability. As scheduling of XMM-Newton observations 
covering a given source is generally random, the data sampling for most sources is sparse and very non-uniform, 
so that only simple measures are employed to characterise potential variability. Figure~\ref{fig:ltv_nsrcs_v_nobs} 
shows the frequency distribution of repeat observations.

The EXTraS LTV catalogue is based on the set of 7781 pointed observations from the 3XMM-DR5 catalogue 
(\cite{Rosen2016}) and a new processing of 2059 available XMM-Newton slews that form 
a large subset of the XMMSL2 slew catalogue\footnote{\url{https://www.cosmos.esa.int/web/xmm-newton/xmmsl2-ug}}. 
The pointed data span 14.9 years, from 03 February 2000 to 20 December 2013, and cover $\approx$2\% of the 
unique sky, while the slew data cover 13.35 years, from 27 August 2001 to 31 December 2014, and image  
around 84\% of unique sky. All photometric measurements are observation-integrated snapshots of 
sources. Pointed observation exposures are generally in the range $\sim$1~ks to $\sim$130~ks, while slew 
observations typically amass an exposure of about 10s. The median 0.2-12.0 keV (total band = band 8) source 
fluxes are  $\sim$2.1$\times 10^{-14}$~erg~cm$^{-2}$~s$^{-1}$ and $\sim$2.7$\times 10^{-12}$~erg~cm$^{-2}$~s$^{-1}$ 
for pointed and slew data, respectively.

\subsection{Slew data processing}\label{sec:LTVslewdata}

Slew data are only obtained with the pn camera in the prime
full window, prime full window extended, or prime large window modes, and
always with the medium filter. The new slew processing exploits improvements 
in both software and calibration since the XMMSL1 catalogue \citep{Saxton2008}, 
still broadly following the approach described there but with 
upgrades to the pipeline that were partly undertaken within the remit of the EXTraS 
project and that incorporate three significant changes to slew processing. We outline them below.



First, a $\sim 0.25^{\circ}$ overlap region is now included between adjacent 
$\sim 1^{\circ}$ x 0.5$^{\circ}$ sub-images along the slew, ensuring that a detected source 
can always be adequately fitted with the PSF in at least one sub-image
(if a 
source is detected in both sub-images, its photometry is taken from the image in which 
the source is closer to the centre).



Second, a new PSF, averaged over the whole FoV, is used in obtaining  
slew photometry, replacing the previous single on-axis point PSF that did not account for the varying 
PSF of a source image as it crosses the FoV (see 
https://www.cosmos.esa.int/web/xmm-newton/xmmsl2-ug). The new PSF matches source 
profiles better than the simple on-axis PSF. 


Third, the filtering of periods of high background is applied on an individual sub-image 
basis; previously, all data were rejected from any slew that was affected by high background. 
This allows including data from many more slews. 


The better slew PSF yields systematic increases in the recovered source counts and 
count rates (by $\sim$2\%) compared to the XMMSL1 catalogue and corresponding 
increases in the detection likelihoods, along with an 
$\sim0\farcs35$ improvement in the astrometric accuracy (by comparing 
to astrometric catalogues).
Photometry (count rates and fluxes) of slew detections is obtained in the broad soft 
(0.2-2.0 keV = band 6), hard (2.0-12.0 keV = band 7), and total (0.2-12.0 keV = 
band 8) energy bands. Fluxes ($F_i$) are computed, in each band, $i$, from 
the count rate, $R_i$, as $F_i=R_i E_i$, where $E_i$ is the energy conversion 
factor (ECF) (see \cite{Saxton2008} and https://www.cosmos.esa.int/web/xmm-newton/xmmsl2-ug\#Fluxes).

\subsection{Pointed data photometry}\label{sec:LTVpointeddata}

Standard pipeline processing of pointed data determines count rates in five energy 
bands by fitting energy-band-dependent PSFs to source images in all five bands 
simultaneously, as described in \cite{Watson2009} (see also \cite{Rosen2016}). 
Because of the much shorter exposure times (thus lower counts), as noted above, 
slew data are binned into three broad bands. Slew processing also fits sources 
separately in each band and uses PSFs extracted at a fixed (1.5 keV) energy in 
all three bands. Consequently, simply combining narrow-band pointed data count rates 
into the broad slew bands for comparison with the slew measurements can yield 
discrepancies of several percent in count rate. Furthermore, the use of ECFs based 
on a fixed spectral profile for sources that often deviate from that profile can 
yield much larger discrepancies in fluxes. The effect is largest in the broadest 
(i.e. hard and total) bands. To maximise the consistency of pointed and slew data 
in long-term light curves, we re-extracted the pointed data in the broad bands in 
the same way as used for slew processing.

The analysis of pointed data was performed for all source detections in the 3XMM-DR5 
catalogue. Images for each available instrument were created directly in bands 
6, 7, and 8, with count rates separately derived in each from fitting of the relevant 
instrument PSF model, extracted at 1.5 keV (hereafter referred to as the direct 
approach). The PSF normalisation (related to the count rate) was fitted for each 
detection, but its position and extent were fixed at its 3XMM-DR5 catalogue values. ECF values used for computing fluxes are given in 
Appendix~\ref{appendix:LTVpointedECFs}.

\subsection{Matching of pointed and slew sources}\label{sect:LTVmatching}

To build long-term light curves, astrometric information was first used to associate pointed and slew detections with unique sources on the sky. Detections from the 3XMM-DR5 catalogue are already matched into unique 
sources as described in \cite{Rosen2016}, exploiting the Bayesian algorithm of \cite{BS2008}. Slew detections were separately matched into unique sources as follows: 
\begin{itemize}
    \item Detections in the same band within 30\arcsec\ of each other in two consecutive slew sub-images are considered to be the same source, and the detection farthest from its sub-image centre is removed from the source list.
    \item Detections in the total and soft bands within 30\arcsec\ of each other in the same sub-image are deemed to be the same source and are merged into one record per source. Any hard-band detection, also within 30\arcsec\ of them and from the same image, is then associated with them. 
    \item The slew source is then identified with the detection in the band that has the highest detection likelihood. 
    \item A further check is made for sources that lie in consecutive images with a separation of <30\arcsec, which have detections in different bands. These are joined. 
    \item Finally, sources seen in two or more slews are combined to set the {\it UNIQUE\_SRCNAME} to the one with the highest detection likelihood (in any band).
\end{itemize}

Pointed source positions reflect the position-error-weighted average of the detections involved. Where possible, slew sources were then matched to existing pointed sources using the above-mentioned Bayesian approach to decide on the association. Some 6.3\% of the slew sources are associated with pointed sources. Where a single match is found (76\% of cases), the constituent slew detections acquire the source identifier ({\it SRCID}) of the pointed source. Where a slew source matches more than one candidate pointed source (i.e. an ambiguity), it is assigned to the pointed source with the highest match probability, but a flag ({\it NPMATCH}) is set, indicating how many other pointed sources it matches. Where a slew source has no match with a pointed source, a new  source is established in the LTV catalogue with the {\it IAUNAME} of the  slew source.

Pointed data astrometry is generally more precise (mean statistical uncertainty, $\sigma$, $\sim 1\farcs4$) than slew data ($\sigma\sim 5\farcs3$) because of (i) better statistics from the longer exposures, (ii) the tighter PSF, and (iii) because pointed detections are generally rectified against an astrometric reference catalogue (see  \cite{Rosen2016} and references therein), which is not possible for slew data. Thus, the position of a source that contains both pointed and slew detections is taken as that of the original pointed source. Sources containing only pointed detections or only slew detections acquire the positions of the respective pointed or slew source.

\subsection{Upper limits}\label{sec:LTVULs}

Where a source is covered by an observation but is not detected, upper-limit data can still provide useful constraints on the source brightness. Upper limits were obtained from all pointed and slew images covering source positions described in 
section~\ref{sect:LTVmatching}, even where the source was detected, broadly following the approach used in the FLIX tool (http://www.ledas.ac.uk/flix/flix\_dr5.html), but tailored 
to the LTV data and energy bands. Upper-limit count rates were extracted for a detection likelihood, $L=10$ (corresponding approximately to a Gaussian sigma of 3.9), for each band and available instrument.

All available count rate and flux upper limits are provided in a dedicated row in the LTV catalogue for any observation of a source where it is not detected in any band or instrument. 
Its detection identifier, {\it DETID}, comprises the observation ID ({\it OBS\_ID}), followed by the relevant source identifier ({\it SRCID}).
For observations where a source is detected but not in all available bands or instruments, upper limits are inserted 
for the bands or instruments with non-detections.  In addition, where the detection likelihood, $L_i$, in band $i$ of a real detection is $<$8.0 for any instrument, we replace the relevant photometric information with the upper limits. Upper-limit values appear as negative numbers in the catalogue, and the error columns for these quantities contain nulls. 
Where an all-EPIC detection likelihood is $<$8.0, the EPIC photometric value is replaced by the highest upper limit from the available individual instruments because the calculation of all-EPIC upper limits is not trivial and the upper-limit software does not compute them. 

\subsection{Long-term variability characterisation}\label{sec:LTVchar}

The generally limited quantity and sparseness of the data for each source and the presence of upper limits makes a detailed systematic analysis difficult. The analysis of long-term 
variability therefore involves some simple variability tests and a set of measurements that characterise the scale and timescale of variability. All quantities are provided for each instrument and for each of bands 6, 7, and 8 where possible.

Three primary measures are computed for variability. One of these is a reduced $\chi^2$ ({\it DRCHISQ} columns), that is,

\begin{equation}
{\frac{1}{(n-1)}} \sum^{n}_{1} {\Bigg( \frac{F_i - \bar{F}}{\sigma_i}\Bigg)^2
}
,\end{equation}

and the associated probability ({\it DPROB} columns) for the null hypothesis of the data being constant, where $F_i$ is the flux of the $i$th data point in a light curve comprising $n$ data points, $\sigma_i$ is its error, and $\bar{F}$ is the variance-weighted mean flux.

The second measure is the largest error-normalised flux change ({\it MDDE} columns) between any pair of points, $i$ and $j$, in the light curve, that is, $max[(F_i - F_j)/\sigma_{ij}],$ where $\sigma_{ij}$ is the quadrature sum of the flux errors on the two points. Both quantities are only based on detections.
 
A third quantity is the ratio of the maximum flux to the 
minimum flux and its error ({\it MR} and {\it MRE} columns). The minimum flux point can be a detection or upper limit value; if it is an upper limit, the result is essentially a lower limit on the {\it MR} ratio. If the maximum value is an
upper limit, it is not used. The error ({\it MRE}) on the maximum flux ratio ($MR = F_{max}/F_{min}$) is simply computed as

\begin{equation}
  MRE = MR \Bigg[ \Bigg( \frac{\Delta{F_{max}}} {F_{max}} \Bigg)^2 + \Bigg(
    \frac{\Delta{F_{min}}} {F_{min}} \Bigg)^2 \Bigg]
,\end{equation}

where $F_{min}$ and $F_{max}$ are the fluxes of the minimum and maximum points, respectively. If the lower point is an upper limit, the flux is taken as the upper limit (see section~\ref{sec:LTVULs}). In these cases, we ascribe a 1$\sigma$ equivalent Gaussian uncertainty to the upper limit of $\Delta{F_{min}} =F_{min}/3.9$. 


While these quantities are provided in the LTV catalogue, no attempt is made to impose a threshold and thus identify sources as variable or not. This is left to the user. 
These three measures are augmented by a number of additional quantities that provide information about the scale and timescales of variations in the long-term light curves. These are listed below.

\begin{enumerate}[i]
\item The time-span ({\it TMDDE}) over which the largest error-normalised flux change ({\it MDDE} see section~\ref{sec:LTVchar} above) occurs.
\item The error-weighted mean flux ({\it WMEAN} and {\it DWMEAN}). Quantities that are preceded by a 'D' involve only detections. 
\item The maximum upward ({\it EMFU}) and downward ({\it EMFD}) flux ratios as measured between $(F - \sigma)_{max}$ and $(F +  \sigma)_{min}$, that is, a conservative measure of the largest change.
\item The timescales over which the largest upward ({\it TEMFU}) and downward ({\it TEMFD}) flux transitions occur.
\item The shortest timescales between two points, $i$ (bright) and $j$ (faint), in which the flux increases ({\it ET2U}) and decreases ({\it ET2D}) by at least a factor 2, that is, where $(F - \sigma)_{i} / (F + \sigma)_{j} > 2.$
\item ({\it ET10U}) and ({\it ET10D}): as for ({\it ET2U}) and ({\it ET2D}), but for changes by a factor $>$10.
\item The significance associated with a runs (Wald-Wolfowitz) test \citep{Bradley1968} to gauge the degree of any systematic variations in a time series ({\it DSIGNIF}). \footnote{The statistic is computed as $Z= (R -\bar{R})/s,$ where $R$ is the number of observed runs, $\bar{R} = 2n_{+}n_{-}/(n+1)$, is the expected number of runs, and $s$ is the standard
deviation of the number of runs ($s^2 = 2n_{+}n_{-}(2n_{+}n_{-} - n)/[n^2(n-1)]$ ): $n_{+}$ and $n_{-}$ are the number of positive and negative
runs, respectively, and $n = (n_{+} + n_{-})$. A run is a sequence of consecutive points above or below the mean flux. It only has relevance to time series with $\sim$ 10 or more points.}
\end{enumerate}

\subsection{LTV catalogue and its basic properties}\label{sect:LTVcat}

The LTV catalogue format is broadly modelled on the 3XMM catalogues but (i) contains only a subset of the most important columns, (ii) includes the aforementioned upper-limit data, and (iii) includes the long-term variability measures  (see section~\ref{sec:LTVchar}) and some additional quality information. Each row of the catalogue refers to one of a pointed detection, a slew detection, or a pointed or slew upper limit, with identifiers for the sources they are associated with.  Information is provided for each instrument (and all-EPIC) and per energy band. 

\begin{figure}[ht]
\centering 
\includegraphics[width=9cm]{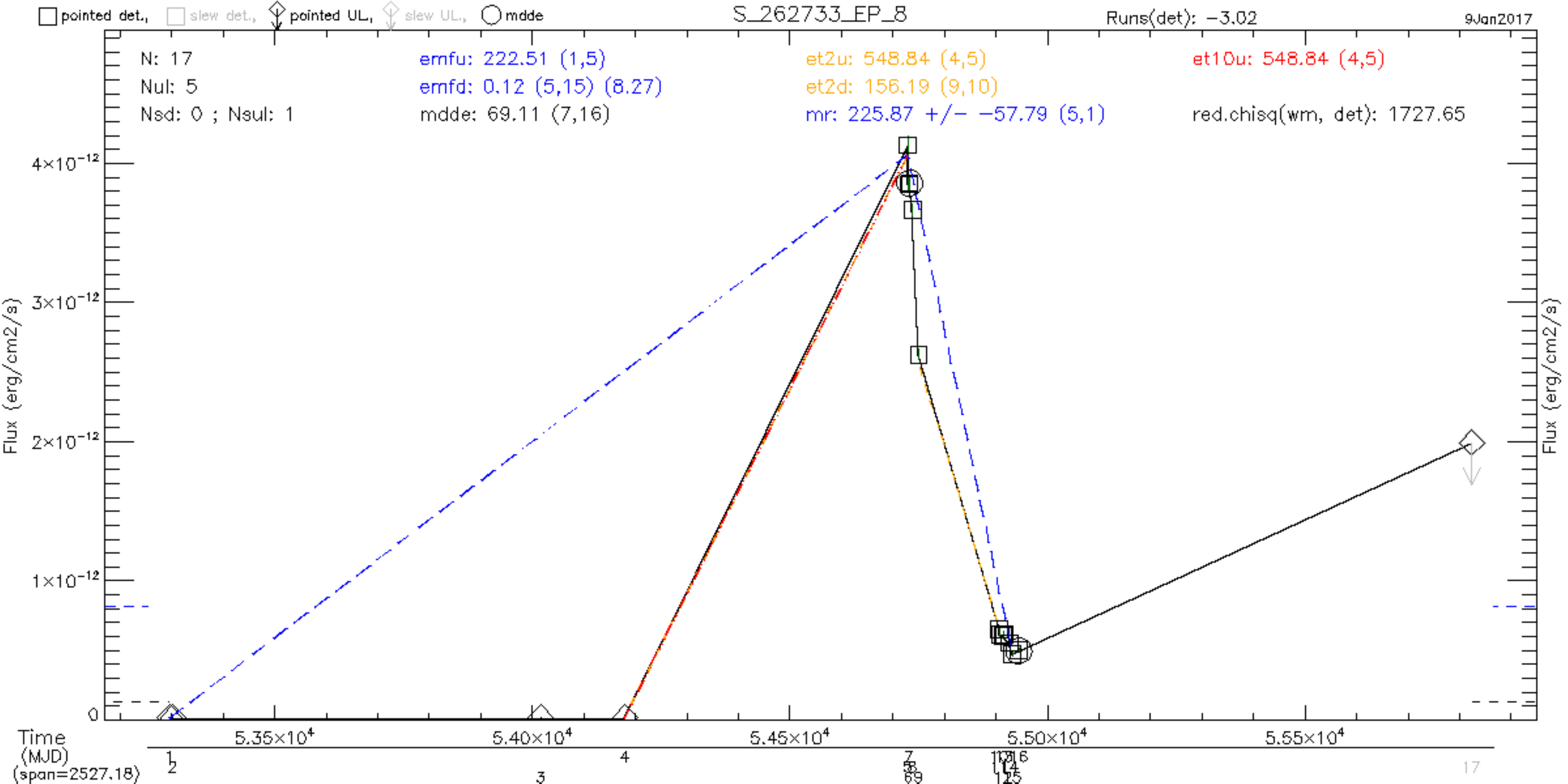}
\caption{Example long-term light curve of a source (SRCID=262733) in the LTV catalogue. The main 
measures and points involved are indicated in the plots by connecting lines, e.g. dashed blue lines join points used in the {\it MR} and {\it EMFU/EMFD} quantities, while dashed orange and red 
lines signify changes by a factor 2 and factor 10 in flux between pairs of points. Some of the key LTV measurement values are printed in the graphic for convenience. In some cases, high slew upper limits are not displayed where they suppress the visibility of low-level changes in other data. } 
\label{fig:ltv_example_lc}
\end{figure}

The contents of the LTV catalogue (410 columns) are described by descriptors in the EXTraS database 
(see~\ref{sec:archive}). 
For each source, a graphic (gif) LTV light-curve product is also created for each instrument and energy band. URL links to the graphics are contained within the FITS LTV catalogue file. They are also accessible within the EXTraS database. An example graphic is shown in Fig.~\ref{fig:ltv_example_lc}.

A summary of the catalogue contents is given in Table \ref{tab:ltvoverview}.

\begin{table}[ht]
\begin{minipage}[ht]{\columnwidth}
\normalsize
\caption{Overview of catalogue properties. The cleanest sources are those in which any or all constituent pointed detections have a summary flag of 0, are point-like and not piled up, and none (nor any constituent slew detections)
  show indications of astrometry problems.} 
\label{tab:ltvoverview}

\small \centering \renewcommand{\footnoterule}{} \tabcolsep 0mm
\begin{tabular}{
l @{\extracolsep{2mm}} c} \hline \hline

\hline 
Detections \& upper limits & 2,030,040 \\ 
Pointed(slew) detections & 565,962 (29,944) \\
Unique sources & 419,240 \\
Unique sources with $>$ 1 EPIC band 8 measurement& 357,178 \\
Unique sources with $>$ 1 EPIC band 8 measurement  & 286,215 \\
\& $\ge$ 1 detection & \\
Unique sources with EPIC band 8 {\it MDDE} $>$ 5 & 10,980 \\
Cleanest unique sources with EPIC band 8 {\it MDDE} $>$ 5 & 2,954 \\ \hline

\end{tabular}
\end{minipage}
\normalsize
\end{table}

\begin{figure}
\centering \includegraphics[width=9cm]{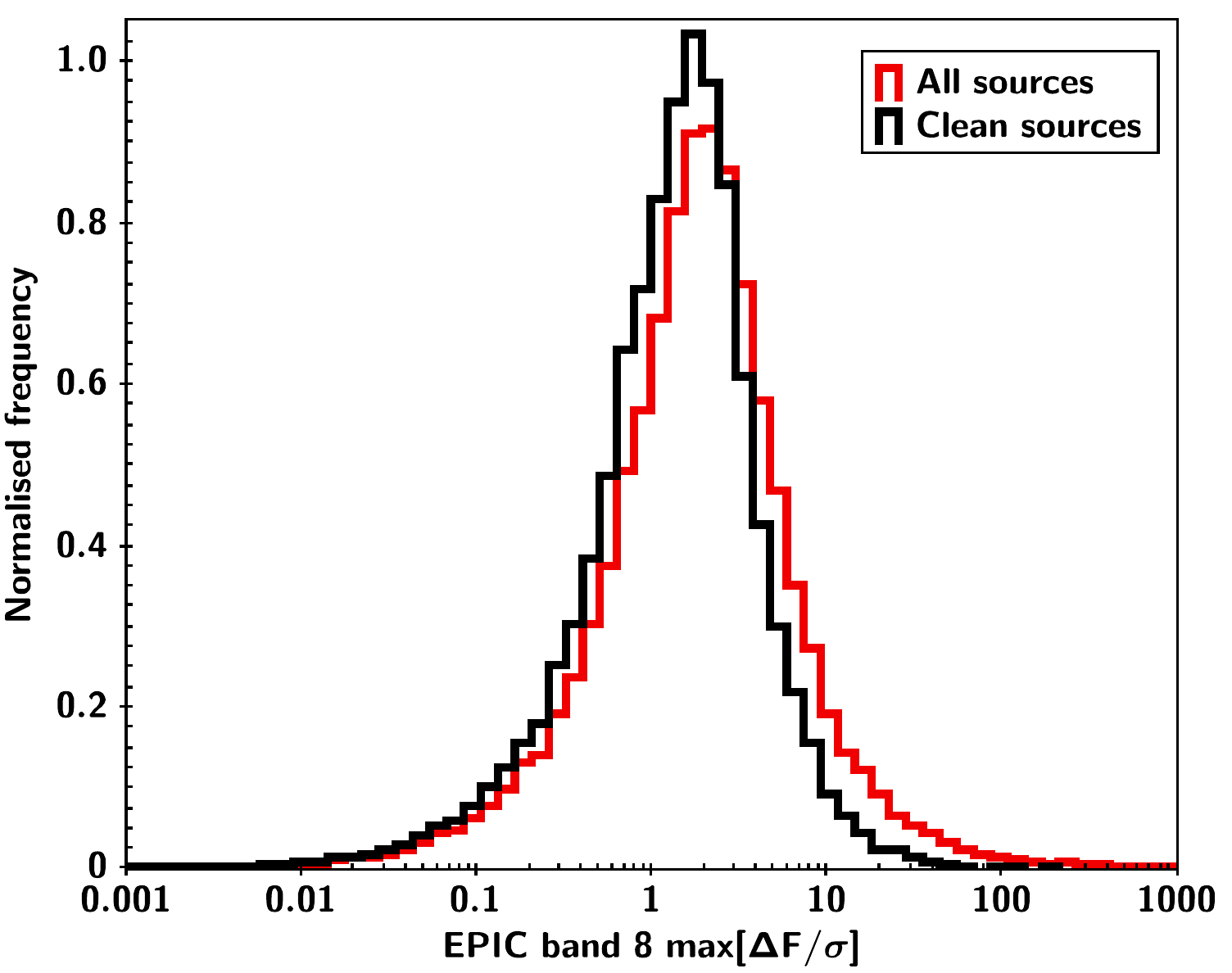}
\caption{Distribution of the EPIC band 8 {\it MDDE} parameter for all the LTV catalogue sources with valid values (red) and for a clean subset (black). Including sources affected by data quality issues shifts the distribution to higher values, suggesting quality issues may lead to increased detections 
of spurious variability. } 
\label{fig:ltvepmdde8}
\end{figure}

In figure~\ref{fig:ltvepmdde8} we show the area-normalised distribution of the EPIC band 8 {\it MDDE} parameter for all 
LTV catalogue data and a clean subset. Including data with quality issues shifts the distribution of each 
quantity in a direction that indicates more sources might be deemed long-term variable, that is, sources whose 
long-term light curves involve lower quality data are more likely to yield spurious detections of 
variability. Quality filtering is thus an important step when the catalogue is used. 

A number of simulations were pursued to gauge the level of spurious variability detection in good-quality sources in the LTV catalogue. A simple approach was adopted in which each source in the catalogue was considered to be constant, with a flux, $F_c$, equal to the error-weighted mean of 
its flux values. Each simulated point took the 1$\sigma$ error associated with the real detection or upper limit, and its flux was randomly drawn from a Gaussian distribution with $F_c$ as its mean and standard deviation, $\sigma$. Points assigned a negative flux, however, were set equal to the upper limit value estimated from the image at the source location. For each source, 10000 simulations were run 
and the simulated data from each run were processed using the same analysis approach as for the real data. 
We find that for clean sources, we expect a false-positive rate of detecting variability of $< 0.1$\% when 
adopting $\Delta{F}/\sigma$ (MDDE) $> 5$ as the definition that a source is variable. 
 

 \subsection{Quality issues}
\label{sect:LTVquality}

A number of issues can affect the quality and reliability of detections and upper limits, such as image artefacts, problems arising from inadequately characterised extended sources, high background levels, imperfections in the PSF model description, astrometry errors, and event pile-up in bright sources. In the worst-case scenarios, these can give rise to spurious detections or reduce the accuracy of photometric data in other cases. It is important to identify sources that contain one or more detection or upper-limit data compromised by such issues because inaccurate photometry can lead to erroneous detections of (or missed) variability.

For pointed data, the 3XMM-DR5 catalogue already contains  multi-element flag sets per detection that reflect issues that are automatically identified by the processing pipeline together with the results of manual screening which sought to identify problems associated with complex regions, high source densities, bright sources, and image artefacts, for example. The multi-element flags are subsequently collapsed into a summary flag {\it SUM\_FLAG}, ranging from 0  (cleanest) to 4. All these flags are explained in \cite{Watson2009}, see also \cite{Rosen2016}. For user convenience, this flag information is propagated to the LTV catalogue for pointed data. For slew data, the main quality issue arises where attitude reconstruction is less reliable, potentially affecting astrometric information. This is reported with the {\it VER\_PSUSP} flag. 

Pile-up in pointed detections is identified by testing whether the total band count rate exceeds a tabulated threshold value \footnote{\url{https://heasarc.gsfc.nasa.gov/docs/xmm/uhb/epicmode.html}} for the instrument or mode. Thresholds for untabulated modes are estimated from tabulated modes by scaling by the inverse of the frame time. Where a mode involves CCDs operating with different frame times,  the adopted threshold applies to the CCD in which the source appears. For slew data, the slew motion reduces pile-up, and a higher count rate of 4.0~cts/s is adopted as the pile-up limit for all slew detections. Where the total-band count rate of a detection exceeds the relevant pile-up threshold in a given instrument ({\it <inst>}
(=PN, M1, M2 or EP)), a logical flag ({\it <inst>\_PU\_FLAG}) is set for that instrument. For a given detection, if any of the instrument pile-up flags are set, a further flag, {\it EP\_PU\_FLAG,} is set.

Together with detection-level flags, the LTV catalogue contains additional flags to highlight sources that contain detections with potential issues. These values are set the same for all rows associated with a source, whether pointed or slew detections or upper limits. We list them below.
\begin{itemize} 
\item {\it WORST\_SF}: The worst {\it SUM\_FLAG} value of any constituent pointed detection.
\item {\it FRAC\_EP\_PU}: Indicates the fraction of detections (pointed and/or slew) with set {\it EP\_PU\_FLAG} . This highlights sources where pile-up may be a causing one or more underestimated detection count rates.
\item {\it FRAC\_EXT}: Quantifies the fraction of detections whose all-EPIC band 8 (or for slew, band 6, 7 or 8) intrinsic extent is > 6\arcsec.
\item {\it SLEW\_FLAG}: Indicates sources containing one or more slew detections where the {\it VER\_PSUSP} flag is set. \footnote{The LTV catalogue excludes all upper limits from slew observations that might be affected by astrometry issues (i.e. those that would have {\it VER\_PSUSP}=T). This is because when they are included, they trigger the setting of the {\it SLEW\_FLAG} in ~7900 sources, but in the vast majority of cases, it is a single slew upper limit entry that causes the {\it SLEW\_FLAG} to be set, but the slew upper limit entry usually adds no useful value to the light curve of the source.}
\item {\it FRAC\_STV}: The fraction of pointed detections showing evidence of short-term variability (identified by their {\it VAR\_FLAG} being set in 3XMM-DR5). Such short-term variability (e.g. short-lived flares) could be a contributing factor to any apparent long-term changes. Because pipeline processing only extracts exposure-level light curves for detections with > 100 EPIC counts, there is no information on short-term variability available from the 3XMM-DR5 catalogue in many cases. More sensitive information on short-term variability can be explored through EXTraS (see section~\ref{sect:stv}), however.
\item {\it N\_NEARSRC}: The number of other sources within 20\arcsec\ of the source. This alerts users to cases with increased risk that the assignment of detections to sources may be suspect.
\item {\it FRAC\_POSCOROK}: Provides the fraction of pointed detections with {\it POSCOROK}=T, that is, where astrometric rectification was considered successful.
\end{itemize}

\subsection{Example usage}\label{LTVexample}

Long-term X-ray variability data can provide key insights into astrophysical sources, such as tidal disruption events, the flaring activity in active galaxy nuclei, the cause of accretion rate changes in X-ray binaries and catalclysmic 
variables, flare frequencies and intensities in active stars, and outbursts from ultra-luminous X-ray (ULX) sources. Here, we briefly illustrate the potential of the LTV catalogue data for studies of ULX sources. 

We take as one example the catalogue of 2139 detections of sources from 3XMM-DR4 that were identified as having non-nuclear associations with bright galaxies (Earnshaw et al. 2019; hereafter E19). Converting the LTV catalogue EPIC band 8 fluxes of these detections into luminosity, $L_8$ (and 1$\sigma$ error, $\Delta{L_8}$), using the same flux-to-luminosity factors as E19, and applying their criteria for selecting ULX candidates (i.e. $L_8 > 10^{39}$\ ergs~s$^{-1}$ or $L_8 + \Delta{L_8}  > 10^{39}$\  ergs~s$^{-1}$), we isolated 351 sources that met the criteria. This compares with 384 found by E19. We find 330 sources in common that meet the criteria, that is, 86\% of the E19 sample. Fifty-six sources in the E19 catalogue have no match in the LTV sources and 23 LTV sources have no match in the E19 sources. The differences stem from the difference in the determination of the count rates and the ECFs used to convert count rates to fluxes. While the EPIC band 8 fluxes used by E19 (from 3XMM-DR4) and the LTV band 8 fluxes broadly follow a one-to-one relation, there is significant scatter due to this difference in method.


Earnshaw, Roberts \& Sathyaprakash (2018) (hereafter E18) selected an initial subsample of 12 candidate transient ULX sources from the 384-source superset of E19 based on those showing at least a factor 10 change in luminosity amongst the detections or upper limits. Subsequently, this was filtered down to 5 sources with a secure factor $>$10 variability, following careful scrutiny of the data.

We used the 351 LTV candidate ULX sources mentioned above to perform a similar selection based on EP\_MR8 $>$10. This selected 15 sources. After we applied a quality threshold to the LTV subset, requiring the lowest summary flag to be $\le$1, 10 sources remained. As in E18, 4 sources are a consequence of a duplication of catalogue identifiers associated with a pair of close sources that are incorrectly identified as a single but different source in two separate observations. These are excluded. The resulting subset of 6 sources are the first 6 entries shown in table \ref{tab:LTVexamplesrcs}. The table includes where measurable the fastest factor 2 and factor 10 upward and downward changes in flux observed in the available LTV data. This subset includes 3 of those in the final subset of 5 ULX transient sources discussed by E18, but does not include NGC 6946 ULX-1 (203500.1+600908) from the E18 final subset of 5 sources because in the LTV catalogue, the maximum/minimum ratio is 7.5, which is below the factor 10 variation threshold. The maximum/minimum of the XMM EPIC band 8 luminosity data provided in the E19 catalogue is also below the threshold. Another of the final 5 sources of E18, M51 ULX-4 (132953.3+471042), passes the EP\_MR8 $>$10 threshold but is also absent from the list because it has three LTV detections with a summary flag of 3. These two E18 cases are shown at the end of table \ref{tab:LTVexamplesrcs}. The 5 sources in the E18 ULX transient list are indicated by a 'Y' in the last column.

Three of the 6 LTV sources are not in the E18 list. The first, 022134.1-053105, is a marginal case of a factor 10 change. In the two observations where it is detected, both are short (2ks) exposures with very few counts in the MOS cameras, while the pn data has a high background. Furthermore, one of the two EPIC band-8 detections is characterised as slightly extended, rendering the flux less reliable. For the second source, 073650.0+653603, the variation (a single detection and two upper limits) is clear and real. This is likely to be one of the sources considered as a blend by E18 as there is a faint source about 20\arcsec\ away. Based on the high ($>$100) max/min ratio of 073650.0+653603, however, accounting for contamination by the faint source would be very unlikely to reduce the ratio to below 10. The third source, 213631.9-543357, is likely to be the other blended case that E18 excluded. Again, this source comprises one EPIC band-8 detection and two upper limits in the LTV catalogue. One of the observations yielding an upper limit is affected by high background in all three cameras, but in the observation where the detection is claimed, the source is clearly present in the available (pn, MOS2) cameras. The LTV catalogue indicates the presence of another source within 20\arcsec\ so that some contamination is likely.

The above very simple process, which mimicks the analysis of the XMM-Newton data performed by E19 and E18 and broadly confirms their sources as long-term variable, shows the merit of the LTV catalogue in quickly finding potential long-term variable X-ray sources in the 3XMM-DR5 catalogue data from  user-defined samples, exploiting the auxiliary global source quality information to filter or check the data. Nevertheless, we urge users always to inspect the data (including the image data) because use of the quality information alone may be not good enough.

\begin{table*}
\label{tab:LTVexamplesrcs}
\centering
\caption{Six clean sources (lowest summary flag $\le$1) in the LTV catalogue with EPIC band-8 luminosities above the threshold adopted by E19 to be considered as ULX candidates, and with maximum-to-minimum flux ratios $>$10 (values in the EP\_MR8 column). All measurement quantities refer to EPIC band-8 data. The EP\_ET quantities are the shortest timescales (in days),  in which factor 2 or factor 10 up (u) or down (d) changes of flux are seen in their LTV data. The last two rows are for two of the five sources from E18 that do not appear in the LTV set. The  reasons are discussed in the text. For 022134.1-053105 and 230457.6+122028, the EP\_ET10u8 and EP\_ET10d8 values are absent because with the more conservative definition of these measures (see section \ref{sec:LTVchar}), which include the errors, the changes are smaller than a factor of 10.}
\begin{tabular}{l|c|c|c|c|c|c}
Source & EP\_MR8 & EP\_ET2u8 & EP\_ET2d8 & EP\_ET10u8 & EP\_ET10d8 & in E18 list\\
\hline
  013636.4+155036 & 13.2   &         &  339.5  &           &     339.46805  & Y\\
  022134.1-053105 & 10.4   & 0.14    & 1078.1  &           &  &  \\
  073650.0+653603 & 115.2  & 133.7   &  367.7  & 133.7 & 367.7 &  \\  
  121847.6+472054 & 34.8   & 156.7   &  1640.0 & 339.3 & 1640.0 & Y \\
  213631.9-543357 & 21.9   & 1239.6  &         & 1239.6  &  &   \\         
  230457.6+122028 & 11.4   &         & 1465.7  &           &  &  Y \\
\hline
  132953.3+471042 & 29.3   & 1839.6  & 4.1     & 1839.6 & 327.1 & Y   \\
  203500.1+600908 & 7.5    & 502.7   & 518.6   &           &
  &   Y \\
\hline
\end{tabular}
\end{table*}


\subsection{Known problems and issues}\label{LTVknownissues}

Whilst the photometric measurements of detections are generally robust, problems can arise in some circumstances.  We discuss some residual points that are relevant to the LTV data and analysis.

\subsubsection{Spatially extended detections of point sources}
\label{LTVissues_extended}

Point sources can sometimes be erroneously characterised as extended, which can yield incorrect photometry. This might lead to spurious identification of variability or failure to detect real variability. While we could forcibly characterise sources as point-like (because the LTV catalogue is mainly about point sources), but this can also produce incorrect 
photometry in some cases. Instead, the {\it FRAC\_EXT} flag is used to indicate the fraction of detections of a source that are characterised as extended. Less than 1.5\% of $\sim$27000 
otherwise clean sources \footnote{clean here means detections whose  
slew flag is not set and whose lowest summary flag 
$\le 1$, which are not piled up and have no other source within 30\arcsec.} comprising two or more pn detections have one or more of those detections (but not all) measured as extended.

\subsubsection{Slew upper limits}\label{LTVissues_slewuls}

Pointed and slew upper limits are estimated by aperture photometry (circular with 28\arcsec\ radius), centred on the source position in each observation covering the position and effectively corrected for the encircled energy fraction (EEF) using an empirical approach \citep{Carrera2007}. The empirical correction factors, however, were derived from pointed sources, but for slew upper limits, should instead reflect the slew-specific PSF discussed in section~\ref{sec:LTVslewdata}. As a result, we estimate that the slew upper limits in the LTV catalogue are $\sim$5\% lower than when correction factors based on slew data were used.

\subsubsection{Spectral effects}\label{LTVissues_spectra}

Because the ECFs used to convert count rates into fluxes (see section~\ref{sec:LTVpointeddata}) assume a fixed spectral profile, spectral changes in a source between epochs can introduce or mask variations in source flux. Simulations of a power-law model whose slope is changed by $\pm$0.6 from the nominal 1.7 used to create ECFs suggest that such spectral changes can leave the count rates unaltered but yield flux changes up to $\sim$ 20\%, 35\%, and 70\% in the soft, hard, and total bands, respectively (see also \cite{Mateos2009}). Furthermore, until the time of creating the 3XMM-DR5 catalogue, while the pn camera sensitivity and thus its ECFs had been deemed stable, the sensitivities of the MOS cameras had evolved, being effectively characterised by 13 time-dependent ECFs. The LTV catalogue MOS fluxes are, like 3XMM-DR5, based on epoch-13 MOS ECFs. \cite{Rosen2016} outlined the effect of using a fixed MOS ECF, but for most sources away from the central $\sim$40\arcsec\ degraded patch, the worst deviations from using the most relevant time-dependent MOS ECF are $<$2.5\%. The band-8 MOS1 long-term light curve of the modestly extended (assumed flux-stable) supernova remnant calibration source, 1ES0102-72.2, shows a declining trend in measured flux of $\sim$6.5\% over the mission duration, supporting this conclusion. As a soft X-ray source, usually observed  within the degraded central patch, it is subject to a greater change in sensitivity than sources outside the patch.

\subsubsection{EPIC upper limits}\label{LTVissues_epiculs}

Although all-EPIC (combined instrument) count rates (the sum of the instrument count rates) and fluxes (the error-weighted average of the instrument fluxes) are computed in each band, computing equivalent all-EPIC upper limits is not straightforward, and they are not calculated by the upper-limit software. EPIC flux upper limits provided in the LTV catalogue for pointed data are instead the highest of the available instrument upper-limit values. This means that they are generally a conservative (high) estimate of the rate and flux
upper limit. All-EPIC upper limits based on more than one instrument would generally be lower due to the lower statistical noise. In addition, instrument upper limits
(where available) also replace EPIC flux and rate values when the all-EPIC detection likelihood value is $<$8.0.

\subsubsection{Systematic uncertainties}\label{LTVissues_systematics}

For the per-instrument, per-band long-term light curves, systematic errors between photometric data from a given instrument are not relevant, other than the ECF issues discussed in sections \ref{sec:LTVslewdata}, \ref{sec:LTVpointeddata}, and \ref{LTVissues_spectra}. Systematics affect the all-EPIC data, however, which is a combination (weighted average) of the available instrument fluxes for a given observation.  Comparing simultaneously observed pn and MOS fluxes for clean sources following a similar approach to \cite{Lin2012}, we estimated systematic errors of 0.13, 0.13, and 0.16 (as fractions of the MOS flux) between pn and MOS (average of MOS1 and MOS2) in bands 6, 7, and 8, respectively.  These systematics, however, are not integrated into EPIC flux error values or used in the LTV analyses. Another potential systematic uncertainty is that between the pointed and slew flux data. This is difficult to estimate, however, because measurements of sources in pointed and slew mode can never be simultaneous. Based on very limited ($<$10) sources observed in a slew and in a pointed observation within a day of each other, it proved impossible to determine any such systematics.


\subsubsection{Short-term variability}\label{LTVissues_stv}

Short-term variability
within an observation can contribute to the appearance of long-term variability. 
To explore this, the subset of LTV sources containing one or more individual pointed detections that are known to show 
variability within the observation (i.e. where the var\_flag is set in the 3XMM-DR5 catalogue), were isolated and the LTV analysis run on their LTV pn band-8 light curves, with any short-term-variable detections excluded.  We find that when all detections are included, the $\Delta{F}/\sigma$ ({\it MDDE}) values are notably shifted to higher MDDE values than when detections affected by short-term variability are excluded. The same effect is seen in clean sources. The median MDDE values are 9.72 (7.58 for the clean subset) when STV detections are included, compared to 4.50 (3.22) when they are excluded. Corresponding medians for the MR parameter are 5.29 (2.64) and 2.66 (1.93), 
and for the DRCHISQ parameter, are 43.68 (32.11) and 9.78 (5.70). When all detections are included, 224 out of 1163 sources (19.3\%) have pn band 8 MDDE > 5, while when short-term variability detections are excluded, 65 out of 758 sources (8.6\%) have pn band 8 MDDE > 5. 

We note that the indication of short-term variability, that is, that the var\_flag is set, does not exclude the possibility that some points in the real data light curves exhibit short-term variability but are not flagged as such. Observation-level light curves are only produced for sources with > 100 EPIC counts in their XMM-Newton light curves, so that any detection that is fainter than this will not have a light curve, hence short-term variability cannot be tested.

\subsubsection{Variability detection: alternative approaches}\label{LTVissues_alternate}

The analysis applied to the LTV data computes the flux ratio, which makes use of detections and upper limits, and the maximum significance and chi-square values, which are restricted to detections. The computations assume a Gaussian
error analysis. Importantly, those involving upper limits effectively treat them as data points with uncertainties (as outlined in section~\ref{sec:LTVchar}), but this is evidently a simplifying approximation. An alternative likelihood approach that more formally takes the non-detections and the Poissonian nature of the data into account, was considered late in the project, broadly following the approach developed as part of the aperiodic variability analysis within the EXTraS project.  This requires raw count information, however, which was not originally envisaged as part of
the LTV work and was not pursued.











%
%

\section{Online resources}\label{sect:resources}

\subsection{The EXTraS public data archive}\label{sec:archive}
The EXTraS public data archive can be accessed from http://www.extras-fp7.eu/index.php/archive. It is the primary online repository for all data generated by the project, supporting a wide range of products such as X-ray light curves, hardness ratios, power spectra, and source catalogues with measures of variability. We summarise the basic functionalities below. For technical details regarding the software implementation, we refer to \cite{dagostino2019}.

The archive is an outgrowth of the existing Leicester Database and Archive Service (LEDAS, https://www.ledas.ac.uk) at the University of Leicester, hosting data from several major X-ray missions. Within the EXTraS project, the core archive system originally developed for LEDAS has been fully rewritten to current software development standards. 
Catalogues and bulk products for all EXTraS data analysis pipelines have been incorporated in LEDAS. 
A total of 18 TB of EXTraS data is currently held in the LEDAS central archival storage.

The main page of the archive provides users with a top-level menu to access EXTraS data products by analysis line: short-term aperiodic variability, search for periodicity, transients, long-term variability. Results of multiwavelength characterisation and classification (not described in this paper, see Sect.~\ref{sect:extras}) are also included. A combined catalogue allowing simultaneous source searching across all EXTraS catalogues and a basic catalogue cross-matching tool are also provided. 
Online help is available for all catalogues.

\begin{figure}[ht]
\centering
\includegraphics[width=9cm]{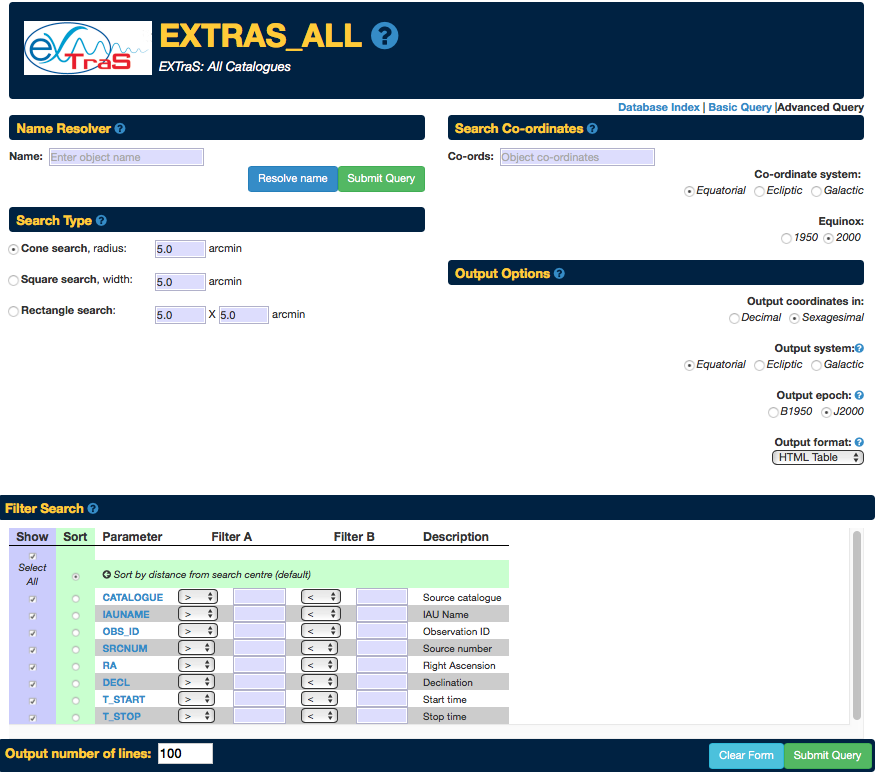}
\caption{Advanced catalogue search form.}
\label{fig_db1}
\end{figure}

The catalogue basic search form allows searches in a given sky region (using either a cone, box, or rectangle search area), or by identifier \citep[resolved by the Simbad database, ][]{sinbad}.
The catalogue advanced search form, shown in Fig. \ref{fig_db1}, allows users in addition to position searches to search for sources by setting filters on any parameter in the EXTraS catalogues. A filter search can be performed either as a match (i.e. selecting database entries where a specific parameter is equal to a desired value) or over a range (i.e. select all database entries where a specific parameter lies in a desired range). Filter searches can be performed as inclusive or exclusive filters by selecting the appropriate option.
In the basic and advanced searches, users can select a minimum or full set of output table columns and display a variety of output formats (HTML, ASCII table, CSV, VOTable etc).

\begin{figure}[ht]
\centering
\includegraphics[width=9cm]{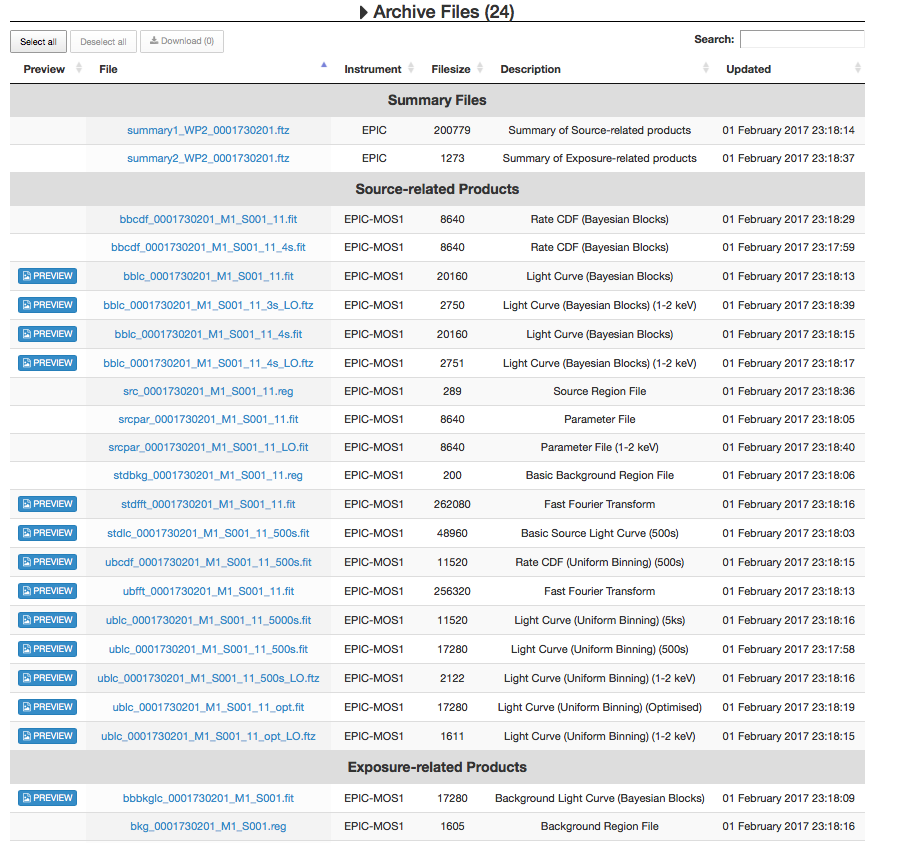}
\caption{Summary page of the catalogue results for each single source.}
\label{fig_db3}
\end{figure}

\begin{figure}[ht]
\centering
\includegraphics[width=9cm]{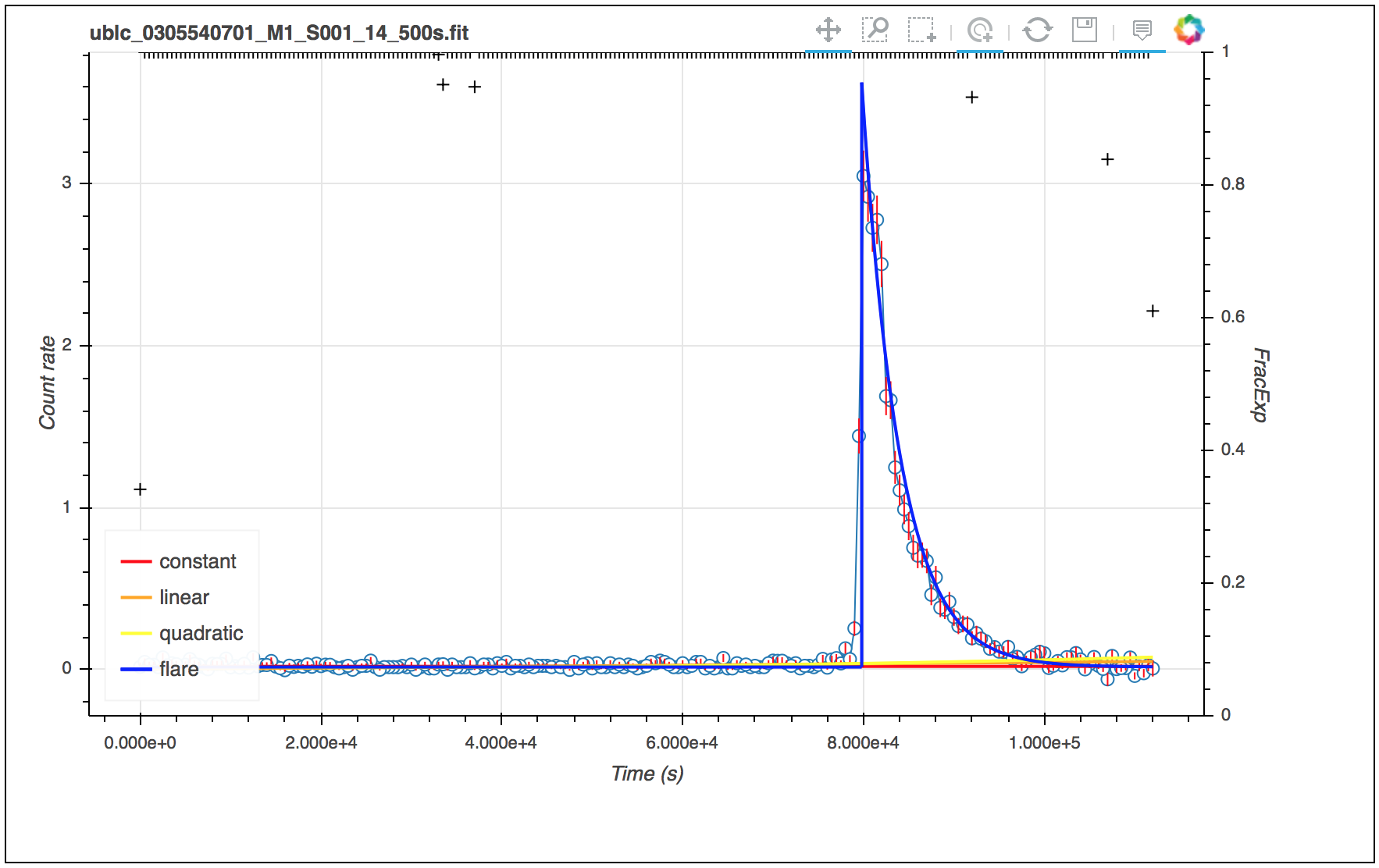}
\caption{Interactive visualisation of products from the short-term aperiodic variability analysis. The case of a light curve with uniform time binning is shown. 
By using command buttons on top of the window, the user can e.g. zoom or pan, read count rate and errors by clicking on data points, and overplot best-fit models.
}
\label{fig_vis1}
\end{figure}

\begin{figure}[ht]
\centering
\includegraphics[width=9cm]{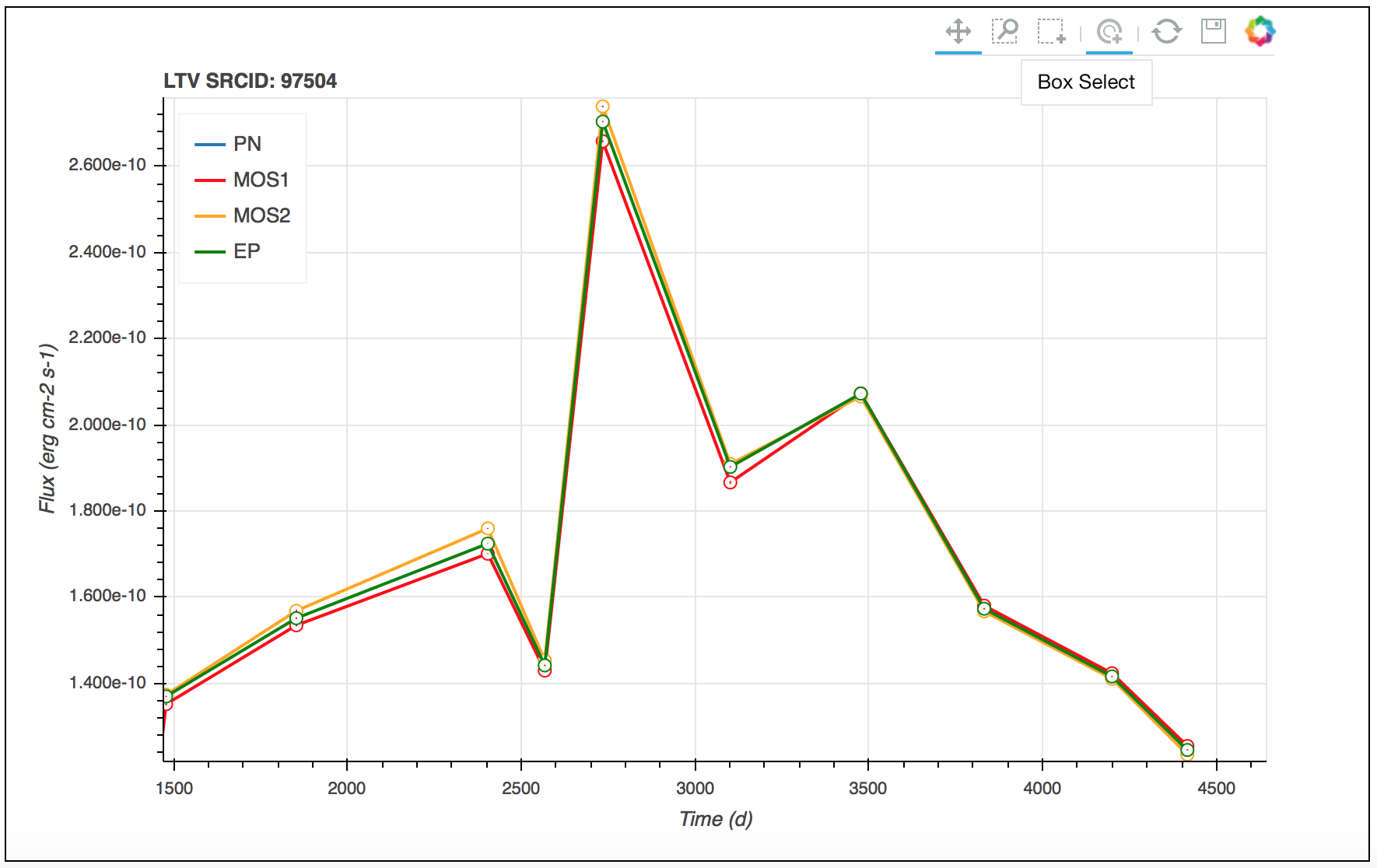}
\caption{Same as Fig.~\ref{fig_vis1} for the case of an LTV light curve.}
\label{fig_vis2}
\end{figure}

The archive interface provides users not only with the possibility to download the results data, but also with an expanding set of dynamic interactive visualisations for the EXTraS catalogue and bulk product data. The visualisations are generated directly in the browser (by clicking on the blue ``PREVIEW'' button shown in the left column of Fig. \ref{fig_db3}) 
and require no additional software installation. 
Figures \ref{fig_vis1} and \ref{fig_vis2} show examples of EXTraS Public Data Archive visualisation output. The user can zoom and pan the plot, read values and uncertainties by clicking on data points, overplot best-fit models, and save the customised plot.


\subsection{The EXTraS portal for online analysis}
\begin{figure}[!hbt]
\centering
\includegraphics[width=9cm]{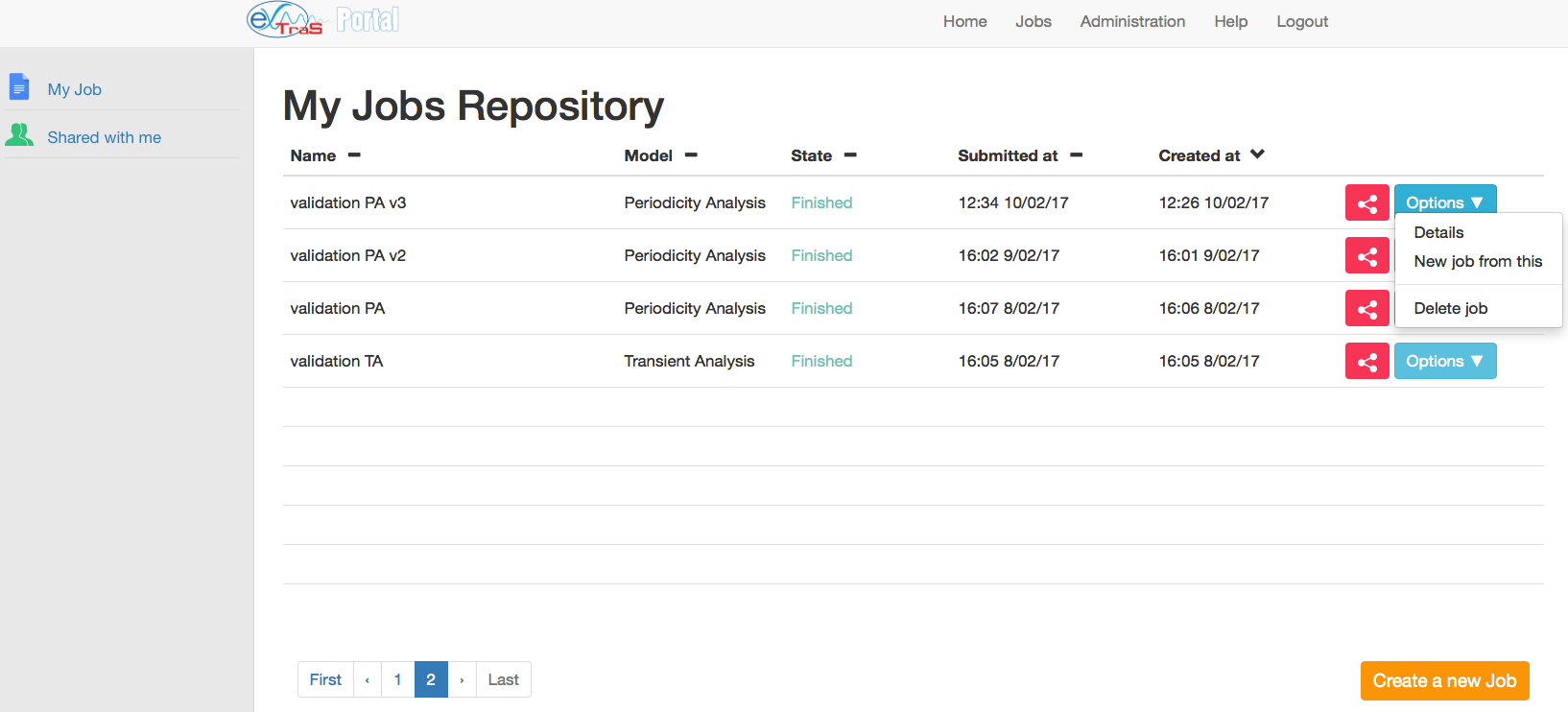}
\caption{Jobs management module interface.}
\label{fig_jm}
\end{figure}

\begin{figure}[!hbt]
\centering
\includegraphics[width=9cm]{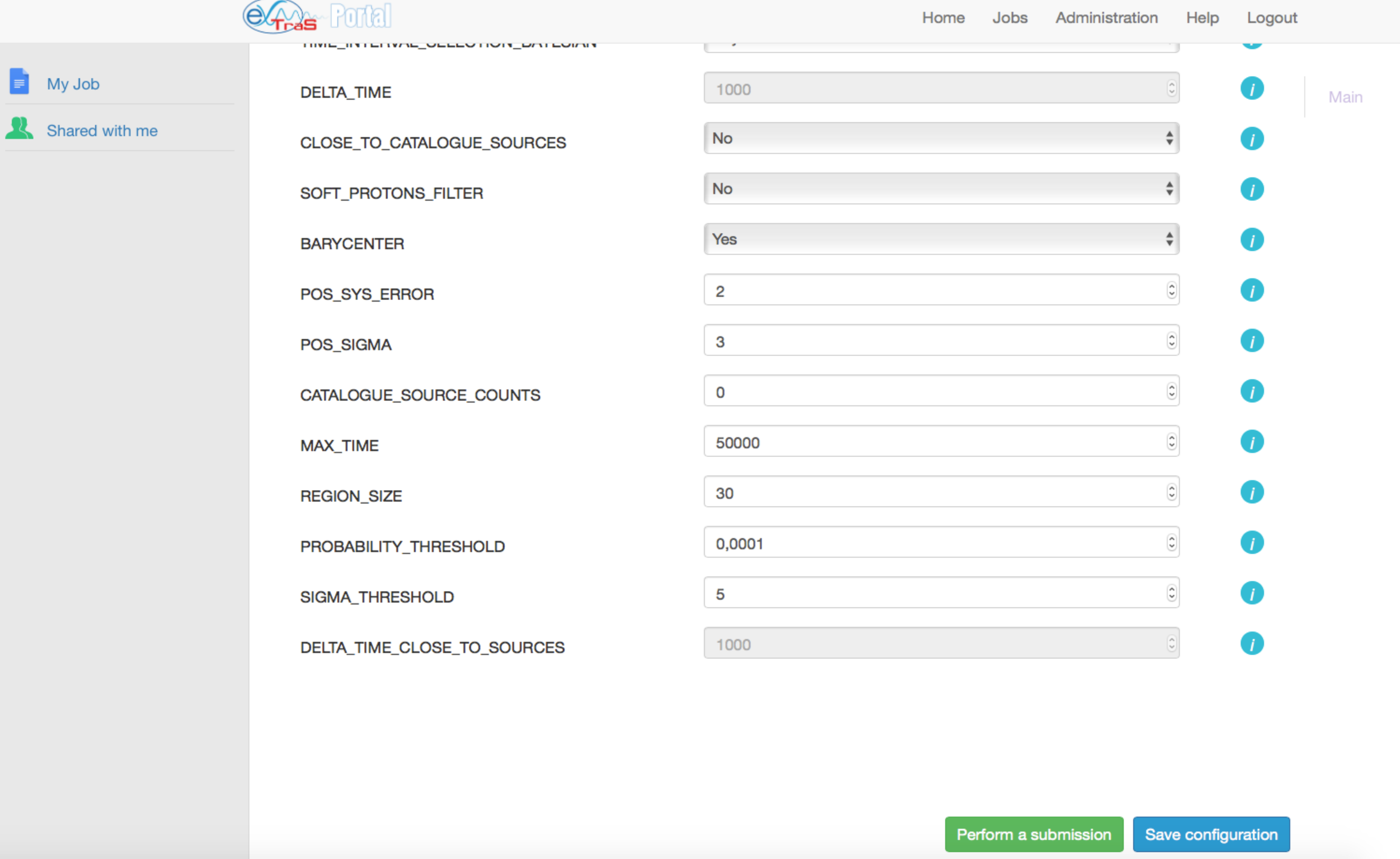}
\caption{Transient analysis user interface shown by the workflow configuration module.}
\label{fig_wc}
\end{figure}

Different strategies can be adopted to provide the scientific community with software tools. A first approach is to release an installer or an archive, including all the files required to compile and run the analysis tool. 
This solution has been adopted for different tools for the astronomical community, such as the XMM-Newton SAS.
A second approach is to make the software available by exporting the corresponding workflows, which can thus be executed using a workflow management system. This solution has commonly been adopted by the astronomical community \citep{AT}. A third approach is to provide the community with a virtual machine having all the software installed on it.  This is possibly the most effective solution for non-expert astronomers and for dissemination purposes, e.g. for educational programs or citizen science. We note that the SAS is also made available as a Linux virtual machine. 
The fourth approach is to make the software available as a set of services through a Web portal designed according to the science gateway paradigm \citep{sg2}. We decided to follow this strategy because science gateways are attracting increasing interest in many communities \citep{sg}, such as the astronomical one \citep{visivo}.

The EXTraS portal \citep{portal} is aimed at providing users with a seamless environment to analyse the observations archived in the XSA with the EXTraS pipelines -- hereafter called 'experiments'. A user-friendly interface is available, with no need to install any software. 

The main page, shown in Fig. \ref{fig_jm}, is a web app \citep{jsongui} allowing users to create, submit, and manage different data analysis sessions, using the software developed by the EXTraS project. All the submitted or configured analyses are shown, with the possibility of creating a new analysis starting from an existing configuration, or sharing results with other users.
After the parameter definition (see Fig. \ref{fig_wc}) and the selection of the observation to be analysed, the analysis job is managed by the portal, based on computing resources provided by EGI Fedcloud \citep{fedcloud} to virtual organisations (VO), i.e. groups of users whose members are engaged in  similar research activities. In particular it can be used also for citizen science activities, as discussed in \citet{dagostino2019}.

All the information regarding a job (e.g. setup,  logs,  results,  ownership or sharing of information as well as possible comments) are stored in the portal database via the Persistence API until it is removed by the user who owns it. 

The EXTraS portal has two further important features: it provides users with the possibility to share an analysis (i.e. the namelist and possibly the results); it also gives support for a discussion (exchange of comments) among the users sharing it.
When a completed job is shared, results of the experiment are visible to other users. The configuration is also shared and can be used to run the same analysis on a new set of data. 
Thus, other users can replicate the execution of a job, to validate the results, or investigate the effects of changing one or a few parameter values. 

Results produced within the portal are not automatically stored in the public data archive. They have to be validated by the project community, who can publicly discuss it using the portal. 

%
%

%
%

\section{Summary and conclusions}\label{sect:summary}

The EXTraS project produced the most sensitive and thorough search for and characterisation of temporal variability in the soft X-ray sky.


We produced a complete characterisation of short-term aperiodic variability (on timescales shorter than the exposure time) for about 420,000 point sources included in the 3XMM catalogue. This was based on modelling of time-averaged properties of point sources in 3XMM and on careful modelling and characterisation of the variable EPIC background noise. For each source we generated (i) background-subtracted light curves with uniform time binning at 500s, optimal, and 5\,ks, (ii) background-subtracted light curves with adaptive time binning based on the Bayesian block approach, with different (sensitive and robust) segmentations, and (iii) power spectra. Starting from these products, we computed a set of synthetic parameters quantifying different aspects of each source's variability. We ran a simplified version of the pipeline to extract light curves for the same set of sources in three energy sub-ranges and to generate hardness ratios. A set of simulations and statistical tests were used to confirm and validate our products and results.

We systematically searched for periodic modulations in more than 300,000 sources in the 3XMM catalogue, running a pipeline based on a generalisation of the FFT approach accounting for non-Poissonian noise components. For each detected signal, a refined search was performed using the Rayleigh technique. Different parameters were computed (e.g. significance level, pulsed fraction) and several products were generated (e.g. light curves, folded light curves, power spectra, periodograms). If no pulsations were found, the $3.5\sigma$ upper limit to the pulsed fraction was evaluated. Statistical tests were performed to confirm the validity of the analysis and its sensitivity. 

We ran a blind search for transients and highly variable faint sources. Two approaches were implemented. In the first, a source detection was run on short time intervals of uniform length. In the second, promising time intervals of optimised duration were spotted by searching for count rate changes (using a Bayesian block approach) in spatially independent portions of the FoV, and a standard source detection was performed on the selected intervals. Different runs were carried out on the whole sample of EPIC observations using different pipelines with different settings. Cross-check and statistical analysis of results together with a complete visual screening allowed us to identify a robust sub-sample of 136 short-duration highly-significant transient sources that are not listed in the 3XMM catalogue. 

We systematically investigated long-term variability (on timescales longer than XMM exposures) in all detected EPIC sources from pointed and slew observations. The analysis was performed in three different energy ranges (total, soft, and hard) and was based on (i) an improved slew data processing pipeline, resulting in an updated slew survey catalogue, (ii) a consistent computation of upper limits in slew and pointed data, (iii) a collation of slew and pointed photometry together with upper limits, and extraction of long- term light curves, and (iv) a search for and characterisation of variability in the resulting typically very sparse time series. Particular attention was devoted to the study of the compatibility of flux measurements in slew and pointed data. The main output was an LTV catalogue including more than two million photometric measurements for about 420,000 unique sources together with meta-data for the observations used, quality information, and a number of variability parameters that gauge the level and timescales of variability.


All results have been released to the community in early 2017 in a public archive, including a database of variability parameters and more than 20 million products. A user-friendly interface for accessing data is operational. A visualisation server was implemented to provide users with a powerful facility for interactive display of all archived data and metadata.
We also released the source code of the software tools developed by EXTraS to perform searches for and characterisation of short-term aperiodic variability, searches for periodicity, search for new transients, and characterisation of long-term variability.
We implemented the EXTraS Science portal, a new science gateway, for providing search for short-term aperiodic variability, search for pulsations, and search for new transients on EPIC data. Users can select their dataset from the XMM-Newton archive and run selected EXTraS pipelines via a simplified interface, with no need to install any software. 
The portal manages all jobs, based on computing resources provided by the European Grid Infrastructure.

EXTraS results and products are proving to be a very rich resource for investigations in almost all fields of astrophysics, with applications ranging from the search for rare events and peculiar objects to the study of the properties of large samples of sources. We encourage the community to explore the EXTraS archive and to develop projects based on our results and tools. 
The outcome of EXTraS will also serve as a learning case for new experiments focusing on the X-ray variable sky, from SVOM to eROSITA to Athena. 

At a different level, our project also offers an extensive test for different data analysis approaches and methods that could be directly applied to the analysis of data from other current and future experiments. We list a few examples below.



First, the overall strategy we devised to measure source and background contributions, including new recipes for (i) optimisation of the source region (Sect.\,\ref{sect:sourceregion}), (ii) modelling the spatial distribution of a constant and of a variable background component (Sect.\,\ref{sec:bkgmod}), and (iii) optimisation of a background region to extract a representative background light curve (Sect.\,\ref{sec:bkgreg}). This might be adopted to compute accurate time-dependent photometry in any imaging photon-counting instrument.


Second, our implementation of the Bayesian block algorithm (Sect.\,\ref{sect:bb}), designed to take the highly variable background rate of the EPIC instrument into account, might easily be used for the production of adaptive binning light curves from any other photon-counting detector, using time-resolved photometric data, on-source (source + background) and off-source (background) measurements.


Third, our periodicity search algorithm that takes the properties of broad-band noise into account (Sect.\,\ref{sec:psearch}, {\em Step 2}), was designed by \citet{{israel96}}. It can be applied to the analysis of power density spectra independent of the detector that was used to collect the time series. It has already proved to be successful in searching for pulsations in a range of different cases \citep[ROSAT, Chandra, XRT, NuSTAR; see e.g.][]{israel2016}.

Fourth, the core of the algorithm we developed to search for new transients is based on the segmentation of the field of view in angularly independent regions and on the Bayesian block analysis of time series from each region to select time intervals displaying deviations from the background count rate (Sect.\,\ref{sec:newtransbb}). This could be used for any imaging photon-counting detector and could also be implemented as a (near) real-time monitor for transients in future experiments.


EXTraS results, products, and tools are also proving to have great potential for the popularisation of science in general and of astronomy in particular, offering excellent opportunities to promote exciting science to students and to a general public audience. With this in mind, an experimental didactic program was designed within the project and was implemented in several workshops for high-school students in Italy, Germany, and the UK. The final goal of the program is to engage the students (and in perspective, citizen scientists) by involving them in a research program. Based on the use of EXTraS online resources, they examine the data and try to select new phenomena, or to characterise already known sources. To do this, they follow the whole validation process. One of these workshops resulted in a very interesting discovery: a peculiar flaring source in the globular cluster NGC6540 \citep{mereghetti18}\footnote{The story was the subject of a Research Highlight News on the Nature magazine, see \url{https://www.nature.com/articles/d41586-018-05959-4}}. Our educational activity will also turn to an experiment of citizen science, allowing us to assess the possibility of involving non-expert (but trained) people in a complex classification task. See \citet{dagostino2019} and references therein for more details.

\begin{acknowledgements}
We thank an anonymous referee and the editor for their constructive and helpful comments and suggestions.
This work is based on observations obtained with XMM-Newton, an ESA science mission with instruments and contributions directly funded by ESA Member States and NASA. This research has made use of data produced by the EXTraS project, funded by the European Union's Seventh Framework Programme under grant agreement no 607452. 
We  acknowledge  the computing  centre  of  INAF–Osservatorio  Astrofisico  di Catania for the availability of computing resources and support under the coordination of the CHIPP project. We acknowledge the computing  centre  of  INAF–Osservatorio  Astrofisico  di Trieste for the availability of computing resources. We acknowledge financial support from ASI under ASI/INAF agreement N.2017-14.H.0. The EXTraS project also  acknowledges extensive use of the ALICE High Performance Computing Facility at the University of Leicester.
{\bf Key roles:} A. Belfiore coordinated the short-term, aperiodic variability analysis; G.L. Israel coordinated the search for pulsations; A. Tiengo coordinated the search for new transients; S. Rosen coordinated the long-term variability analysis; A.M. Read produced the updated slew survey; D. Law-Green implemented the data archive; D. D'Agostino implemented the science gateway.
\end{acknowledgements}

%
%

\bibliographystyle{aa}
\bibliography{EXTraS_paper}

%
%

\begin{appendix}

\section{XMM Science Analysis Software counting-mode issue} \label{appendix:SAS_issue}

We describe the bug that is present in old versions of the SAS software. It produces incorrect time-tagging of events. The problem was found to affect a significant fraction of PPS event files at the time of the EXTraS project.
The recent bulk reprocessing of XMM data in late 2019 fixed the problem in all PPS files.

At the beginning of the project, we compared event files in PPS products to event files generated using SAS v14, starting from observation data files (ODF) in order to select the starting point for
the analysis.
We recall that ODF files are level-0 products that require a time-consuming pre-processing to obtain a level-1 event file. We used SAS v14.0.
PPS files instead that contain level-1 event files processed with a specific configuration (i.e. specific version of SAS tools) that is usually updated on a yearly basis\footnote{See\url{https://www.cosmos.esa.int/web/xmm-newton/pipeline-configurations}}. 
We found large differences between event files from PPS (hereafter 'PPS') and event files generated from ODF (hereafter 'reprocessed') for a number of test cases.

\begin{figure*}[ht]
\centering \includegraphics[width=15cm]{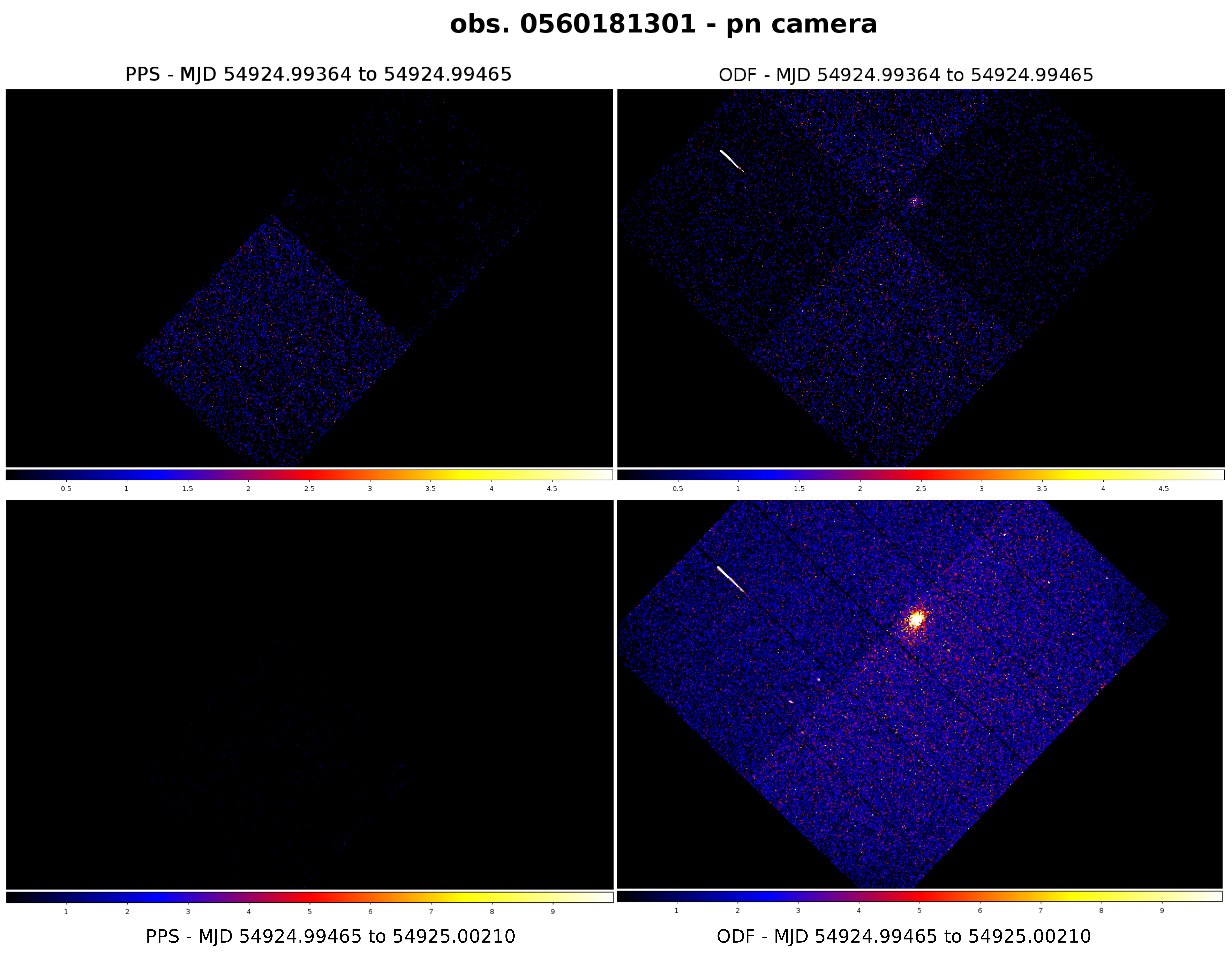}
\caption{Comparison between pn images obtained for the same time period starting from PPS and ODF event files (left and right panels, respectively). In the upper panels
we extracted events from the last former GTI listed in the PPS event file, while in the lower panels we extracted events after that time period.} 
\label{fig:cmimage}
\end{figure*}

\begin{figure*}[ht]
\centering \includegraphics[width=15cm]{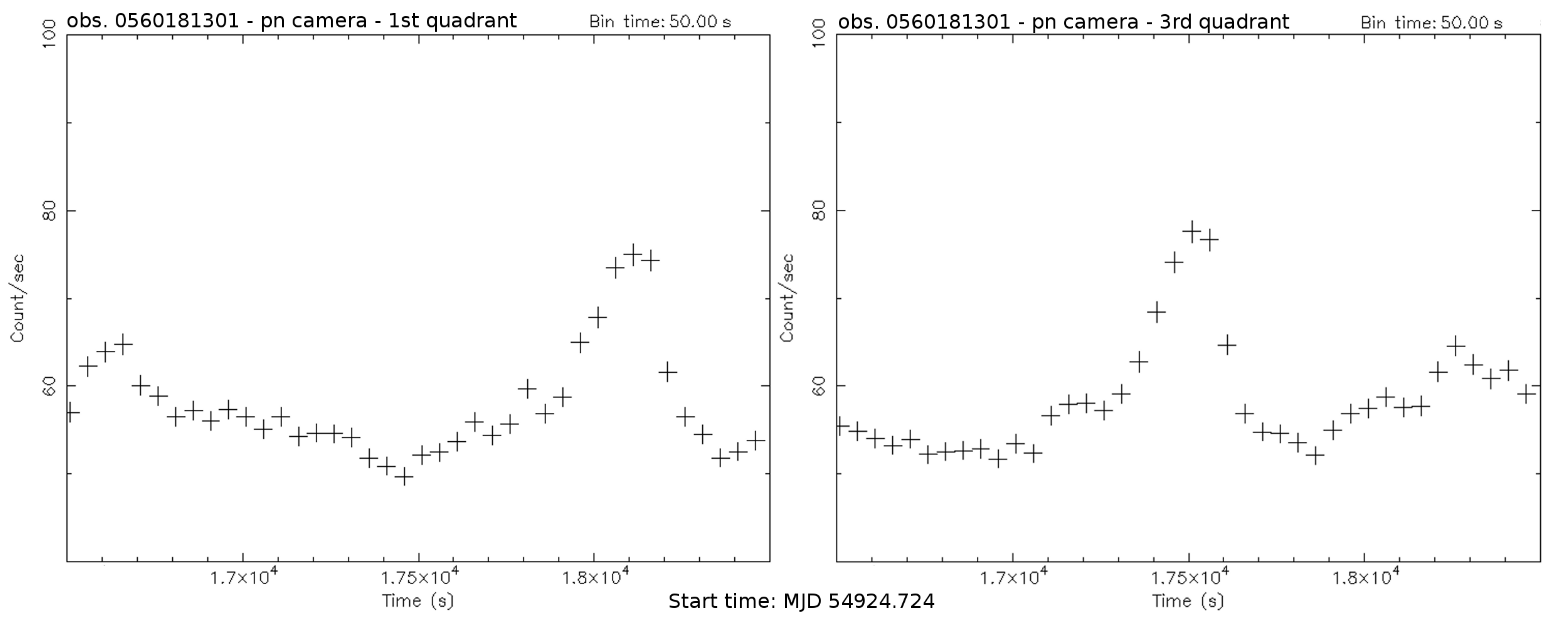}
\caption{Comparison between two light curves of different pn quadrants ({\it left panel:} first,$^{}$ {\it right panel:} third$^{}$) of the same observation (obs.id 0560181301),
obtained from the PPS event file.} 
\label{fig:cmlightcurve}
\end{figure*}

Good time intervals in PPS and reprocessed files can be different, with reprocessed files listing
additional GTI and events after the last ones in PPS (see Fig. \ref{fig:cmimage}). Moreover, in PPS event files we found time intervals that were included in GTIs with no recorded photons, with durations up to a kilosecond.
Finally, the background light curves taken from different CCDs (quadrants for pn) from the PPS event file
can display a progressive shift of flares with time (see Fig. \ref{fig:cmlightcurve}) up to few kiloseconds at the end of long exposures.
These differences are only seen in event files that are affected by high-background time intervals.
When the telemetry rate is exceeded by the data rate in one CCD (or quadrant in the case of the pn), the so-called counting mode is triggered and that CCD (quadrant) stops recording individual events for a few time intervals. Only the number of dropped events is then transmitted.
It is apparent that the software used to produce PPS can make an incorrect reconstruction of time-of-arrival of events after counting-mode occurrences. 
Then, events with incorrect time of arrivals that fall outside GTIs are deleted.

This software problem has a potentially strong effect on the timing analysis of {\it XMM-Newton} data, which is the main focus of EXTraS.
Analysis of aperiodic variability is mostly impacted by an incorrect characterisation and subtraction of the background. The search for pulsations can be hampered by
 incorrect times of arrival. The absolute time at which new X-ray transients are detected can be incorrect.\\

We evaluated the number of exposures that are affected by this problem. Starting from PPS event files, we produced light curves with 50 s time bins for each CCD
of MOSs and with 20 s time bins for each quadrant of pn. We selected these time bins in order to have at least 25 counts s$^{-1}$ per bin from the
quiescent background, and thus a lower than a 5$\sigma$ probability of zero-count bins. When the flaring background and the celestial sources
that are not subtracted in this exercise are also taken into account, the probability of a zero-count bin is negligible. Thus, we produced new bad time intervals from time bins with zero counts. These were compared with the GTIs
reported in the PPS event files. We obtained differences for at least one quadrant in 27\% of pn exposures and 3\% of MOSs exposures. For pn, time shifts
of $>$100 s (up to few kiloseconds) at the end of the observation are registered for at least one quadrant in 15\% of exposures. These figures were computed for the PPS archive at the epoch of the EXTraS project. As already stated, the current archive is free from this problem.

The summary of changes from SAS v13.0 to 13.5 reports a change in the FIFO resets that would cause an underestimated deadtime due to FIFO losses and resets in the {\tt epframes} package, which is part of the pre-processing pipeline. This can explain the incorrect time reconstruction for SAS versions before 13.5. 

\section{Catalogue of new transients}
\label{appendix:transients}
We give in Table~\ref{table:translist} the full list of the 136 new transients discovered by the dedicated analysis described in Section~\ref{sec:transients}.

\clearpage
\onecolumn

\begin{longtable}{l|l|l|l|l|l|l}
\caption{\label{table:translist}List of 136 new transient sources discovered by the dedicated analysis described in Sect.~\ref{sec:transients}. Basic properties of each source are shown: the EXTraS transient name, the XMM Observation ID, Galactic longitude (l) and latitude (b) in degrees, overall uncertainty on the position (arcsec), transient duration (s), and EPIC counts (0.1-12 keV). Sources are sorted by increasing duration of the transient. See Sect.~\ref{sec:transients} for more details. A full version of the catalogue is available online at \url{https://www88.lamp.le.ac.uk/extras/archive}. }\\
\hline
Transient & Observation & l & b & Error & Duration & EPIC counts \\
\hline
ID & ID  & (degrees) & (degrees) & (arcsec) & (s) & \\
\hline
\endfirsthead
\multicolumn{7}{c}%
{\tablename\ \thetable\ -- \textit{Continued from previous page}} \\
\hline
\hline
Transient & Observation & l & b & Error & Duration & EPIC counts \\
\hline
ID & ID  & (degrees) & (degrees) & (arcsec) & (s) & \\
\hline
\endhead
\hline \multicolumn{7}{r}{\textit{Continued on next page}} \\
\endfoot
\hline
\endlastfoot
   EXMM J023135.0-603743 & 0675010401 & 283.0184 & -52.4543 & 1.4 & 315.1         &        54$\pm$9\\
  EXMM J083215.8-452454 & 0672040101 & 263.4212 & -3.3647 & 2.7 & 322.5         &        18$\pm$5\\
  EXMM J061723.5+225537 & 0600110101 & 188.7580 & 3.2281 & 1.5 & 359.3  &        38$\pm$7\\
  EXMM J070900.0-492415 & 0653510501 & 260.0442 & -17.5718 & 2.3 & 388.9         &        13$\pm$4\\
  EXMM J174033.7-310504 & 0301730101 & 357.5970 & -0.1976 & 1.8 & 431.3         &        21$\pm$5\\
  EXMM J215653.6-114708 & 0103860501 & 44.7440 & -46.3584 & 2.1 & 494.8         &        22$\pm$5\\
  EXMM J164340.4-542138 & 0603220201 & 332.7182 & -5.5218 & 1.5 & 600.2         &        45$\pm$8\\
  EXMM J003954.6+401810 & 0402560601 & 120.5529 & -22.5156 & 2.3 & 717.7         &        12$\pm$4\\
  EXMM J174535.5-285929 & 0674601101 & 359.9495 & -0.0234 & 1.0 & 741.1         &        76$\pm$11\\
  EXMM J161510.5-224401 & 0555650301 & 352.5542 & 19.9434 & 1.6 & 761.6         &        32$\pm$6\\
  EXMM J174628.4-290617 & 0202670701 & 359.9531 & -0.2471 & 1.5 & 784.4         &        42$\pm$7\\
  EXMM J171042.4-280452 & 0206990401 & 356.4635 & 6.8619 & 2.1 & 797.6  &        23$\pm$6\\
  EXMM J173613.1-353035 & 0606200101 & 353.3705 & -1.7951 & 1.3 & 808.6         &        38$\pm$7\\
  EXMM J181008.1-194543 & 0301270501 & 10.7321 & -0.2319 & 1.8 & 833.3  &        23$\pm$5\\
  EXMM J111245.2-603617 & 0051550101 & 291.1095 & -0.0183 & 1.3 & 856.6         &        60$\pm$9\\
  EXMM J203347.9+601124 & 0401360101 & 95.6707 & 11.8029 & 1.5 & 859.6  &        69$\pm$12\\
  EXMM J111653.2+440231 & 0651330301 & 165.0809 & 64.5643 & 1.5 & 903.1         &        38$\pm$7\\
  EXMM J104620.4+524822 & 0200480201 & 156.4753 & 55.3929 & 1.1 & 921.5         &        80$\pm$11\\
  EXMM J151033.9+333059 & 0303930101 & 53.5763 & 59.4535 & 1.1 & 922.2  &        61$\pm$12\\
  EXMM J163547.9-472914 & 0502140101 & 337.0444 & -0.0340 & 2.2 & 939.6         &        28$\pm$7\\
  EXMM J182806.2+063510 & 0201730301 & 36.1774 & 8.1730 & 1.1 & 997.1   &        89$\pm$11\\
  EXMM J182903.1+003008 & 0402820101 & 30.8220 & 5.2091 & 1.0 & 1000.0  &        59$\pm$9\\
  EXMM J174544.9-290504 & 0202670601 & 359.8882 & -0.1012 & 1.7 & 1000.0         &        85$\pm$11\\
  EXMM J162721.5-244146 & 0305541101 & 352.9841 & 16.5582 & 1.1 & 1000.0         &        66$\pm$10\\
  EXMM J162714.7-245135 & 0305540701 & 352.8380 & 16.4694 & 0.7 & 1000.0         &        253$\pm$18\\
  EXMM J154227.2-522431 & 0152780201 & 327.4138 & 2.1399 & 1.4 & 1000.0         &        52$\pm$11\\
  EXMM J141328.4-651755 & 0111240101 & 311.3692 & -3.7796 & 1.3 & 1000.0         &        72$\pm$9\\
  EXMM J031659.2-663214 & 0405090101 & 283.5192 & -44.7121 & 0.9 & 1000.0         &        116$\pm$11\\
  EXMM J180041.1-224343 & 0135742601 & 7.0675 & 0.2362 & 1.4 & 1064.7   &        62$\pm$10\\
  EXMM J092441.0-213122 & 0065940501 & 251.8552 & 20.2413 & 2.5 & 1134.9         &        42$\pm$8\\
  EXMM J183205.9-191433 & 0404720201 & 13.6242 & -4.5572 & 2.1 & 1138.2         &        23$\pm$5\\
  EXMM J212805.1-651052 & 0670380101 & 328.2836 & -40.5255 & 1.0 & 1142.5         &        99$\pm$12\\
  EXMM J103154.4-142301 & 0203770101 & 259.4147 & 36.4357 & 1.9 & 1143.0         &        28$\pm$7\\
  EXMM J002115.2+592518 & 0693390101 & 119.0952 & -3.2250 & 1.2 & 1150.1         &        74$\pm$10\\
  EXMM J124840.7-055437 & 0153450101 & 301.6735 & 56.9547 & 1.3 & 1203.8         &        58$\pm$9\\
  EXMM J203222.9+414045 & 0305560201 & 80.4210 & 1.1353 & 1.9 & 1258.1  &        36$\pm$8\\
  EXMM J111939.8-611834 & 0672790201 & 292.1454 & -0.3745 & 1.1 & 1292.5         &        94$\pm$11\\
  EXMM J025737.3+132247 & 0112260201 & 164.0056 & -39.2247 & 1.9 & 1328.5         &        27$\pm$6\\
  EXMM J232545.2+613150 & 0404720301 & 112.9056 & 0.3380 & 1.7 & 1362.0         &        28$\pm$6\\
  EXMM J174553.3-290445 & 0604300801 & 359.9087 & -0.1246 & 0.7 & 1396.0         &        243$\pm$20\\
  EXMM J224259.1+530613 & 0654030101 & 104.2452 & -5.0432 & 0.6 & 1419.1         &        238$\pm$19\\
  EXMM J213452.0+473048 & 0650591701 & 92.1561 & -3.2663 & 1.9 & 1428.4         &        41$\pm$7\\
  EXMM J080344.1-400619 & 0159360501 & 256.0783 & -4.7350 & 1.4 & 1433.2         &        48$\pm$7\\
  EXMM J180452.2-274315 & 0305970101 & 3.1866 & -3.0451 & 0.7 & 1435.7  &        185 $\pm$15\\
  EXMM J084839.3-453548 & 0159760301 & 265.3207 & -1.2019 & 1.3 & 1486.7         &        118$\pm$18\\
  EXMM J063553.4+054141 & 0146870401 & 206.0927 & -0.8473 & 1.0 & 1497.9         &        97$\pm$14\\
  EXMM J104439.0-593700 & 0112560201 & 287.5195 & -0.5937 & 1.4 & 1515.9         &        61$\pm$10\\
  EXMM J230201.7+584917 & 0057540301 & 109.1685 & -1.0956 & 1.4 & 1520.5         &        59$\pm$11\\
  EXMM J171420.6-381830 & 0670330101 & 348.6015 & 0.2541 & 0.7 & 1536.9         &        218$\pm$19\\
  EXMM J141157.0-651343 & 0111240101 & 311.2392 & -3.6635 & 1.4 & 1622.7         &        55$\pm$8\\
  EXMM J165415.0-415314 & 0109490401 & 343.4259 & 1.1329 & 1.5 & 1724.0         &        64$\pm$10\\
  EXMM J164709.7-455034 & 0505290201 & 339.5665 & -0.4111 & 1.0 & 1724.4         &        113$\pm$13\\
  EXMM J131233.2-624631 & 0510980101 & 305.3441 & -0.0024 & 1.3 & 1863.9         &        61$\pm$9\\
  EXMM J161132.6-603430 & 0550451101 & 325.2074 & -6.6767 & 1.9 & 1874.0         &        39$\pm$8\\
  EXMM J224401.5+531513 & 0654030101 & 104.4542 & -4.9850 & 1.7 & 1907.4         &        36$\pm$7\\
  EXMM J144350.6-621945 & 0504810301 & 315.5771 & -2.2495 & 1.2 & 1923.5         &        95$\pm$12\\
  EXMM J065442.8-240004 & 0652250601 & 234.8727 & -10.0120 & 0.7 & 1929.5         &        195$\pm$16\\
  EXMM J170213.4-295801 & 0205580201 & 353.8250 & 7.2331 & 1.0 & 1940.3         &        96$\pm$13\\
  EXMM J022701.8-053144 & 0404964801 & 173.7029 & -58.6311 & 2.8 & 1951.0         &        23$\pm$6\\
  EXMM J191119.6+045739 & 0694870201 & 39.6172 & -2.1445 & 1.9 & 1972.5         &        25$\pm$6\\
  EXMM J215645.3-074944 & 0404910701 & 49.6538 & -44.4155 & 2.1 & 1973.8         &        40$\pm$8\\
  EXMM J181836.6-134818 & 0605130101 & 16.9337 & 0.8349 & 1.2 & 2000.0  &        69$\pm$12\\
  EXMM J164707.0-455158 & 0410580601 & 339.5437 & -0.4203 & 1.3 & 2000.0         &        88$\pm$14\\
  EXMM J053546.1-051051 & 0134531701 & 208.8725 & -19.1798 & 0.7 & 2000.0         &        215$\pm$19\\
  EXMM J053521.8-055403 & 0112660101 & 209.5052 & -19.5948 & 0.8 & 2000.0         &        272$\pm$20\\
  EXMM J142517.6+225545 & 0143652301 & 26.9391 & 68.3630 & 1.0 & 2012.2         &        148$\pm$17\\
  EXMM J203304.8+410048 & 0165360101 & 79.9632 & 0.6342 & 1.6 & 2035.8  &        40$\pm$9\\
  EXMM J023126.0-712906 & 0510181701 & 292.2680 & -43.5134 & 1.1 & 2177.0         &        144$\pm$18\\
  EXMM J203254.4+410638 & 0505110401 & 80.0220 & 0.7184 & 0.6 & 2225.2  &        513$\pm$31\\
  EXMM J020825.7+352826 & 0084140101 & 140.1353 & -24.8127 & 1.6 & 2228.2         &        40$\pm$7\\
  EXMM J161753.9-505650 & 0113050701 & 332.5270 & -0.3394 & 1.2 & 2325.9         &        39$\pm$8\\
  EXMM J181243.3-104054 & 0500030101 & 19.0002 & 3.5845 & 1.1 & 2361.5  &        82$\pm$11\\
  EXMM J051723.2-685921 & 0113000501 & 279.6914 & -33.5711 & 2.2 & 2379.1         &        28$\pm$6\\
  EXMM J061751.0-325214 & 0092360101 & 240.1367 & -20.8940 & 1.2 & 2443.7         &        66$\pm$9\\
  EXMM J104421.4-593453 & 0112560201 & 287.4701 & -0.5800 & 1.2 & 2474.6         &        89$\pm$13\\
  EXMM J203400.9+412801 & 0505110401 & 80.4322 & 0.7634 & 1.4 & 2493.1  &        45$\pm$8\\
  EXMM J042225.1+281148 & 0101440701 & 169.5137 & -15.0331 & 1.5 & 2503.4         &        56$\pm$10\\
  EXMM J170208.5-485246 & 0204730301 & 338.7994 & -4.2920 & 1.1 & 2545.8         &        74$\pm$13\\
  EXMM J230219.5+583338 & 0057540101 & 109.0973 & -1.3493 & 1.5 & 2603.8         &        51$\pm$9\\
  EXMM J151552.5+561021 & 0673920301 & 91.3784 & 51.1752 & 1.7 & 2615.8         &        31$\pm$7\\
  EXMM J183630.3-064816 & 0503320601 & 25.1775 & 0.2015 & 1.6 & 2723.9  &        41$\pm$8\\
  EXMM J180614.4-212650 & 0673690101 & 8.8161 & -0.2533 & 2.0 & 2763.5  &        34$\pm$8\\
  EXMM J150230.1-413335 & 0555630301 & 327.7098 & 14.9210 & 1.8 & 2804.4         &        57$\pm$11\\
  EXMM J053219.8-072932 & 0690200201 & 210.6656 & -20.9786 & 0.8 & 2813.7         &        142$\pm$16\\
  EXMM J190757.0-205142 & 0671850301 & 15.8578 & -12.8410 & 0.7 & 2818.3         &        263$\pm$22\\
  EXMM J070206.1-111429 & 0654880301 & 224.1490 & -2.7677 & 1.9 & 2902.3         &        54$\pm$9\\
  EXMM J170759.6-410042 & 0406580101 & 345.6993 & -0.3605 & 1.6 & 2962.7         &        30$\pm$7\\
  EXMM J104450.1-594208 & 0160160901 & 287.5800 & -0.6585 & 1.1 & 3000.0         &        39$\pm$8\\
  EXMM J203317.5+411303 & 0200450201 & 80.1510 & 0.7238 & 1.6 & 3016.7  &        52$\pm$10\\
  EXMM J201744.3+372759 & 0670480401 & 75.3524 & 1.0094 & 1.6 & 3030.8  &        42$\pm$8\\
  EXMM J180152.2-231706 & 0145970401 & 6.7192 & -0.2759 & 1.3 & 3075.0  &        40$\pm$9\\
  EXMM J171924.9+264033 & 0500670201 & 49.2190 & 31.0319 & 1.6 & 3100.6         &        52$\pm$12\\
  EXMM J072837.7+674629 & 0302400301 & 148.0559 & 28.4566 & 1.5 & 3106.6         &        41$\pm$8\\
  EXMM J083916.3-454613 & 0603510701 & 264.4420 & -2.5964 & 1.3 & 3206.9         &        65$\pm$11\\
  EXMM J175954.5-240928 & 0503850101 & 5.7387 & -0.3190 & 1.4 & 3239.6  &        49$\pm$10\\
  EXMM J070238.9-114145 & 0654880401 & 224.6152 & -2.8557 & 2.0 & 3250.2         &        32$\pm$8\\
  EXMM J151819.8-615757 & 0555690901 & 319.2918 & -3.8536 & 0.6 & 3252.3         &        381$\pm$31\\
  EXMM J173602.1-444555 & 0146420101 & 345.5071 & -6.7156 & 1.2 & 3290.6         &        70$\pm$11\\
  EXMM J083833.5-355215 & 0303230301 & 256.4936 & 3.3111 & 2.0 & 3301.4         &        36$\pm$8\\
  EXMM J113835.0+170650 & 0066950201 & 239.9104 & 70.4156 & 1.3 & 3410.4         &        55$\pm$10\\
  EXMM J235822.8+563209 & 0553510301 & 115.6285 & -5.5817 & 0.8 & 3507.2         &        183$\pm$18\\
  EXMM J004449.9+415244 & 0109270301 & 121.6149 & -20.9759 & 1.0 & 3531.4         &        114$\pm$14\\
  EXMM J113622.2-613751 & 0201160401 & 294.1390 & -0.0474 & 1.2 & 3615.3         &        74$\pm$12\\
  EXMM J082521.5+261559 & 0603500301 & 197.2481 & 31.3173 & 1.0 & 3641.7         &        69$\pm$10\\
  EXMM J184100.9-053819 & 0604820301 & 26.7275 & -0.2600 & 1.6 & 3740.9         &        51$\pm$9\\
  EXMM J104520.8-593254 & 0311990101 & 287.5656 & -0.4922 & 1.7 & 3760.1         &        36$\pm$8\\
  EXMM J035849.7+541255 & 0112200301 & 148.1860 & 0.8005 & 2.3 & 3815.0         &        29$\pm$6\\
  EXMM J174617.8-291150 & 0505670101 & 359.8540 & -0.2621 & 1.4 & 3977.8         &        62$\pm$12\\
  EXMM J053508.2+095532 & 0402050101 & 195.0594 & -11.9997 & 0.6 & 4000.0         &        554$\pm$37\\
  EXMM J100422.2-701215 & 0099020301 & 289.6087 & -11.7843 & 1.4 & 4030.8         &        52$\pm$8\\
  EXMM J173432.0-255552 & 0202680101 & 1.2433 & 3.6790 & 0.9 & 4037.2   &        163$\pm$18\\
  EXMM J174537.2-285500 & 0506291201 & 0.0167 & 0.0102 & 1.3 & 4128.7   &        55$\pm$ 9\\
  EXMM J203138.2+413027 & 0305560201 & 80.2001 & 1.1458 & 1.3 & 4131.8  &        59$\pm$10\\
  EXMM J203352.8+412516 & 0505110301 & 80.3804 & 0.7564 & 0.9 & 4174.3  &        144$\pm$18\\
  EXMM J165430.4-415455 & 0109490601 & 343.4343 & 1.0781 & 1.2 & 4193.9         &        57$\pm$10\\
  EXMM J132724.8-620703 & 0036140201 & 307.1246 & 0.4610 & 2.1 & 4215.2         &        23$\pm$6\\
  EXMM J104435.8-593120 & 0112580601 & 287.4694 & -0.5135 & 1.3 & 4275.8         &        66$\pm$11\\
  EXMM J004322.3+413432 & 0690600401 & 121.3132 & -21.2708 & 1.5 & 4313.4         &        53$\pm$10\\
  EXMM J045638.4+302913 & 0671960101 & 172.6676 & -7.8720 & 1.8 & 4439.7         &        30$\pm$7\\
  EXMM J162705.9-244015 & 0305540601 & 352.9629 & 16.6193 & 1.2 & 4538.7         &        60$\pm$11\\
  EXMM J180542.6-211847 & 0405750201 & 8.8730 & -0.0799 & 1.4 & 4622.2  &        69$\pm$12\\
  EXMM J111844.6-612232 & 0150790101 & 292.0649 & -0.4749 & 1.4 & 4647.0         &        66$\pm$12\\
  EXMM J103528.6+631021 & 0403760401 & 144.9605 & 47.7256 & 1.8 & 4660.3         &        157$\pm$22\\
  EXMM J070810.2-492944 & 0653510301 & 260.0817 & -17.7308 & 1.7 & 4677.2         &        45$\pm$8\\
  EXMM J172020.7-290720 & 0552002601 & 356.8339 & 4.5330 & 0.6 & 4745.0         &        319$\pm$23\\
  EXMM J165201.9-415313 & 0602020201 & 343.1648 & 1.4526 & 1.4 & 4746.8         &        51$\pm$9\\
  EXMM J111116.9-602649 & 0051550101 & 290.8826 & 0.0603 & 1.6 & 4761.5         &        68$\pm$11\\
  EXMM J144707.3-622053 & 0504810201 & 315.9137 & -2.4283 & 1.5 & 4799.5         &        57$\pm$10\\
  EXMM J001930.4+591440 & 0693390101 & 118.8525 & -3.3741 & 1.0 & 4809.4         &        120$\pm$14\\
  EXMM J084638.5-525906 & 0201910101 & 270.8901 & -6.0828 & 0.7 & 4861.8         &        223$\pm$20\\
  EXMM J203412.5+602046 & 0691570101 & 95.8314 & 11.8507 & 2.0 & 4879.6         &        48$\pm$8\\
  EXMM J203323.8+411847 & 0200450501 & 80.2396 & 0.7648 & 1.1 & 4916.7  &        61$\pm$12\\
  EXMM J191400.9+045016 & 0075140401 & 39.8175 & -2.7953 & 1.4 & 4942.2         &        55$\pm$11\\
  EXMM J214407.1+382511 & 0602310101 & 87.3095 & -11.1691 & 0.9 & 4985.2         &        96$\pm$13\\
  EXMM J053928.4-691943 & 0113020201 & 279.7217 & -31.5785 & 1.5 & 4993.5         &        37$\pm$8\\
  EXMM J162729.5-243917 & 0305540701 & 353.0372 & 16.5628 & 0.7 & 5000.0         &        151$\pm$14\\
\hline
\end{longtable}

\section{Energy conversion factors for pointed data for the LTV analysis}
  \label{appendix:LTVpointedECFs}

The ECFs used for the pointed data analysis for the LTV
catalogue, defined as $F_i=R_i/E_i,$ where $F_i$ is the flux, $R_i$ is the
count rate, and $E_i$ is the ECF (each in band i), are shown in
table~\ref{tab:ecfs}.

\begin{table}[ht]
\begin{minipage}[ht]{\columnwidth}
\normalsize
\caption{Energy conversion factors (in units of
  $10^{11}$~cts~cm$^2$~erg$^{-1}$) used to convert pointed data count rates into
  fluxes for each instrument, filter, and energy band.}
\label{tab:ecfs}
\small \centering \renewcommand{\footnoterule}{} \tabcolsep 0mm
\begin{tabular}{
l @{\extracolsep{2mm}} c @{\extracolsep{5mm}} c @{\extracolsep{6mm}} c
@{\extracolsep{6mm}} c } \hline \hline & & \multicolumn{3}{c}{Filters}
  \\ Camera & Band & \multicolumn{1}{c}{Thin} & \multicolumn{1}{c}{Medium} &
  \multicolumn{1}{c}{Thick} \\
\hline pn & 6 & 7.3868 & 7.030 & 5.4091 \\ & 7 & 1.1089 & 1.0992 & 1.0561 \\ &
8 & 3.3245 & 3.1924 & 2.5929 \\ MOS1 & 6 & 1.9237 & 1.8492 & 1.5293 \\ & 7 &
0.3745 & 0.3713 & 0.3585 \\ & 8 & 0.9232 & 0.8949 & 0.7736 \\ MOS2 & 6 &
1.9286 & 1.8536 & 1.5316 \\ & 7 & 0.381 & 0.3775 & 0.3644 \\ & 8 & 0.9292 &
0.9004 & 0.7782 \\ \hline
\end{tabular}
\end{minipage}
\normalsize
\end{table}

\end{appendix}

\end{document}